\RequirePackage{fix-cm}
\documentclass[twocolumn]{svjour3}          
\journalname{Astronomy Astrophysics Review}
\smartqed  
\usepackage{graphicx}
\usepackage{url}
\usepackage{txfonts}
\usepackage[round]{natbib}

\newcommand{\apj}{Astrophys J }
\newcommand{\apjl}{Astrophys J Lett }
\newcommand{\aj}{Astronom J }
\newcommand{\apjs}{Astrophys J Suppl }
\newcommand{\mnras}{MNRAS }
\newcommand{\aap}{Astron Astrophys }
\newcommand{\aaps}{Astron Astrophys Suppl }
\newcommand{\aapr}{Astron Astrophys Rev }
\newcommand{\nat}{Nature }
\newcommand{\nar}{New Astronom Rev }
\newcommand{\pasa}{Publications Astron Soc Australia }
\newcommand{\pasj}{Publications Astron Soc Japan }
\newcommand{\pasp}{Publications Astron Soc Pacific }
\newcommand{\araa}{Annu Rev Astron Astr }
\newdimen\digitwidth
\setbox0=\hbox{2}
\digitwidth=\wd0
\catcode `#=\active
\newcommand#{\kern\digitwidth}
\begin{document}

\title{Active Galactic Nuclei: what's in a name?
}


\author{P. Padovani \and D. M. Alexander \and R. J. Assef \and B. De Marco \and P. Giommi \and 
R. C.  Hickox \and G. T. Richards 
\and V. Smol{\v c}i{\'c} \and E. Hatziminaoglou \and V. Mainieri \and M. Salvato}


\institute{P. Padovani \and E. Hatziminaoglou \and V. Mainieri \at
              European Southern Observatory, Karl-Schwarzschild-Str. 2,
D-85748 Garching bei M\"unchen, Germany \\
              \email{ppadovan@eso.org}           
              \and 
              D. M. Alexander \at
              Centre for Extragalactic Astronomy, Department of Physics, Durham University, UK
              \and
              R. J. Assef \at 
              N\'ucleo de Astronom\'ia de la Facultad de Ingenier\'ia, 
              Universidad Diego Portales, Santiago, Chile
              \and
              B. De Marco  \at 
              Max Planck Institute for extraterrestrial Physics, Garching bei M\"unchen, Germany 
              \at 
              Nicolaus Copernicus Astronomical Center,  PL-00-716 Warsaw, Poland (current address)
              \and
              P. Giommi \at 
              Italian Space Agency, ASI, via del Politecnico snc, 00133 Roma, Italy 
              \and 
              R. C. Hickox \at 
              Department of Physics \& Astronomy, Dartmouth College, Hanover, NH, USA 
              \and 
              G. T. Richards \at
              Dept. of Physics, Drexel University, Philadelphia, PA, USA
              \and
              M. Salvato \at 
              Max Planck Institute for extraterrestrial Physics, Garching bei M\"unchen, Germany 
              \and
              V. Smol{\v c}i{\'c} \at 
              Department of Physics, Faculty of Science, University of Zagreb,  Bijeni\v{c}ka cesta 32, 10000  Zagreb, Croatia
}

\date{Received: June 12, 2017 / Accepted: July 21, 2017}

\maketitle

\begin{abstract}
Active Galactic Nuclei (AGN) are energetic astrophysical sources powered by accretion onto  supermassive black holes in galaxies, and present unique observational signatures 
that cover the full electromagnetic spectrum over more than twenty orders of magnitude in frequency. The rich phenomenology of AGN has resulted in a large number of different ``flavours'' in the literature that now comprise a complex and confusing AGN ``zoo''. It is increasingly clear that these classifications are only partially related
to intrinsic differences between AGN, and primarily reflect variations in a relatively small number of astrophysical 
parameters as well the method by which each class of AGN is selected. Taken together, observations in different electromagnetic bands as well as variations over time provide complementary windows on the physics of different 
sub-structures in the AGN. In this review, we present an overview of AGN multi-wavelength properties with the aim of painting their ``big picture'' through observations in each electromagnetic band from radio to $\gamma$-rays as well as AGN variability.
We address what we can learn from each observational method, the impact of selection effects, the physics 
behind the emission at each wavelength, and the potential for future studies. To conclude we use these observations to piece together the basic  
architecture of AGN, discuss our current understanding of unification models, and highlight some open questions that present opportunities for future observational and theoretical progress. 

\end{abstract}

\tableofcontents

\section{The Active Galactic Nuclei zoo}
\label{sec:intro}

The discovery of quasars \citep{Schmidt_1963} opened up a whole new branch of astronomy 
\citep[e.g.][for historical 
details]{Donofrio_2012,Kellermann_2015}. Twenty years earlier \cite{Seyfert_1943}
had reported the presence of broad and strong emission lines in the nuclei of six spiral nebulae 
(including some by now ``classical'' AGN, like NGC 1068 and NGC 4151).
However, his work remained largely ignored until \cite{Baade_1954} pointed out the 
similarities between 
the spectra of the galaxies studied by Seyfert and that of the galaxy they had associated with the 
Cygnus A radio source. 

As implicit in the name, AGN are stronger emitters than the nuclei of 
``normal'' galaxies. This ``extra'' component is 
unrelated to the nuclear fusion powering stars and is now universally accepted to be
connected instead to the presence of an
actively accreting central supermassive ($\gtrsim 10^6~M_{\odot}$) black hole (SMBH). 

AGN have {\it many} interesting properties. These include: 
(1) very high luminosities (up to $L_{\rm bol} \approx 10^{48}$ erg s$^{-1}$), which make
them the most powerful non-explosive sources in the Universe and therefore visible up to 
very high redshifts \citep[currently $z=7.1$:][]{Mortlock_2011}; (2) small emitting regions in most bands,
of the order of a milliparsec, as inferred from their rapid variability  \citep[e.g.][]{Ulrich_1997}, 
implying high energy densities; (3) strong evolution of their luminosity 
functions \citep[LFs; e.g.][]{Merloni_2013}; (4) detectable emission covering the whole 
electromagnetic spectrum (this review). 

\begin{figure*}
\centering
\hspace{-1.18truecm}
\includegraphics[width=13.0cm]{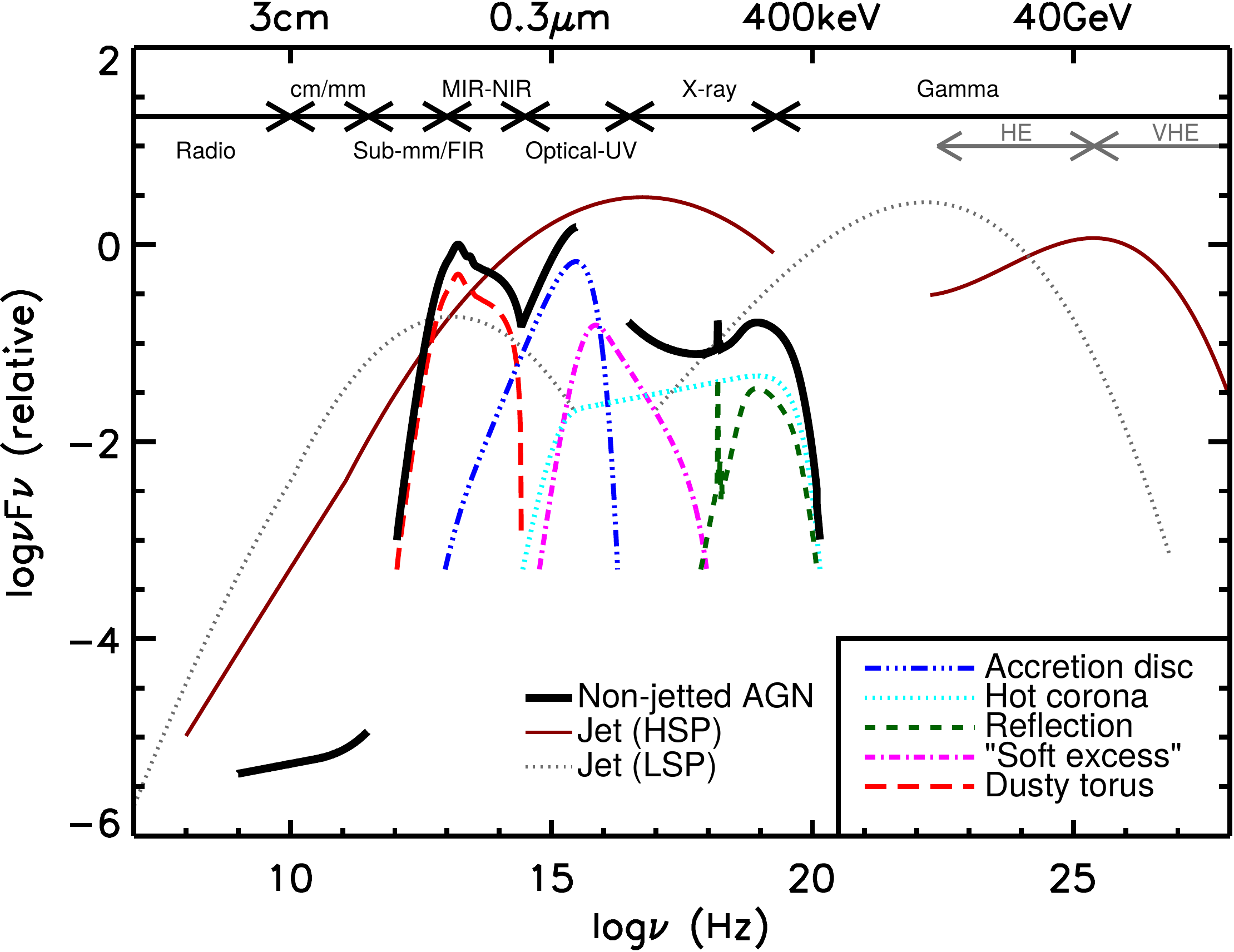}
\caption{A schematic representation of an AGN spectral energy distribution (SED), loosely based on the observed SEDs of non-jetted
quasars \citep[e.g.][]{Elvis_1994,Richards_2006a}. The black solid curve represents the total emission and the various coloured 
curves (shifted down for clarity) represent the individual components. The intrinsic shape of the 
SED in the mm-far infrared (FIR) regime 
is uncertain; however, it is widely believed to have a minimal contribution (to an overall galaxy SED) compared to 
star formation (SF), except in the most intrinsically luminous quasars and powerful jetted AGN. The primary emission 
from the AGN accretion disk peaks in the UV region. The jet SED is also shown for a high synchrotron peaked blazar (HSP, based
on the SED of Mrk 421) and a low synchrotron peaked blazar (LSP, based on the SED of 3C 454.3; see Sect. \ref{sec:gamma_AGN}). 
Adapted from \cite{Harrison_2014}. Image credit: C. M. Harrison.}
\label{fig:SED}       
\end{figure*}

The latter property means that AGN are being discovered 
in all spectral bands. Different methods are employed 
in different bands to identify these sources 
but, most importantly, the various wavelength regimes
provide different
windows on AGN physics. Namely, the infrared (IR) band is mostly
sensitive to obscuring material and dust, the optical/ultraviolet(UV) band is related to
emission from the accretion disk, while the X-ray band traces the emission
of a (putative) corona. $\gamma$-ray and (high flux density) radio samples,
on the other hand, preferentially select AGN emitting strong non-thermal
(jet [or associated lobe] related) radiation (see Fig. \ref{fig:SED})\footnote{The mm/sub--mm band 
is missing from this paper because it mostly probes molecular gas that resides 
in the AGN host galaxy. However, with the high resolution capabilities 
of the Atacama Large Millimeter/submillimeter Array (ALMA) we are starting  
to resolve the innermost parts of AGN down to parsec scales (Sect. \ref{sec:summary_future}).}. The surface densities of AGN also vary strongly across the
electromagnetic spectrum due to a complex mix of physical processes, selection effects,
and technological limitations (as shown in Sect. \ref{sec:discussion}). 

\begin{table*}
\begin{center}
\caption{The AGN zoo: list of AGN classes.}
\begin{tabular}{lll}
\hline 
Class/Acronym & Meaning & Main properties/reference\\
\hline 
Quasar &Quasi-stellar radio source (originally) & Radio detection no longer required \\
Sey1 & Seyfert 1 & FWHM $\gtrsim 1,000$ km s$^{-1}$ \\
Sey2 & Seyfert 2 & FWHM $\lesssim 1,000$ km s$^{-1}$  \\
QSO &Quasi-stellar object & Quasar-like, non-radio source \\
QSO2 &Quasi-stellar object 2 & High power Sey2\\
RQ AGN & Radio-quiet AGN & see ref. 1\\
RL AGN & Radio-loud AGN & see ref. 1\\
Jetted AGN & & with strong relativistic jets;  see ref. 1\\
Non-jetted AGN & & without strong relativistic jets;  see ref. 1\\
Type 1 & & Sey1 and quasars\\
Type 2 & & Sey2 and QSO2\\
FR I &Fanaroff-Riley class I radio source & radio core-brightened (ref. 2)\\ 
FR II &Fanaroff-Riley class II radio source & radio edge-brightened (ref. 2)\\ 
BL Lac & BL Lacertae object & see ref. 3\\
Blazar & BL Lac and quasar & BL Lacs and FSRQs\\
\hline 
BAL & Broad absorption line (quasar) & ref. 4\\
BLO &Broad-line object & FWHM $\gtrsim 1,000$ km s$^{-1}$   \\
BLAGN &Broad-line AGN & FWHM $\gtrsim 1,000$ km s$^{-1}$  \\
BLRG &Broad-line radio galaxy & RL Sey1 \\
CDQ & Core-dominated quasar & RL AGN, $f_{\rm core} \ge f_{\rm ext}$ (same as FSRQ)\\
CSS &Compact steep spectrum radio source & core dominated, $\alpha_{\rm r} > 0.5$ \\
CT & Compton-thick & $N_{\rm H} \ge 1.5 \times 10^{24}$ cm$^{-2}$  \\
FR 0 &Fanaroff-Riley class 0 radio source & ref. 5 \\
FSRQ &Flat-spectrum radio quasar & RL AGN, $\alpha_{\rm r} \le 0.5$  \\
GPS &Gigahertz-peaked radio source & see ref. 6\\
HBL/HSP & High-energy cutoff BL Lac/blazar &  $\nu_{\rm synch~peak} \ge 10^{15}$ Hz  
(ref. 7)\\
HEG &High-excitation galaxy & ref. 8\\ 
HPQ & High polarization quasar & $P_{\rm opt} \ge 3 \%$ (same as FSRQ)\\
Jet-mode & & $L_{\rm kin} \gg L_{\rm rad}$ (same as LERG); see ref. 9\\
IBL/ISP & Intermediate-energy cutoff BL Lac/blazar & $10^{14} \le \nu_{\rm synch~peak} \le 10^{15}$ Hz  
(ref. 7)\\
LINER & Low-ionization nuclear emission-line regions &  see ref. 9\\
LLAGN & Low-luminosity AGN & see ref. 10\\
LBL/LSP & Low-energy cutoff BL Lac/blazar & $\nu_{\rm synch~peak} < 10^{14}$ Hz  
(ref. 7)\\
LDQ & Lobe-dominated quasar & RL AGN, $f_{\rm core} < f_{\rm ext}$\\
LEG &Low-excitation galaxy & ref. 8\\
LPQ & Low polarization quasar & $P_{\rm opt} < 3 \%$ \\
NLAGN &Narrow-line AGN & FWHM $\lesssim 1,000$ km s$^{-1}$   \\
NLRG &Narrow-line radio galaxy & RL Sey2 \\
NLS1 & Narrow-line Seyfert 1 & ref. 11 \\
OVV &Optically violently variable (quasar) & (same as FSRQ)\\
Population A & & ref. 12\\
Population B & & ref. 12\\
Radiative-mode & & Seyferts and quasars; see ref. 9\\
RBL & Radio-selected BL Lac & BL Lac selected in the radio band \\
Sey1.5 & Seyfert 1.5 & ref. 13 \\
Sey1.8 & Seyfert 1.8 &  ref. 13\\
Sey1.9 & Seyfert 1.9 &  ref. 13\\ 
SSRQ &Steep-spectrum radio quasar & RL AGN, $\alpha_{\rm r} > 0.5$   \\
USS & Ultra-steep spectrum source & RL AGN, $\alpha_{\rm r} > 1.0$ \\
XBL & X-ray-selected BL Lac & BL Lac selected in the X-ray band\\
XBONG & X-ray bright optically normal galaxy & AGN only in the X-ray band/weak lined AGN\\
\hline 
\multicolumn{3}{l}{\footnotesize The top part of the table relates to major/classical classes. The last column describes the 
main properties.}\\
\multicolumn{3}{l}{\footnotesize When these are too complex, it gives a reference to the first paper, which defined the 
relevant class or, when}\\ 
\multicolumn{3}{l}{\footnotesize  preceded by ``see'', a recent paper, which gives up-to-date details on it. 
Reference key: 1. \cite{Padovani_2016};}\\
\multicolumn{3}{l}{\footnotesize 2. \cite{Fanaroff_1974}; 3. \cite{bsv1}; 4. \cite{Weymann_1981};}\\ 
\multicolumn{3}{l}{\footnotesize 5. \cite{Ghisellini_2010}; 6. 
\cite{Odea_1991}; 7. \cite{Padovani_1995};}\\ 
\multicolumn{3}{l}{\footnotesize 8. \cite{Laing_1994}; 9. \cite{Heckman_2014};  10. \cite{Ho_2008}; 
11. \cite{Osterbrock_1985}; }\\
\multicolumn{3}{l}{\footnotesize 12. \cite{Sulentic_2002}; 13. \cite{Osterbrock_1981}}\\
\end{tabular}
\label{tab:agn_zoo}
\end{center}
\end{table*}

The past years have seen a proliferation of AGN classes, which outsiders to the field (but insiders
as well!) find mesmerising. A (possibly incomplete) list is 
given in Tab. \ref{tab:agn_zoo}, which gives the class or acronym in col. (1),
its meaning in col. (2), and the main properties or a reference to a relevant paper in col. (3).

Reality is much simpler, however, as we know that most of these seemingly different classes 
are due to changes in only a small number of parameters, namely: orientation 
\citep[e.g.][]{Antonucci_1993,Urry_1995,Netzer_2015}, accretion rate \citep[e.g.][]{Heckman_2014}, 
the presence (or absence) of strong jets \citep[e.g.][]{Padovani_2016}, and possibly the host galaxy and the 
environment. Sorting out these issues is a pre-requisite to understand AGN physics and the role AGN play in 
galaxy evolution \citep[e.g.][]{alexander2012}. 

To go beyond taxonomy and paint the AGN ``big picture'', which
comes out of multi-wavelength surveys, and understand the truly
intrinsic and fundamental properties of AGN,  
the workshop ``Active Galactic Nuclei: what's in a name?'' 
was organised at ESO, Garching, between June 27 and July 1, 2016. 
This was done by discussing AGN
selection and physics in {\it all} bands and by addressing: 

\begin{itemize}
\item the different types of AGN selected in the various spectral bands;
\item the similarities and differences they display; 
\item the impact of selection effects on the interpretation of the
results;
\item the physical mechanism(s) behind emission in a given band;
\item the effective range of black hole (BH) mass ($M_{\rm BH}$) and Eddington 
ratios\footnote{The ratio between the observed
  luminosity and the Eddington luminosity, $L_{\rm Edd} = 1.3 \times
  10^{46}~(M/10^8 \rm M_{\odot})$ erg/s, where $\rm M_{\odot}$ is one solar
  mass. This is the maximum isotropic luminosity a body can achieve when there is
  balance between radiation pressure (on the electrons) and gravitational
  force (on the protons).} ($L/L_{\rm 
Edd}$) probed by 
each selection method; 
\item the possible limitations of current
observations and/or facilities.
\end{itemize}

The workshop consisted of seven different sessions: radio, IR, optical, X-ray, $\gamma$-ray, 
variability, and multi-frequency. All of the sessions (with the exception of the multi-frequency one) 
were introduced by a review talk which 
set the scene, followed by 
contributed talks, for a total of eighty-six speakers, $48\%$ of whom were women. Sixty-seven posters completed the programme. 
A summary talk and a discussion session were held at the end of the workshop\footnote{Most 
presentations and posters can be found at \url{www.eso.org/sci/meetings/2016/AGN2016.html}.}. 
The workshop was very well attended, with 165 participants, covering five continents and thirty-one 
different countries; sixty of the participants were students. This review was inspired by the
workshop. 

In this paper we review our progress in addressing these key issues by discussing radio-, IR-, 
optical-, X-ray-, $\gamma$-ray-, and variability-selected AGN (Sect. 2 -- 7). Section \ref{sec:discussion}
summarizes our understanding of AGN and examines some open issues. 
Throughout this paper, spectral indices are defined by $S_{\nu} \propto \nu^{-\alpha}$
and the values $H_0 = 70$ km s$^{-1}$ Mpc$^{-1}$, $\rm \Omega_{\rm m} = 0.27$, 
and $\rm \Omega_{\rm \Lambda} = 0.73$ have been used. The acronyms of the AGN classes mentioned in
this paper are defined in Tab. \ref{tab:agn_zoo}. 

\section{Radio-selected AGN}\label{sec:radio}

In this section we describe radio-selected AGN. The types of, and physical mechanisms powering, radio-AGN are  outlined in Sect.~\ref{sec:radio_class}. Selection effects are discussed in Sect.~\ref{sec:radio_bias}, while the evolution of radio-AGN and future prospects are addressed in Sect.~\ref{sec:radio_evolv} and \ref{sec:radio_future}, respectively. 
Typical observing frequencies in the radio regime range from about 10~MHz ($\lambda \sim 30$~m) up to a few tens of GHz (e.g. 30~GHz, $\lambda = 1$~cm). 
For other recent reviews of radio-selected AGN we refer to \citet{Heckman_2014,Padovani_2016,Tadhunter_2016,Smolcic_2016a}. 

\subsection{Physical mechanism(s) and types of AGN selected in the radio}
\label{sec:radio_class}

The dominant emission process in the radio band is synchrotron emission, i.e. radiation by charged particles gyrating at relativistic velocities through magnetic fields\footnote{Free-free emission originating in H II regions may substantially contribute to the overall radio spectrum of SFGs. This, however, is expected to occur at rest-frame frequencies higher than $\sim20$~GHz (e.g.\ Fig.~1 in \citealt{Condon_1992}). The contribution of this emission process to the overall radio spectrum is taken to be negligible for galaxies dominated in the radio regime by an AGN.}. 
Being non-thermal in origin, this emission is usually parametrized by a power law of the form $S_\mathrm{\nu}\propto\nu^{-\alpha}$ where $S_\mathrm{\nu}$ is the flux density [expressed in Jy, mJy, etc.] at frequency $\nu$, and $\alpha$ is the spectral index.
Supernova remnants and processes related to the central SMBH are the main sources of synchrotron radiation in galaxies, resulting in two dominant galaxy populations identified in extragalactic radio continuum surveys, namely star forming galaxies (SFGs) and AGN (e.g. \citealt{Miley_1980}; \citealt{Condon_1992}). The former, being intrinsically weaker radio sources, become more prominent at the faintest radio flux densities (e.g. \citealt{Wilman_2008,Padovani_2015,Padovani_2016,Smolcic_2017a}). We outline here the main classes of radio emitting AGN. 

\begin{figure}[h!]
\centering
\includegraphics[bb=0 0 489 520, scale=0.4]{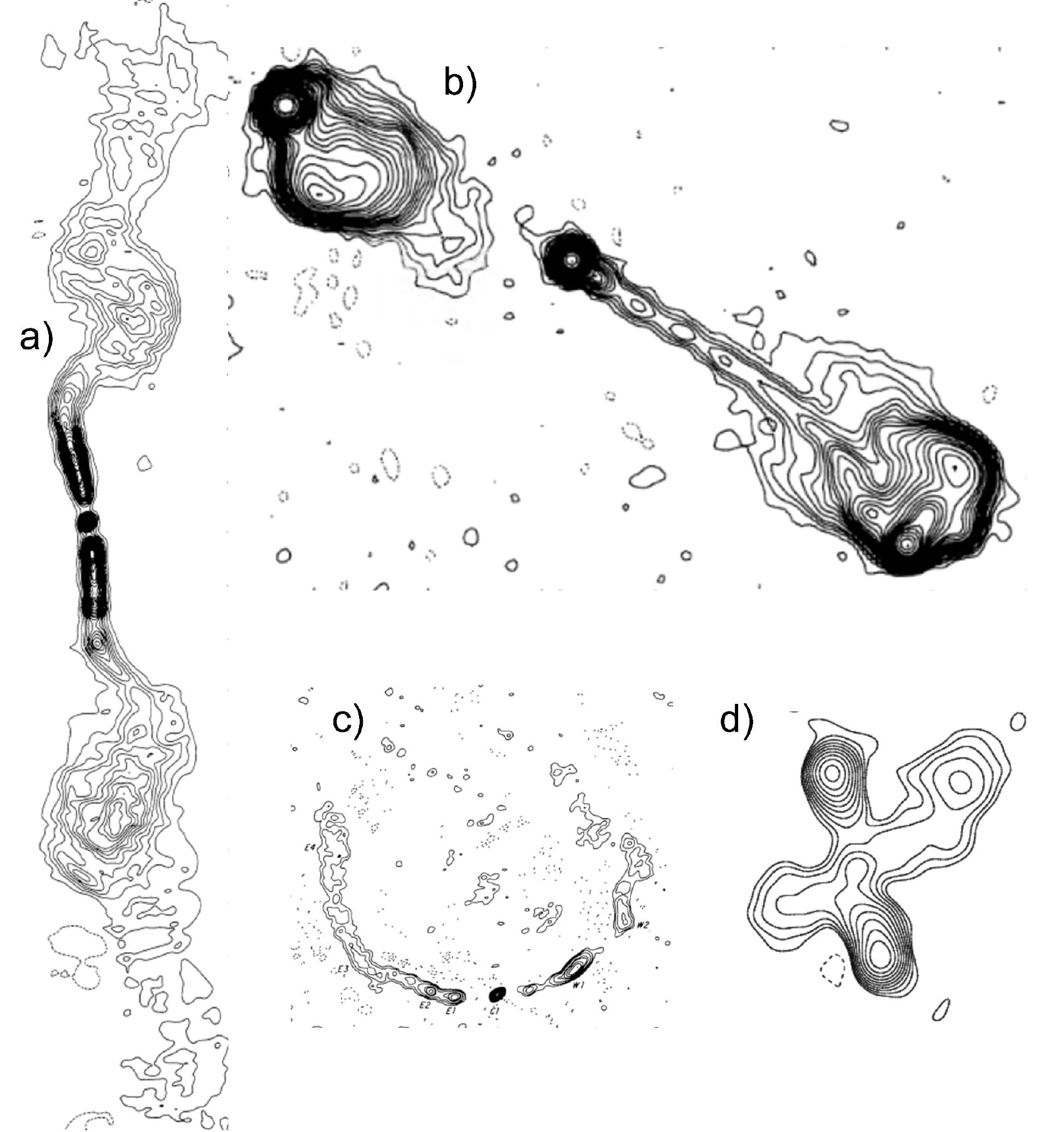}
\caption{Examples of FR~I (panel 'a'; 3C~449; \citealt{Perley_1979}), FR~II (panel 'b'; 3C~175; \citealt{Bridle_1994}), wide-angle tail (panel 'c'; 3C83.1B; \citealt{Owen_1978}), and X-shaped (panel 'd'; 3C~223.1; \citealt{Dennett_2002}) RGs.}
\label{fig:RGs}       
\end{figure}

\subsubsection{Radio galaxies}
\label{sec:RGs}

\citet{Fanaroff_1974} differentiated radio galaxies (RGs) into type I and II based on the ratio of the separation of the highest surface brightness regions  on opposite sides of the central galaxy, and 
the extent of the source measured from the lowest surface brightness contour\footnote{Any compact component coincident with the central galaxy was not taken into account.} (see Fig.~\ref{fig:RGs}). These classes are commonly referred to in the literature as FR~I (ratio $<0.5$) and FR~II (ratio $>0.5$) RGs, where the first (second) are often described as core- (edge-) brightened.  They also separate in the radio luminosity vs.\ optical host galaxy luminosity plane (\citealt{Ledlow_1996}; see also Fig.~11 in \citealt{Buttiglione_2010}). Recently a third class, FR~0, was suggested by \citet[][see also \citealt{Baldi_2010}]{Baldi_2015}, which includes 
 RGs sharing the properties of FR~Is but lacking prominent extended radio emission, being a factor of $\sim30$ more core-dominated. Further morphological classes often found in the literature refer to bent or warped appearances of the RGs (e.g. wide-angle tail, narrow-angle tail, X-shaped RGs; see Fig.~\ref{fig:RGs}; see also \citealt{Miley_1980} for a review, and Figs.~6 and 7 therein). 

\subsubsection{Flat- and steep-spectrum sources}
\label{sec:FS_SS}

When multiple radio continuum frequency observations are available, radio sources are usually separated into two main groups of steep ($\alpha\geq0.5$) and flat ($\alpha<0.5$) radio spectrum sources (e.g. \citealt{Wall_1975, Peacock_1981, Willott_2001, Kimball_2008}), which roughly correspond, with some exceptions, to extended and compact sources.  
They can further be classified as, e.g. USS (e.g. \citealt{DeBreuck_2001}), GPS, and CSS sources (see \citealt{Sadler_2016} and \citealt{O'Dea_1998} for reviews). FSRQs and BL Lacs, both of which are flat spectrum sources, make up the blazar class, which includes 
AGN hosting jets oriented at a very small angle ($\lesssim 15 - 20^{\circ}$) with respect to the line of sight (e.g. \citealt{Urry_1995,Giommi_2013}; see also Sect. \ref{sec:gamma}).
The two sub-classes main difference lies in their optical spectra, with FSRQs displaying strong,
broad emission lines just like standard quasars, and BL Lacs instead showing at most weak emission lines, sometimes exhibiting absorption features, and in many cases being completely featureless. 
Blazars dominate the bright radio \citep[e.g.][]{Padovani_2016} and the $\gamma$-ray sky (Sect. \ref{sec:gamma_AGN}). 
In this respect, we note that a strong radio flux density is one of the most efficient (albeit incomplete) AGN selection 
criteria: out of the 527 sources with 5 GHz flux density $>1$ Jy and $|b_{\rm II}| \ge 10^{\circ}$
\citep{Kuehr_1981} only one, M 82, is not an AGN (and only two do not belong to the RG, radio 
quasar, or blazar classes: M 82 and NGC 1068).

\subsubsection{RL and RQ AGN}
\label{sec:RL_RQ}

One of the most used classifications of radio AGN is their division into RL and RQ AGN. Initially, this distinction was defined in the context of quasars ($M_\mathrm{B}<-23$) with the threshold between the two classes set either in: (1) radio flux density or luminosity (e.g. \citealt{Peacock_1986}); (2) or the ratio of radio-to-optical flux density or luminosity (e.g. \citealt{Schmidt_1970}). Radio loudness has since then been defined in different ways in the literature
but in any case these ``classical'' definitions apply only to type 1 AGN \citep{Padovani_2011,Bonzini_2013}. Following \citet{Balokovic_2012} the parametrization of radio loudness for type 1 AGN can be summarized as 
$R_\mathrm{K} = \log{L_\mathrm{radio}} - \mathrm{K} \cdot \log{L_\mathrm{\Delta\lambda}}$. $K=0$ for a simple radio flux density or luminosity threshold (where L$_\mathrm{radio}$ may refer to  flux density or luminosity measurements at frequencies $\sim1-6$~GHz in the observed- or rest-frame, respectively; \citealt{Peacock_1986, Miller_1990, Ivezic_2002}), and $K=1$ for a threshold in the logarithm of the ratio of flux density or luminosity in the radio band and that within a wavelength 
range  $\Delta\lambda$, which can either be in the  optical (e.g. B-band; \citealt{Kellermann_1989}; r-, i-, or z-band; \citealt{Ivezic_2002}) or IR (e.g. 24~$\mu$m; \citealt{Padovani_2011, Bonzini_2013}). 

\cite{Padovani_2016} \citep[see also][]{Padovani_2017} has argued that the distinction between these two types of AGN is not
simply a matter of semantics but rather that the two classes represent intrinsically
different objects, with RL AGN emitting a {\it large fraction} of their
energy non-thermally and in association with powerful relativistic jets,
while the multi-wavelength emission of RQ AGN is {\it dominated} by thermal
emission, directly or indirectly related to the accretion
disk (see Fig. \ref{fig:SED}). Moreover, he pointed out that the ``radio-loud/ra\-dio-quiet'' classes 
are obsolete, misleading, and inappropriate. Since the major physical difference between these two classes is 
the presence or lack of {\it strong} relativistic jets, which also implies that the two classes reach
widely different maximum photon energies (see his Sect. 2.3 and Fig. \ref{fig:SED}), 
we will be using in this review the terms ``jetted''
and ``non-jetted'' instead of RL and RQ AGN. We discuss this further in Sect. \ref{sec:discussion}. 

Note that, although we know that jetted AGN represent a minority, their exact fraction is still not well determined. 
The oft-quoted value of $\approx 15\%$ comes from optically selected samples of quasars \citep[e.g.][]{Kellermann_1989}. 
\cite{Padovani_2011a}, by integrating the radio LFs of jetted and non-jetted AGN, has suggested a much smaller
fraction ($< 1\%$).  

\subsubsection{Low- and high-excitation AGN}
\label{sec:LEG_HEG}

\begin{figure}
\begin{center}
\vspace{-5em}
\hspace*{-2.0em}
\includegraphics[scale=0.55]{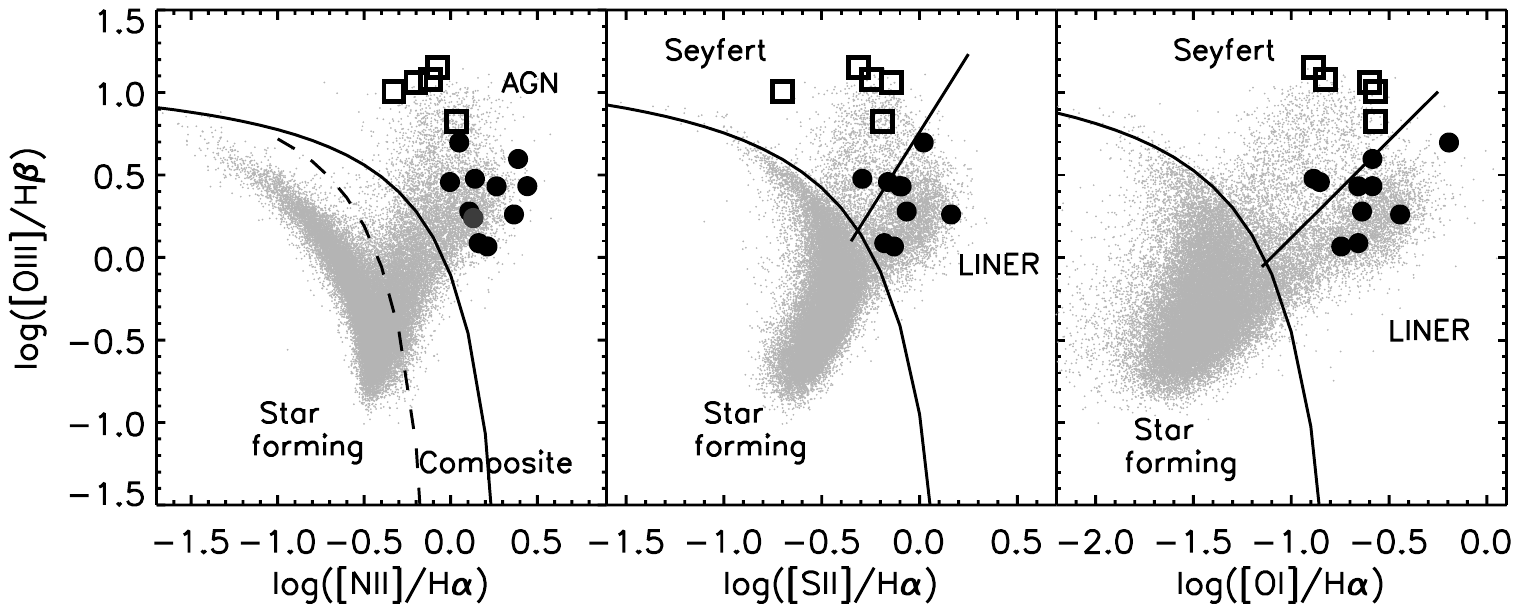}
\vspace{-30.5em}
\caption{Optical spectroscopic diagnostic diagrams (see \citealt{Kewley_2001,Kauffmann_2003b, Kewley_2006}) that separate emission-line galaxies into star forming,
composite galaxies, and Seyfert and LINER AGN.
Small grey dots represent galaxies from the Sloan Digital Sky Survey (SDSS) DR4 ``main'' spectroscopic
sample. Large open squares (filled dots) denote $z < 0.1$ Revised Third Cambridge Catalogue of Radio Sources \citep[3CRR;][]{Laing_1983} RGs 
independently classified based on their core X-ray emission as systems with
radiatively efficient (inefficient) BH accretion \citep{Evans_2006}. 
\copyright~AAS. Figure reproduced from \citet{Smolcic_2009b}, Fig. 1, with permission.}
\label{fig:radio_bpt}
\end{center}
\end{figure}

The classification of radio AGN into two main classes based on their optical spectroscopic properties 
goes back to \citet{Hine_1979}. They divided the Third Cambridge Catalogue of Radio Sources (3CR) into objects characterized by strong emission lines in their spectra ([O\,\textsc{ii}]~$\lambda3727$, [O\,\textsc{iii}] $\lambda5007$, [Ne\,\textsc{ii}]~$\lambda3867$), and sources which exhibited either absorption line spectra typical of giant elliptical galaxies (absorption line galaxies hereafter) or weak 
[O\,\textsc{ii}]~$\lambda3727$ emission lines. Since then this has been expanded and refined by using also [O\,\textsc{iii}]~$\lambda5007$ equivalent widths (EWs; e.g. \citealt{Tadhunter_1998}) or high/low excitation/ionization emission line criteria (\citealt{Laing_1994,Kewley_2006, Buttiglione_2009, Buttiglione_2010}; see Fig~\ref{fig:radio_bpt}). In general, objects without and with 
high-excitation emission lines in their optical spectra are referred to as LEGs and HEGs, 
respectively. The LEG/HEG classification holds not only for radio selected AGN but 
applies also to AGN selected in other bands (for which an optical spectroscopic classification 
is available). More specifically,  quasars and Seyferts belong to the HEG category, while LINERs and absorption line galaxies are classified as LEGs (see Fig.~\ref{fig:radio_bpt} and Sect. \ref{sec:optical}; \citealt{Baldwin_1981,Veilleux_1987, Kewley_2001,Kauffmann_2003a, Kewley_2006}. But see \citealt{Sarzi_2010} for evidence that the nebular emission of most objects in the LINER part of Fig. \ref{fig:radio_bpt} is not powered by an AGN). 

\begin{figure}[ht!]
\begin{center}
\includegraphics[bb=  55 628 612 752, scale=0.5]{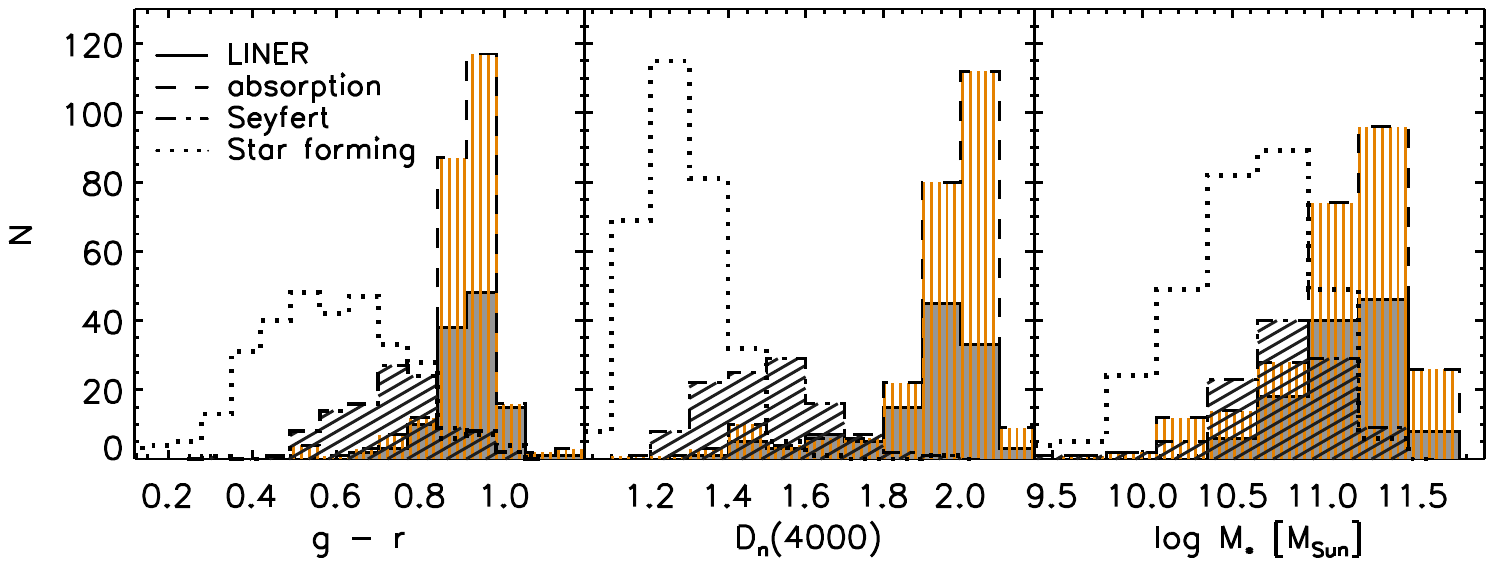}
\includegraphics[bb= 90 400 612 622, scale=0.5]{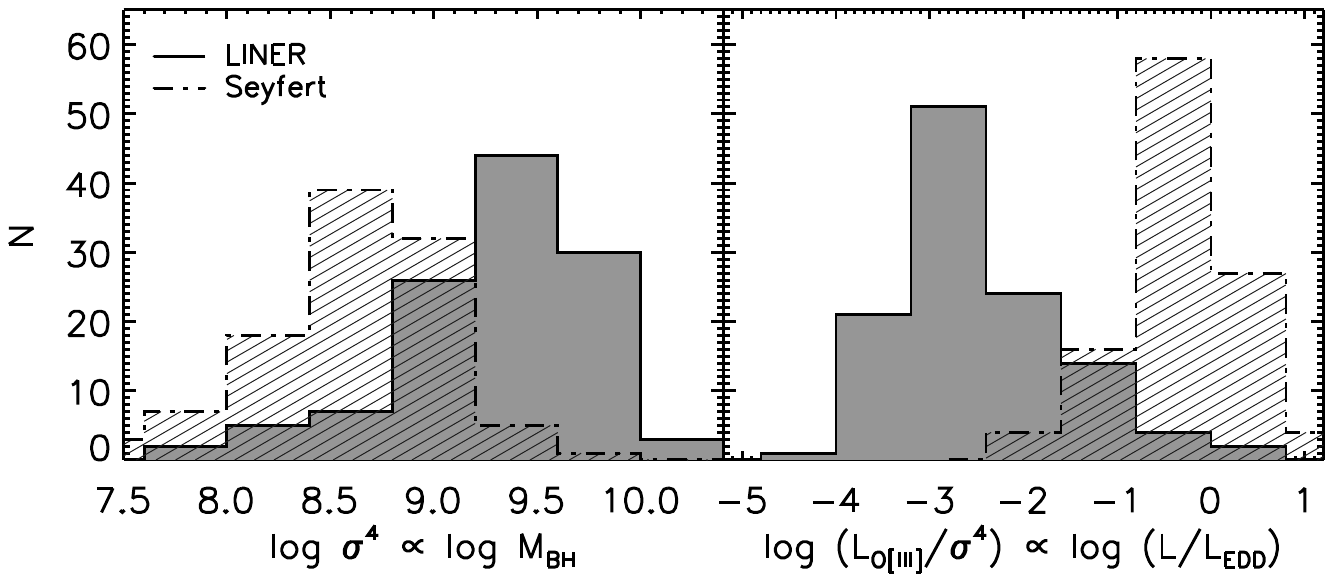}
\caption{From left to right the top panels show the distribution of the observed $g-r$ colour, 4000~\AA \ break strength, and stellar mass, while the bottom panels show the distribution of the velocity dispersion, $\sigma^4$,
proportional to BH mass \citep{Tremaine_2002}, and $\log (L_{\rm{[O\,\textsc{iii}]}}/ \sigma^4)$ proportional to BH accretion rate (in Eddington units;
\citealt{Heckman_2004}) for various galaxy populations  drawn from the FIRST-NVSS-SDSS sample with $0.04 < z < 0.1$ (SFGs: dotted lines and empty histograms; Seyferts: dash-dotted lines and diagonally hatched histograms; absorption line galaxies: dashed lines and vertically hatched histograms; LINERs: full lines and filled histograms). \copyright~AAS. Figure reproduced from \citet{Smolcic_2009b}, Fig. 2 and 3, with permission.}
\label{fig:radio_props}
\end{center}
\end{figure}

As illustrated in Fig.~\ref{fig:radio_props} fundamental physical differences between these two types of radio AGN (at $z<1$) have been found (e.g. \citealt{Nagar_2005, Evans_2006, Hardcastle_2007, Smolcic_2009a, Smolcic_2009b, 
Buttiglione_2010, Best_2012, Padovani_2015}; see also Fig.~6 in \citealt{Smolcic_2016a}). Namely, LEGs (LINERs and absorption line systems) have on average redder optical $g-r$ colours, larger values of the 4000~\AA \ break strength, and higher stellar masses than HEGs (Seyferts; top panel). LEGs are also shown (bottom panel) to exhibit radiatively inefficient accretion related to low $L/L_{\rm Edd}$ ($\lesssim 0.01$), possibly fuelled by the hot phase of the inter-galactic medium (IGM), and are typically highly efficient in collimated jet production. HEGs accrete in a radiatively efficient manner, at high Eddington rates ($0.01 \lesssim L/L_{\rm Edd} \lesssim 1$), are fuelled by the cold IGM phase, and (on average) 
less likely to launch collimated jets. 
From a theoretical aspect, the observed difference in $L/L_{\rm Edd}$ can be related to the switch between a standard accretion, i.e. radiatively efficient, geometrically thin (but optically thick) disk accretion flow (Shakura \& Sunyaev 1973), and a radiatively inefficient, geometrically thick (but optically thin) disk accretion flow \citep{Esin_1997,Narayan_1998}. The switch occurs at accretion rates below a certain $L/L_{\rm Edd}$ ($\approx 0.01$; \citealt{Rees_1982, Narayan_1994, Meier_2002, Fanidakis_2011}) as then the decreased accreting gas density lowers the cooling rate, and a substantial amount of heat can be carried along, i.e. advected, rather than irradiated\footnote{The switch in $L/L_{\rm Edd}$ is not to be taken as sharp but as a transition in a statistical sense. The fundamental physical separation of the various AGN types may be a function of more parameters (such as spin and BH mass) and one should keep in mind that the observational data used to constrain this separation are subject to measurement and computational uncertainties and biases (e.g. the role of environment in kinetic luminosity determinations, contamination of selection proxies by stellar, rather than AGN related processes, etc.; see, e.g. \citealt{Mingo_2014}).}. 

\subsubsection{Radio-selected AGN classes and unification}
\label{sec:radio_unif}

There is some overlap amongst the classes described above. For example, almost all FR Is are LEGs, while most  FR IIs
usually display strong emission lines and would thus be classified as HEGs. However, there is no one-to-one correspondence 
between FR class and emission lines as quite a few FR IIs \citep[$\gtrsim 20$\%: e.g.][]{Gendre_2013} have been found to be LEGs
(see also \citealt{Evans_2006, Buttiglione_2010, Mingo_2014}). Powerful radio quasars and RGs are generally of the HEG type, while
the less powerful (and more common) RGs are mostly LEGs \citep[e.g.][]{Padovani_2016}. 
``Classical'' non-jetted AGN (i.e. Seyferts and quasars) are always HEGs. 
 
Moreover, radio quasars are intrinsically the same sources as some RGs. Namely 
they are simply FR II/HEG RGs with their jets at an angle $\lesssim 45^{\circ}$
with respect to the line of sight
\citep{Orr_1982,Barthel_1989,Antonucci_1993,Urry_1995}. The fact that radio
quasars display strong and Doppler broadened lines in their optical spectra (with
full width half maximum [FWHM] $\gtrsim 1,000$ km s$^{-1}$), unlike RGs, requires also
the presence of dust in a flattened configuration roughly perpendicular to the jet
(see also Sect. \ref{ssec:IR_SED}). 
This so-called ``unification model'' explains in a natural way why the (projected) 
sizes of the jets of RGs are larger than those of quasars \citep{Barthel_1989}. 

With regard to FR I/LEG RGs, obscuration towards their nuclei appears to
be much smaller than that of their FR II/HEG relatives
\citep[e.g.][]{Chiaberge_2002,Evans_2006}, which indicates that dust 
might be not present (see also Sect. \ref{sssec:IR_hdps}). This applies also to the population of FR II/LEG
RGs. LEG RGs, therefore, irrespective of their radio
morphology, are ``unified'' with BL Lacs \citep[e.g.][]{Giommi_2013}. 

\subsubsection{Classification of radio-selected AGN}
\label{sec:radio_classif}

The classification of radio-selected AGN is a complex matter, especially so for the faint radio 
sources routinely studied these days. Getting optical spectra for their counterparts is very 
time consuming, prohibitively so for the very faint tail [$R \gtrsim 26$] even with 8-10m class telescopes. But even if we had optical spectra for all radio sources, optical-based classification is well 
known to be prone to obscuration biases (Sect. \ref{sec:opt_bias}) and for this reason spectra need to be complemented with information from multi-wavelength ancillary data.
Multi-wavelength methods used in the literature to identify AGN detected also in the radio band include: (1) rest-frame optical 
colours of the host galaxies, which have been shown to correlate with the galaxies spectral emission line properties 
(\citealt{Strateva_2001, Smolcic_2006, Smolcic_2008, Smolcic_2009b, Ilbert_2010});
(2) X-ray luminosity ($L_\mathrm{X}>10^{42}$~erg~s$^{-1}$; e.g. \citealt{Szokoly_2004}; see also Sect. \ref{sec:X-ray_sel}); (3) mid-IR (MIR) colours (e.g. 
\citealt{Lacy_2004, Stern_2005, Donley_2012}; see also Sect. \ref{ssec:MIR_broad_MIR}); (4) multiple component SED fitting using sets of galaxy {\em and} 
AGN templates (\citealt{Berta_2013,Delvecchio_2014,Delvecchio_2017}); (5) excess of radio luminosity relative to a tracer 
of the SF rate (SFR) in the host galaxy (e.g. far or total IR luminosity; \citealt{Condon_1992, DelMoro_2013,Delvecchio_2017}).
A detailed discussion of these methods (and others) can be found in \cite{Padovani_2016}. 

\subsection{Selection effects: SF contribution to the total radio luminosity output}
\label{sec:radio_bias}

In the local universe LEGs occupy the red sequence of galaxies, while HEGs are hosted by bluer  galaxies, populating the so-called green valley (see Fig.~4 in \citealt{Smolcic_2009b}). As, generally, bluer host galaxy colours imply higher SFRs (e.g. \citealt{Ilbert_2009}), in the latter case 
the observed radio emission could be partially or entirely due to synchrotron radiation generated by supernova remnants, rather than by the central SMBH.  As the classification of radio AGN often relies on non-radio properties, such as the optical emission line properties or multi-wavelength proxies (Sect. \ref{sec:radio_classif}), an {\em a-posteriori} assessment of the (SF- or AGN-related) origin of the observed radio emission is needed as X-ray, optical, or IR signatures of an accreting central SMBH do not necessarily imply AGN-related radio emission. Thus, one of the most severe selection effects in the radio band is the contribution of SF-related processes to the total radio power\footnote{Note that, 
while it might be safe to assume that anything above $P_{\rm 1.4GHz} \approx 10^{24}$ W Hz$^{-1}$ has nothing to do with SF, this is only valid at low redshifts given the strong evolution of SFGs \citep[e.g.][]{Padovani_2016}.}
 (especially given the limited angular resolution available in many radio continuum surveys). 

Detailed high angular resolution radio continuum studies often reveal a mixture of SF- and 
accretion-related radio emission in non-jetted HEGs (e.g. \citealt{Chi_2013, Maini_2016, 
Herrera_2016}). For example, 
\citet{Maini_2016} used the Australian Long Baseline Array to search for compact radio cores in four non-jetted AGN located in the Extended {\it Chandra Deep Field}-South (E-CDFS). They find that some such sources contain an active AGN that can contribute significantly ($>50$\%) to the total radio emission.
\citet{Herrera_2016} studied  three non-jetted quasars ($M_B<-23$) in the COSMOS field observed with the Very Large Array (VLA) and the Very Long Baseline Array (VLBA) at $1.5^{\prime\prime}$ and $<20$~mas resolutions, respectively.  
Comparing the core (VLBA) and total (VLA) radio flux densities they infer that 50-75\% of the  radio emission in these sources is powered by AGN activity. Such high-angular resolution studies are, however, still limited to rather small samples, and to-date the main source 
of radio emission in the most powerful AGN (i.e. $M_i<-22$) is still debated. While agreement exists that radio emission in jetted quasars is powered by SMBH accretion-related processes (e.g. \citealt{Miller_1990}), opposing (statistical) evidence can be found in  the literature related to the origin of radio emission  in non-jetted quasars (the so-called quasar radio loudness dichotomy problem):  SF in the host galaxies  \citep{Kimball_2011, Condon_2013}, or AGN activity (\citealt{White_2015}, \citealt{Zakamska_2016}; see also \citealt{Padovani_2016} for a detailed discussion of this topic).

From a statistical point of view, studies of radio AGN drawn from large radio-continuum surveys, combined with spectroscopic and/or multi-wavelength data find that  ($z<1$) HEGs are much more likely to be associated with SF in their host galaxies, with SFRs at least a factor $3-4$ higher than those in LEGs (\citealt{Hardcastle_2013}; see also \citealt{Gurkan_2015}). The inferred statistical contribution of SF-related processes to the observed radio emission in HEGs and LEGs ($0.04<z\lesssim0.2$) is estimated to be $\sim60\%$ ($\sim10\%$) for HEGs (LEGs; \citealt{Moric_2010}). Similarly, \citet{Bonzini_2015} find that the radio luminosities of non-jetted AGN at $z\sim1.5 - 2$ in the E-CDFS survey are consistent with the galaxies' SFRs inferred from their FIR luminosities. Identifying AGN in a slightly different way,
\cite{Delvecchio_2017} observe that for about 70\% of their AGN the radio luminosities are consistent (within $3\sigma$ from the average) with those expected based on the SFR in the host galaxies inferred from the total IR emission (corrected for the AGN component). 

\subsection{Cosmic evolution of radio-selected AGN}
 \label{sec:radio_evolv}
  
Past research has shown that radio AGN evolve via a ``downsizing'' effect, i.e. low radio luminosity sources evolve less strongly than high-luminosity ones (e.g. \citealt{Longair_1966, Willott_2001, Rigby_2015}; see also Sect. 
\ref{sec:X-ray_types} for the X-ray perspective). 
Studies of powerful radio AGN (L$_\mathrm{1.4GHz}\gtrsim 2 \times 10^{26}$~W~Hz$^{-1}$) have found a strong positive density evolution at $z\lesssim2$, beyond which their comoving volume density declines \citep{Dunlop_1990, Willott_2001}. A substantially slower evolution, with a  lower redshift ($z\sim 1-1.5$) comoving volume density turnover, has been found for weaker radio AGN (L$_\mathrm{1.4GHz}> 2 \times 10^{25}$~W~Hz$^{-1}$; \citealt{Waddington_2001}). Studies of even lower luminosity radio AGN (L$_\mathrm{1.4GHz}\lesssim10^{25}$~W~Hz$^{-1}$) find a mild evolution out to $z\sim1$ (e.g. \citealt{Smolcic_2009a, Sadler_2007, Donoso_2009, Padovani_2011,Smolcic_2017b}). 

The local radio LFs derived separately for HEGs and LEGs have been presented by e.g.\ \citet[][]{Filho_2006}, \citet[][]{Best_2012}, \citet[][]{Gendre_2013}, \citet[][]{Pracy_2016}. While it is clear that HEGs dominate the volume densities at high radio luminosities (L$_\mathrm{1.4GHz}>10^{26}$ W~Hz$^{-1}$; \citealt{Heckman_2014, Pracy_2016}), the slope of the low-luminosity end of the HEG radio LF is still somewhat unclear. \cite{Best_2012} find a significantly flatter slope than \cite{Pracy_2016}\footnote{Derived for FIRST-SDSS AGN at $z<0.3$ from the $\sim900$~deg$^2$ LARGESS survey and spectroscopically classified.}, and \citet[][]{Filho_2006}\footnote{Derived for Seyfert galaxies using $1^{\prime\prime}$ angular resolution radio continuum data; see also Padovani et al.\ (2015).} (see Fig.~8 in \citealt{Pracy_2016}). As discussed in \citet{Pracy_2016} this is likely due to the difficulty of disentangling the real contribution of AGN-related radio emission in HEGs within the faint radio luminosity regime, dominated by SFGs. Regardless of the faint end slope, studies out to $z\sim1$ consistently find that HEGs evolve more rapidly than LEGs. For example, \citet{Pracy_2016} find that their LEG population displays little or no evolution over the observed redshift range ($0.005<z<0.75$), evolving as $(1+z)^{0.06^{+0.17}_{-0.18}}$ [$(1+z)^{0.46^{+0.22}_{-0.24}}$] assuming  pure density [pure luminosity] evolution, while their HEG population evolves more rapidly, as $(1+z)^{2.93^{+0.46}_{-0.47}}$ [$(1+z)^{7.41^{+0.79}_{-1.33}}$]  assuming  pure density [pure luminosity] evolution. 

Constraining the cosmic evolution of the two dominant AGN types detected in the radio band beyond redshift 1 ($z\lesssim6$) is not trivial given  the difficulty of: (1) separating the two distinct AGN types so that they can quantitatively be related to the HEG/LEG populations identified via optical spectroscopy at $z<1$; (2) isolating the fraction of radio luminosity arising from the AGN (rather than SF in the host galaxy), which is challenging even in the lower ($z<1$) redshift universe. 
Recently, \citet{Padovani_2015} have constrained the cosmic evolution of their selected non-jetted and jetted AGN identified in the E-CDFS survey out to $z\sim4$. These two populations can qualitatively  be roughly related to the HEG and LEG 
populations respectively (see \citealt{Bonzini_2013, Padovani_2015}). Padovani et al. find a strong evolution, similar to that for SFGs, of their non-jetted sample throughout the redshift range probed, and a peak at $z\sim0.5$ in the 
number density of their jetted AGN, with a decline at higher redshifts. The first results using the VLA-COSMOS 3~GHz Large Project (Ceraj et al., in prep.), combined with the COSMOS multi-wavelength data \citep{Laigle_2016,Marchesi_2016}, yield 
a stronger cosmic evolution for AGN with the highest bolometric (radiative) luminosities throughout the entire redshift range ($z\lesssim6$; and qualitatively broadly consistent with HEG samples), 
relative to that of the AGN sample with lower bolometric (radiative) luminosities (and qualitatively broadly consistent with LEG samples)\footnote{The former class has been selected through 
X-ray, IR, and SED-criteria, while the latter has been identified via $>3\sigma$ radio-excess relative to the host galaxies' IR-based SFRs, and red rest-frame optical colours, and lacking X-ray, IR, and SED-based signatures of AGN activity \citep{Smolcic_2017a,Delvecchio_2017}. In this study the SF related contribution to the total radio luminosity was statistically subtracted.}. We stress however that both the E-CDFS and COSMOS fields are not large enough to constrain the broad radio luminosity range, encompassing also the rare, highest luminosity AGN (detectable in shallower, wide-area surveys). Hence, for a full, quantitative assessment of the evolution of the AGN radio LF separated into the two types out to high redshift and over a broad luminosity range a combination of  surveys with various areal coverages is needed, with access to a robust AGN classifier, and methods to isolate AGN-related radio emission in (low radio luminosity) AGN, that can be uniformly applied throughout the entire redshift range considered. 

\subsection{The future of AGN studies in the radio band}\label{sec:radio_future}

Studies such as those discussed in the previous section 
are becoming feasible only now, and will be invigorated with the onset of the Square Kilometre Array (SKA\footnote{ \url{www.skatelescope.org}}), 
offering an observing window between 50 MHz and 20 GHz 
extending well into the nanoJy regime with unprecedented versatility, 
in combination with contemporaneous  projects over the entire electromagnetic spectrum (such as, e.g., 
the Large Synoptic Survey Telescope [LSST], {\it Euclid}, the Extended ROentgen Survey with an Imaging Telescope Array (eROSITA), and the {\it James Webb Space Telescope} [JWST]).

A revolution has in fact started in radio astronomy, which has entered an era
of large area surveys reaching flux density limits well below current
ones. 
The Jansky Very Large Array (JVLA\footnote{\url{science.nrao.edu/facilities/vla}}), 
the LOw Frequency ARray (LOFAR\footnote{\url{www.astron.nl/radio-observatory/astronomers/lofar-astronomers}}), 
the Murchison Widefield Array\footnote{\url{www.mwatelescope.org}}, are already taking data, and are being joined by the
Australian Square Kilometre Array Pathfinder
(ASKAP\footnote{\url{www.atnf.csiro.au/projects/askap/}}), 
Meer\-KAT\footnote{\url{www.ska.ac.za/meerkat}}, e-MERLIN\footnote{\url{www.e-merlin.ac.uk}}, 
and A\-PERTIF\footnote{\url{www.astron.nl/general/apertif/apertif}}.  
These projects will survey the sky vastly faster than it 
is possible with existing radio telescopes producing surveys covering large
areas of the sky down to fainter flux densities than presently available,
as fully detailed in \cite{Norris_2013}. This, amongst other things, will revolutionise AGN studies. 
The Evolutionary Map of the Universe \citep[EMU;][]{Norris_2011},
one of the ASKAP surveys, for example, is expected to detect $\sim
70$ million sources, about half of which will likely be AGN unaffected by the 
problems of obscuration, which plague 
the optical (Sect. \ref{sec:opt_bias}) and soft X-ray (Sect. \ref{sec:X-ray_sel}) bands. 
Identifying AGN in these new radio surveys, however, will not be straightforward and will require 
many synergies with facilities in other bands \citep[e.g.][]{Padovani_2016}. 

\section{Infrared-selected AGN}\label{sec:IR}

IR studies have had a strong impact on our understanding of AGN
structure, their evolution through cosmic time, and their role in galaxy
evolution. In Sect. \ref{ssec:IR_SED} we discuss the mechanisms that give
rise to the IR emission in AGN and the advantages of AGN identification in the MIR wavelengths. In
Sect. \ref{ssec:MIR_phot_sel} we examine in detail how the MIR
selection of AGN works and what  the characteristics of the selected samples
are. In Sect. \ref{ssec:other_ir} and 
\ref{ssec:IR_spec} we explore additional AGN selection criteria
that rely on IR observations. Finally, in Sect. \ref{ssec:IRfuture} we 
analyse the future of AGN studies in the IR in light of upcoming facilities.
For the purpose of this section, we divide the IR SED of AGN in three wavelength regimes: the near-IR (NIR; $1 - 3~\mu$m), the MIR
($3 - 50~\mu$m), and the FIR ($50 - 500~\mu$m).

\subsection{Physical mechanism behind IR emission}\label{ssec:IR_SED}

Despite many drawbacks, the ``dusty torus'' paradigm has been quite successful in 
explaining the appearance of a wide variety of AGN. The basis of this paradigm is the presence of dust surrounding
the accretion disk on scales larger than that of the broad line
region (BLR), with an inner boundary set by the sublimation temperature of the dust grains \citep{Barvainis_1987}. 
This dust reprocesses the emission of the accretion disk into
the IR and dominates the AGN SED from wavelengths longer than
$\sim$1~$\mu$m up to a few tens of micron (see Fig. \ref{fig:SED}). It
plays a fundamental role in the AGN unification scheme \citep[][see also Sect. \ref{sec:radio_unif}]{Antonucci_1993,Urry_1995}, as through polarisation studies it was
established that the difference between type 1 and 2 AGN is simply an effect of
orientation with respect to the dust. In type 2 AGN the dust obscures
the line of sight towards the accretion disk and the BLR and 
only narrow emission lines can be observed in
the optical spectrum \citep[e.g.][although see \citealt{elitzur16}
  for a discussion about possible {\it{real}} type 2 AGN where the
  difference is not caused by dust
  obscuration; see also Sect. \ref{sec:X-ray_types}]{antonucci85,Antonucci_1993}. 
  
There is a significant debate in the literature over whether the dust is
smoothly distributed in the torus \citep{pierkrolik92,dullemond05,fritz06}, whether it is clumpy in the form of optically and geometrically thick
clouds \citep{krolik88,nenkova02,nenkova08,Elitzur_2006,tristram07}, or
a combination of the two
\citep{stalevski12,assef13}. Observations of the strength of
the silicate feature at 9.7~$\mu$m in AGN, for example, seem to favour 
models where the dust is most prominently clumpy
\citep{nenkova08,nikutta09,hatzimi15}, but \citet{feltre12} has argued
that observations are not yet able to discriminate between the
different models. Recent ground-based MIR interferometric observations, on the other hand, suggest 
that a large proportion of the dust might, instead, reside in the walls of the ionization cone \citep[][and references therein]{asmus16}, 
at least in a fraction of nearby AGN \citep{lopez16}. 
For simplicity and compatibility
with the rest of the literature, we will refer to this structure
as the ``dusty torus'' throughout this section, despite evidence that this dust component may have a 
significantly more complex distribution \citep[e.g.][]{nenkova08}. 

A number of authors have studied the fraction of lines-of-sight that are
obscured by the dusty torus, either by comparing the relative fraction
of type 1 and 2 AGN at a given redshift, or by modelling the SED of individual
objects. Average obscured fractions of 40\% to 75\% are reported in
the literature \citep[see,
  e.g.][]{treister04,hatzimi09,assef13,roseboom13}. However, a single number
does not englobe the diversity of AGN in Nature. Some authors have
found that the fraction of obscured lines-of-sight diminishes with
increasing luminosity of the accretion disk
\citep{ueda03,hasinger04,simpson05,hatzimi08,assef13,mateos16}, an
effect typically referred to as the receding torus
\citep{lawrence91}, although others have found no evidence of such an
effect
\citep[][see also Sect. \ref{sec:X-ray_types}]{wang06,lawrence10,honig11,lacy13,stalevski16}. Additionally, some
authors have found a significant variance in the amount of dust in AGN
\citep[e.g.][]{roseboom13}, with some showing little to no dust (see Sect.
\ref{sssec:IR_hdps}) and some showing very large amounts
\citep[e.g.][]{mateos16}. Furthermore some authors have found a
larger fraction of obscured sources at the highest luminosities
\citep{banerji12,assef15}, suggesting a more complex scenario, and
possibly consistent with models where AGN dust obscuration evolves
through time \citep{sanders88,hopkins08}.

As already mentioned, the emission of the dusty torus is very prominent in the MIR for both type 1 and type 2 AGN. 
Dust emission from SF can rival in luminosity the AGN
but with typically much cooler temperatures $\lesssim 40$~K \citep[e.g.][]{magnelli12}. As significant
SF activity is regularly ongoing in the host galaxies of many 
AGN, 
it is more likely to dominate the FIR \citep[e.g.][]{hatzimi10}. 
At rest-frame NIR wavelengths, where the AGN emission 
has a local minimum at the cross-over between the dropping accretion disk emission and the rising
dust emission, the stellar
1.6~$\mu$m peak can severely hamper AGN identification. As the stellar
emission drops steeply longward of the 1.6~$\mu$m peak, stellar
contamination is less of an issue in the MIR (although see
Sect. \ref{ssec:MIR_phot_ledd}). Hence, the MIR wavelengths are the 
optimal IR wavelengths for AGN identification.

\subsection{AGN in the MIR}\label{ssec:MIR_phot_sel}

\subsubsection{Broad-band MIR AGN identification}\label{ssec:MIR_broad_MIR}

The large sky background and water absorption by the Earth's atmosphere make ground-based MIR observations challenging. Spaceborne telescopes are, therefore, better suited for the identification of large AGN samples. In what follows, we focus solely on selection using
space-based broad-band photometry, as they account for the great
majority of MIR identified AGN, although most implications and many of
the caveats also apply to ground-based and to spectroscopic
observations. We explore those further in Sect. \ref{ssec:IR_spec}.

A number of AGN MIR selection criteria have been proposed over the years.
The first
ones were already developed for the Infrared Astronomical Satellite (IRAS) mission
\citep{degrijp85,degrijp87,leech89}, and their number has grown
enormously since, with the advent, in the past decade, of the {\it{Spitzer Space Telescope}} (\citealt{werner04}; see, e.g. \citealt{Lacy_2004,lacy07,lacy13,Stern_2005,hatzimi05,Donley_2012}), 
{\it{AKARI}} \citep{murakami07,oyabu11}, and
the Wide-field Infrared Survey Explorer (WISE: \citealt{wright10};
see, e.g. \citealt{stern12,mateos12,wu12b,assef13}). These
selection criteria have typically been calibrated against independent
AGN selection methods and rely primarily on colours to separate AGN
from stars or galaxies with inactive nuclei, as AGN are expected to be
significantly redder in the shorter wavelength MIR bands
\citep[e.g.][]{Stern_2005}. We note that the latter is not necessarily true
for AGN at redshifts where the H$\alpha$ emission line contaminates
the shortest wavelength channels, which can lead to significant biases
against $4 \lesssim z \lesssim 5$ AGN in some cases
\citep{Richards_2009b,Assef_2010}, but is generally true otherwise.

\begin{figure*}
  \begin{center}
    \includegraphics[width=13cm]{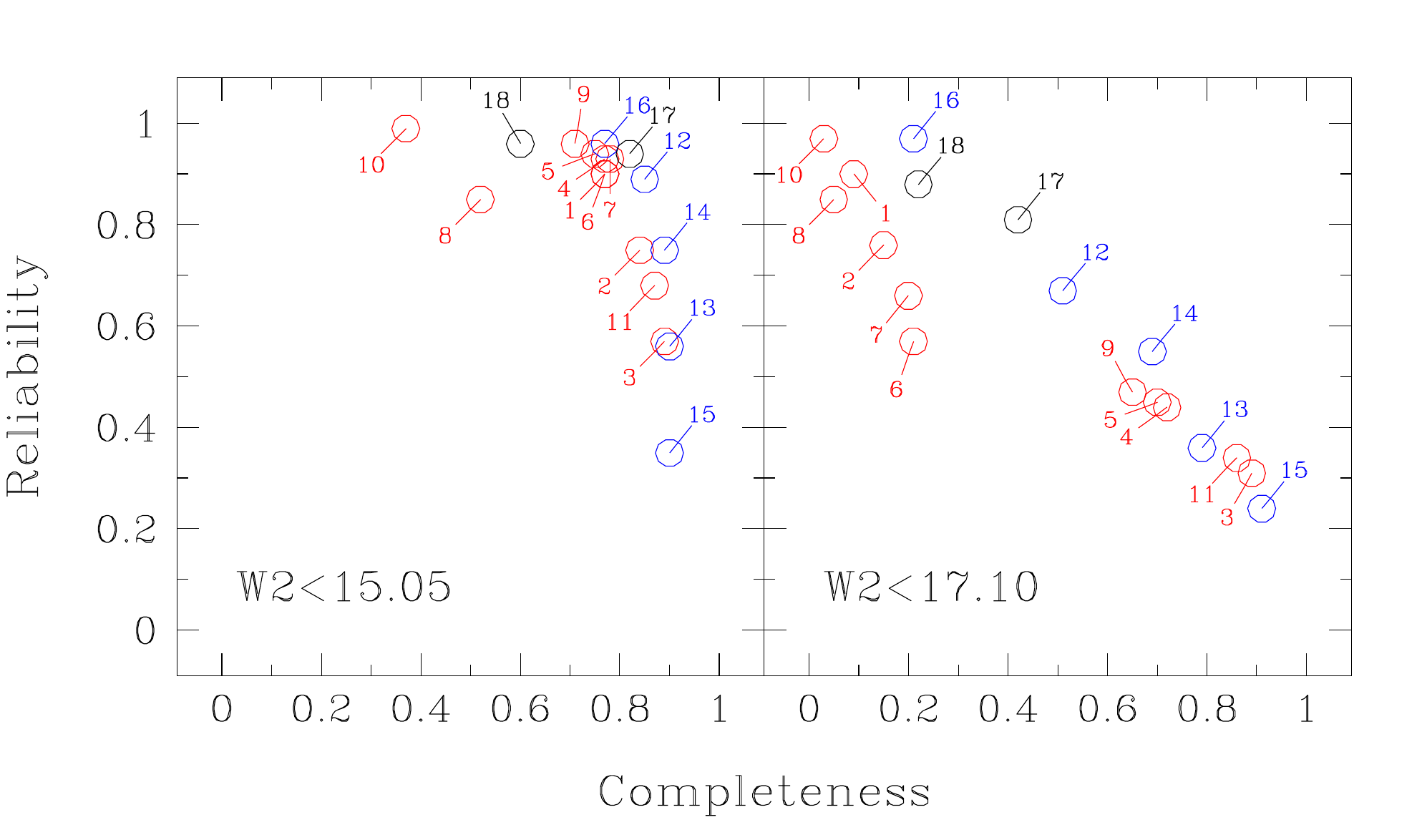}
    \caption{Comparison of completeness and reliability of popular photometric MIR AGN 
    selection criteria, assessed by the test developed by \citet{assef13} and discussed in 
    Sect. \ref{ssec:MIR_broad_MIR}. The left and right panels show the case of a shallow and deep survey,
    respectively. 
    The criteria shown correspond to the references listed in Table \ref{tab:mir_sel}. The red 
    points show the WISE only selection criteria, the blue points show the {\it{Spitzer}} only 
    selection criteria, and the black points show the NIR+{\it{Spitzer}} criteria.}
    \label{fig:MIR_phot_comp_rel}
  \end{center}
\end{figure*}

\citet{assef13} presented an interesting comparison between the WISE- and
{\it{Spitzer-}}based\footnote{For {\it{Spitzer}} we refer specifically to
  the four broad bands of the IRAC instrument \citep{fazio04} centred
  at 3.6, 4.5, 5.8 and 8~$\mu$m (and referred to as [3.6], [4.5],
  [5.8] and [8.0] respectively), and to the 24~$\mu$m band of the MIPS
  instrument \citep{rieke04}. For WISE we refer to all its four bands,
  centred at 3.4, 4.6, 12 and 22~$\mu$m, usually referred to as
  W1--W4.} selection criteria. These are listed in Table \ref{tab:mir_sel} and shown in Fig.
\ref{fig:MIR_phot_comp_rel}\footnote{Adapted from \citet{assef13} to include the 
criteria of \citet{Donley_2012} and \citet{lacy13}. Note that the WISE selection criteria still use the data from the All-Sky
  data release \citep{cutri12}.}.
Using a sample of large, multi-wavelength, spectroscopically identified AGN they
determined how reliable and complete each criterion is. The results are
shown for two W2 limiting Vega magnitudes,
namely $W2<15.05$ on the left, representative of the shallowest fields in
the WISE mission, and $W2<17.1$ on the right, that probes down to a 3$\sigma$
depth. Figure \ref{fig:MIR_phot_comp_rel} shows that 
shallow and deep surveys need to be analysed separately.
For the shallow surveys, most selection criteria recover samples with high
reliability and completeness. 
For deeper surveys, however, the situation is different. Most selection criteria line up 
diagonally in the diagram trading completeness for reliability. 
One should also keep in mind that 
due to their lower sensitivity, the completeness of selection criteria relying on the W3 and W4 bands is lower than the completeness obtained using criteria derived based on a W2-limited
sample. A selection relying on these longer wavelengths can be of particular use in
the WISE fields closer to the ecliptic poles, where the survey scan
pattern is denser and the shorter wavelength bands reach the confusion
limit \citep[e.g.][]{jarrett11}.

\begin{table*}

  \caption{AGN MIR selection criteria shown in Fig. \ref{fig:MIR_phot_comp_rel}}
  \label{tab:mir_sel}

  \begin{tabular}{l l l}
    \noalign{\smallskip}\hline\noalign{\smallskip}
    ID & Reference & Criteria\\
    \noalign{\smallskip}\hline\noalign{\smallskip}
    \multicolumn{3}{l}{WISE only criteria}\\
    \noalign{\smallskip}
    
    \hphantom{1}(1) & \citet{assef13} -- R90$^{\dagger}$     & $\rm W1-\rm W2>0.662\ \exp\{0.232\ (\rm W2-13.97)^2\}$\\
    \noalign{\smallskip}

    \hphantom{1}(2) & \citet{assef13} -- R75$^{\dagger}$     & $\rm W1-\rm W2>0.530\ \exp\{0.183\ (\rm W2-13.76)^2\}$\\
    \noalign{\smallskip}

    \hphantom{1}(3) & \citet{assef13} -- C90$^{\dagger}$     & $\rm W1-\rm W2>0.50$\\
    \noalign{\smallskip}

    \hphantom{1}(4) & \citet{assef13} -- C75$^{\dagger}$     & $\rm W1-\rm W2>0.77$\\
    \noalign{\smallskip}

    \hphantom{1}(5) & \citet{stern12}$^{\ddagger}$           & $\rm W1-\rm W2>0.80$\\
    \noalign{\smallskip}

    \hphantom{1}(6) & \citet{jarrett11}$^{*}$               & $\rm W2-\rm W3>2.2$ $\land$ $\rm W2-\rm W3<4.2$ $\land$\\
      &                            & $\rm W1-\rm W2>(0.1\ (\rm W2-\rm W3)+0.38$ $\land$\\
      &                            & $\rm W1-\rm W2<1.7$, {\it except} \\
      &                            & (i) $\rm W1<10.5$ $\land$ $\rm W2-\rm W3<1.5$ $\land$\\
      &                            &\hphantom{(i) }$\rm W1-\rm W2<0.4$ {\it or}\\
      &                            & (ii) $\rm W3-\rm W4<1.2$\\
    \noalign{\smallskip}

    \hphantom{1}(7) & \citet{mateos12}$^{\star}$ -- 3-band & $\rm W1-\rm W2>-3.172\ (\rm W2-\rm W3) + 7.624$ $\land$\\
      &                            & $\rm W1-\rm W2<0.315\ (\rm W2-\rm W3) + 0.796$ $\land$\\
      &                            & $\rm W1-\rm W2>0.315\ (\rm W2-\rm W3) - 0.222$\\
    \noalign{\smallskip}

    \hphantom{1}(8) & \citet{mateos12}$^{\star}$ -- 4-band & $\rm W1-\rm W2>-2.00\ (\rm W3-\rm W4) + 4.33$ $\land$\\
      &                            & $\rm W1-\rm W2< 0.50\ (\rm W3-\rm W4) + 0.979$ $\land$\\
      &                            & $\rm W1-\rm W2> 0.50\ (\rm W3-\rm W4) - 0.405$\\
    \noalign{\smallskip}

    \hphantom{1}(9) & \citet{Assef_2010} -- 2-band  & $\rm W1-\rm W2>0.85$\\
    \noalign{\smallskip}

    (10)& \citet{Assef_2010} -- 4-band  & $\rm W3-\rm W4>2.1$ $\land$ $\rm W1-\rm W2>0.85$ $\land$\\
      &                            & $\rm W1-\rm W2>1.67\ (\rm W3-\rm W4) - 3.41$\\
    \noalign{\smallskip}

    (11)& \citet{wu12b}              & $\rm W1-\rm W2>0.57$\\
    \noalign{\smallskip}

    \hline\noalign{\smallskip}
    \multicolumn{3}{l}{{\it{Spitzer}} only criteria}\\    
    \noalign{\smallskip}

    (12)& \citet{Stern_2005}            & $\rm [5.8]-\rm [8.0] > 0.6$ $\land$\\
      &                            & $\rm [3.6]-\rm [4.5] > 0.2\ (\rm [5.8]-\rm [8.0]) + 0.18$ $\land$\\
      &                            & $\rm [3.6]-\rm [4.5] > 2.5\ (\rm [5.8]-\rm [8.0]) - 3.5$\\
    \noalign{\smallskip}

    (13)& \citet{Lacy_2004}             & $\log f_{8.0}/f_{4.5} > -0.2$ $\land$ $\log f_{5.8}/f_{3.6} > -0.2$ $\land$\\
      &                            & $\log f_{8.0}/f_{4.5} < 0.8\ \log f_{5.8}/f_{3.6}+0.5$\\
    \noalign{\smallskip}

    (14)& \citet{lacy07}             & $\log f_{8.0}/f_{4.5} > -0.2$ $\land$ $\log f_{5.8}/f_{3.6} > -0.1$ $\land$\\
      &                            & $\log f_{8.0}/f_{4.5} < 0.8\ \log f_{5.8}/f_{3.6}+0.5$\\
    \noalign{\smallskip}

    (15)& \citet{lacy13}             & $\log f_{8.0}/f_{4.5} > -0.3$ $\land$ $\log f_{5.8}/f_{3.6} > -0.3$ $\land$\\  
      &                            & $\log f_{8.0}/f_{4.5} < 0.8\ \log f_{5.8}/f_{3.6}+0.5$\\
    \noalign{\smallskip}

    (16)& \citet{Donley_2012}           & $\log f_{5.8}/f_{3.6} \geq 0.08$ $\land$ $\log f_{8.0}/f_{4.5} \geq 0.15$ $\land$\\
      &                            & $\log f_{8.0}/f_{4.5} \geq 1.21\ \log f_{5.8}/f_{3.6} - 0.27$ $\land$\\
      &                            & $\log f_{8.0}/f_{4.5} \leq 1.21\ \log f_{5.8}/f_{3.6} + 0.27$ $\land$\\ 
      &                            & $f_{4.5}>f_{3.6} \land f_{5.8}>f_{4.5} \land f_{8.0}>f_{5.8}$\\

    \hline\noalign{\smallskip}
    \multicolumn{3}{l}{NIR + {\it{Spitzer}} criteria}\\
    \noalign{\smallskip}

    (17)& \citet{messias12} -- KI    & $K_s -\rm [4.5] > 1.42$ $\land$ $\rm [4.5]-\rm [8.0] > 1.14$\\    
    \noalign{\smallskip}
    (18)& \citet{messias12} -- KIM   & $K_s -\rm [4.5] > 1.42$ $\land$ $\rm [8.0]-\rm [24] > 2.87$ $\land$\\ 
      &                            & $\rm [8.0]-\rm [24] > -2.9\ (\rm [4.5]-\rm [8.0]) + 8.47$\\    
    \noalign{\smallskip}

    \noalign{\smallskip}\hline

  \end{tabular}
  \bigskip\\

  {\footnotesize{In all the criteria above, the name of a band
      represents its magnitude in the Vega system, while $f_X$
      represents the flux density of band $X$. Note that [24] refers
      to the MIPS 24~$\mu$m band Vega magnitude, for which we have
      assumed a flux density zero point of 7.14~Jy. For empirically
      calibrated WISE selection criteria we indicate the magnitude
      range of the calibration sample if one was applied, as the
      comparison in Figure \ref{fig:MIR_phot_comp_rel} may be extrapolated to fainter
      magnitudes in some cases.\\
      \smallskip\\
      $^{\dagger}$Calibrated for galaxies with $\rm W1<18.50$ and $\rm W2<17.11$.\\
      \smallskip\\
      $^{\ddagger}$Calibrated for galaxies with $\rm W2<15.05$.\\
      \smallskip\\
      $^{*}$Calibrated for galaxies with $\rm W1<18.1$, $\rm W2<17.2$, $\rm W3<13.4$.\\
      \smallskip\\
      $^{\star}$Calibrated for galaxies detected with $S/N>5$ in all WISE bands used.
  }}
     
\end{table*}

\subsubsection{Comparison with general AGN identification at other wavelengths}

MIR AGN identification is considerably less sensitive to obscuration of the central engine by dust compared to optical identification, as dust opacity is lower at longer wavelengths,
and is therefore better for the selection of obscured AGN than optical identification, although its sensitivity to obscured AGN decreases with increasing redshift due to the K-correction 
\citep[e.g.][]{Assef_2011}. 

As discussed in Sect. \ref{sec:X-ray_sel}, X-rays are also sensitive to
obscured sources, especially in the harder X-ray energies, which are less affected by neutral
hydrogen absorption. 
The main advantage of the MIR over the X-rays is that the integration times needed for AGN identification are much shorter, and
hence allow for faster survey speeds. For example, \citet{gorjian08}
finds that 97.5\% of all sources in the 5~ks integrations of
{\it{Chandra X-ray Observatory}} of the XBo\"otes survey
\citep[0.5 -- 7.0 keV flux $> 8 \times 10^{-15}$ erg cm$^{-2}$ s$^{-1}$;][] 
{murray05} have counterparts detected in the 90~s integrations
of the IRAC Shallow Survey \citep[$f_{\rm 3.6 ~\mu m} > 12.3~\mu$Jy;][]{eisenhardt04}. Of the X-ray sources
without an IR match, 40\% might be
spurious. Furthermore, MIR identification can find AGN that are hard
to detect in the X-rays, either due to obscurations or to intrinsic
X-ray faintness \citep[e.g.][see also Sect. \ref{sec:X-ray_sel}]{stern12}. 

However, MIR identification is affected by contaminants and biases that are only 
marginally relevant to X-ray or optical selections. In the next sections we discuss these issues,
which need to be taken into account when drawing statistical
conclusions about the AGN population from MIR selected samples.

\subsubsection{Contaminants}

As mentioned earlier, AGN selection using MIR broad-bands mostly
relies on the typically redder colours of AGN, particularly in the observed
3--5~$\mu$m wavelength range. However, there are a number of different
populations that can mimic the colours of AGN in these bands and will
affect most selection criteria, although the extent will depend on
each specific selection. At redshift $\sim 0.2$, strongly SFGs 
with powerful polycyclic aromatic hydrocarbon (PAH) emission can have red enough colours to
be confused with AGN in some identification schemes
\citep[see e.g.][]{Stern_2005,Assef_2010,hainline16}. Such galaxies can
appear as contaminants in shallow and deep observations (see previous
section). As they are uncommon and the co-moving volume is low enough
at the respective redshift range, they are typically only a minor
contaminant. However, \citet{hainline16} has recently pointed out that
such contaminants can be confused with AGN hosted in dwarf galaxies and hence represent a major problem for their
identification \citep[e.g.][]{satyapal14,satyapal16}.

For deeper surveys, the most serious contaminants are high redshift
($z\gtrsim 1$) massive galaxies. At those redshifts the 1.6~$\mu$m
stellar bump is shifted into the MIR, and their colours can become red
enough to mimic those of AGN in many selection schemes
\citep[see][]{donley07,Donley_2012,yun08,Assef_2010,assef13,mendez13}. 
The selection technique developed by \citet{Donley_2012} for {\it{Spitzer}}
observations is specifically aimed at avoiding these galaxies,
resulting in high reliability at faint fluxes although at the cost of
completeness. Using W2 magnitude dependent prescriptions, 
as described in \citet{assef13}, has the same effect resulting also in high reliability at faint fluxes with a very low completeness level (see Fig. \ref{fig:MIR_phot_comp_rel}).

In addition to extragalactic contaminants, there are a number of
Galactic sources that can mimic the colours of AGN in the MIR, such as brown dwarfs or young stellar objects. Brown dwarfs are rare and typically only account for a small fraction of the
contaminants but might still affect the identification of high-redshift ($z>5$) AGN \citep{stern07}. 
Young stellar objects also have MIR colours that can be confused with those of AGN \citep[see, e.g.][]{koenig12}, and they can be numerous contaminants 
close to the Galactic plane. 

\subsubsection{Dust-free AGN and hot dust poor quasars}\label{sssec:IR_hdps}

Broad-band MIR AGN selection
primarily relies on the detection of the hot dust emission at low and
intermediate redshifts. While hot dust emission is rather prominent in most AGN, 
its ratio to the accretion disk emission
(usually described as the torus' apparent covering fraction) shows a wide
distribution \citep[see e.g.][]{roseboom13,mateos16,hernan16}. Therefore, AGN with
low hot dust emission relative to that of their host could escape identification, especially 
if they reside within luminous hosts. 
Such objects account for $\sim 10\%$ of the quasar population selected in the X-rays, optical or MIR \citep[][but see also \citealt{Lyu_2017}]{hao10,hao11} and their fraction may be increasing with redshift \citep{hao10,mor11,Jun_2013}. 
According to \cite{hao11}, their small amount of dust seems to be sufficient to place them to the borders of the \citet{Lacy_2004} criteria, but they would be systematically missed by a
more stringent selection.

MIR selections would of course completely miss dust-free AGN, such as the local analogues 
of the two sources 
identified by \citet{jiang10} among a sample of 21 $z\sim 6$ quasars
observed with {\it{Spitzer}}. Although such analogues may not necessarily exist, 
they might occur more often at lower luminosities. 
In fact, \cite{Elitzur_2006} showed
that if the torus is populated by outflowing clouds of material from
the accretion disk, at $L\lesssim 10^{42}~\rm erg~\rm s^{-1}$ the
accretion would be too low to sustain the required outflow rate resulting in the disappearance of the torus. 
This may apply to some LEG RGs (see also Sect. \ref{sec:radio_unif}). A clear case among the observational evidence  \citep[e.g.][]{chiaberge99,maoz05,muller13} is
the nearby LEG M87, that has been shown to lack  the thermal radiation of the
torus \citep{whysong04, perlman07}. Such objects would be missed systematically by all MIR selection criteria at low redshifts.

\subsubsection{Eddington Ratios and BH Masses}\label{ssec:MIR_phot_ledd}

The luminosity of the
spheroidal component of a galaxy correlates with the mass of its SMBH
\citep[at least at relatively low redshifts: e.g.][]{marconi03,gultekin09} and
hence with its Eddington limit. The ratio of the specific luminosity
of the AGN to that of the host, $L_{\nu, \rm AGN}/L_{\nu, \rm Host}$,
therefore directly depends on the AGN $L/L_{\rm Edd}$. In other words,
the incompleteness due to host dilution directly translates into a bias of MIR AGN
selection against low $L/L_{\rm Edd}$. Such a bias has been discussed
by, e.g., \cite{hickox2009} (see also Fig. \ref{fig:schematic3}) and \cite{mendez13}, who showed that 
MIR AGN selection only probes the upper end ($\gtrsim 1\%$) of the $L/L_{\rm Edd}$ 
distribution compared to X-ray selection (this applies also to optically selected AGN: 
see Sect. \ref{sec:opt_Edd_masses}). 
Quantifying this bias is difficult, as it strongly
depends on the selection function being used, but it needs to be taken into account and the complete selection function needs to be modelled in order to be able to give a physical and statistical interpretation of results based on MIR-selected AGN.

Although in an indirect manner, this selection effect can also bias
the distribution of BH masses in MIR identified AGN. As
BH masses correlate only with the luminosity of the host
galaxy's spheroidal component, the selection effect discussed above
will be stronger in galaxies with important disk or irregular
components, as the starlight from them will increase $L_{\nu, \rm
  Host}$ for a fixed BH mass. In other words, AGN in galaxies
with disk components need to accrete at higher $L/L_{\rm Edd}$ to be
identified by MIR selection techniques. As such non-bulge
components are typically more prominent in lower mass galaxies, and
these host lower mass BHs \citep{magorrian98}, the latter will
be systematically underrepresented in MIR AGN samples.

\subsection{Red, reddened and high-redshift AGN}\label{ssec:other_ir}

NIR photometry has also been used to select AGN \citep[e.g.][]{warren00,francis04,kozuma10}. Such methods, however, offer little gain compared to MIR and optical selection. Nevertheless, NIR has been useful in the identification of red AGN samples, e.g. 
\citet{glikman07} using 2MASS, FIRST and $R-$band photometry, as well as
the heavily reddened quasars of \citet{banerji12,banerji15} 
found in the UKIDSS survey.

Populations of heavily reddened AGN have also been found by means of
MIR photometry, often in combination with optical observations. For
example, using the colour between the  optical $R-$band and the MIPS 24~$\mu$m band, 
\citet{dey08} found a new class of objects, dubbed Dust-Obscured Galaxies or DOGs, many of which host heavily reddened AGN
\citep[see e.g.][]{melbourne12}. 
Recently \citet{ross15} presented a
selection of Extremely Red Quasars relying on SDSS and WISE
data. Similar is the case of the Hot Dust Obscured galaxies or
Hot DOGs \citep{eisenhardt12,wu12},
selected based purely on their WISE colours, for which a number of
studies suggest they probe an important stage of galaxy evolution
\citep[see e.g.][]{jones14,assef15,tsai15,diaz16}. 

Finally, NIR and MIR wavelengths are very important for the identification
of the earliest quasars in the Universe, which are not observable in the
optical due to Ly$\alpha$ forest absorption and the Lyman break. For example, the
highest redshift quasar currently known at $z=7.1$ was found by
\citet{Mortlock_2011} relying on the IR coverage of the UKIDSS
survey, and \citet{banados16} has recently used NIR and MIR
observations from 2MASS, VHS and WISE to help in the identification of
$z>5.6$ quasars candidates selected from the optical PanSTARRS\footnote{\url{panstarrs.stsci.edu/}}
survey. Upcoming NIR surveys such as those that will be provided by {\it{Euclid}} and the Wide-Field Infrared
    Survey Telescope ({\it WFIRST}) will allow
to better probe the quasar population at the earliest cosmic times (see Sect. \ref{ssec:IRfuture}). 

\subsection{MIR spectroscopy}\label{ssec:IR_spec}

MIR spectroscopy, particularly with the InfraRed Spectrograph (IRS; 
\citealt{houck04}) on board the {\it Spitzer} Space Telescope, provided new
insights into the physics and classification of AGN. The unambiguous
observations of the silicate feature at 9.7~$\mu$m in emission in many known 
AGN \citep{hao05, siebenmorgen05, sturm05, buchanan06, shi06} came 
as the long sought confirmation of the unified scheme. At the same time,
however, IRS observations indicated that in some cases the source of
obscuration resides in the host rather than the torus \citep[e.g.][]{goulding12, hatzimi15}.

Identification through MIR spectroscopy is very powerful, allowing
to detect obscured AGN components even when the MIR is dominated by
the host galaxy. Several classification diagrams have been developed
to determine the AGN contribution to an observed spectrum based on
certain spectral features, such as high ionisation emission lines like
[Ne\,\textsc {v}], [Ne\,{\sc ii}] and [O\,{\sc iv}], the EW of PAH
features and the strength of the silicate feature at 9.7~$\mu$m \citep[see,
  e.g.][]{spoon07,armus07,veilleux09,hernan11}. A number of techniques have
also been developed to model the observed MIR spectra and constrain
the AGN and starburst contributions \citep[see e.g.][]{schweitzer08,
  nardini08,deo09,feltre13}. 
  
Although MIR spectroscopy has had a great impact on our
understanding of AGN, the number of objects studied through these
techniques is limited when compared to photometric studies, as
spectroscopic observations require significantly longer integration
times. Ground-based observations are generally limited to the
brightest targets due to the effects of the Earth's atmosphere \citep[e.g.][]{Alonso-Herrero_2016}, 
while
deeper observations were possible with the IRS
during its cryogen-cooled phase. For the most part, such observations
were limited to $z\lesssim 1$ luminous IR galaxies (LIRGs), ultraluminous IR galaxies (ULIRGs), and quasars
\citep[][and references therein]{hernan11} 
although a number of higher redshift ULIRGs were also studied by
IRS \citep[see e.g.][]{kirkpatrick12}. The impact of these techniques
will be greatly expanded by the upcoming JWST (\citealt{gardner06}) and {\it{Space Infrared Telescope for
    Cosmology and Astrophysics}} (SPICA; \citealt{Nakagawa_2015}), that will probe
significantly fainter targets and will allow us 
to select new, currently inaccessible, sets of objects, as discussed next.

\subsection{The future of AGN studies in the IR band}\label{ssec:IRfuture}

The upcoming generation of ground-based giant telescopes will
significantly expand upon the current NIR and MIR capabilities, as
most of them will have significant focus on these wavelengths. The
Giant Magellan Telescope (GMT\footnote{\url{www.gmto.org}}) is planning on first generation
instruments with imaging capabilities from 0.9 to 2.5~$\mu$m and
spectroscopic capabilities in the 1--5~$\mu$m range, and its first
light is currently expected for 2021. The Thirty Meter Telescope (TMT\footnote{\url{www.tmt.org}})
is planning on first-light photometric and spectroscopic instruments
in the 0.8--2.5~$\mu$m range, with the goal of extending this range to 
28~$\mu$m within its first decade of operations. 
The first generation instrumentation of the largest of the upcoming giant
telescopes, the Extremely Large Telescope (ELT\footnote{\url{www.eso.org/sci/facilities/eelt}}), will
allow for imaging and spectroscopy all the way to 19~$\mu$m. Finally, the University of
Tokyo Atacama Observatory (TAO\footnote{\url{www.ioa.s.u-tokyo.ac.jp/TAO/en}}) 6.5m telescope
will focus on the
IR, with planned first light instrumentation probing wavelengths
as long as 38~$\mu$m.

The only space-based observatories  with MIR imaging
capabilities currently in operation are {\it{Spitzer}}, whose operations 
have recently been
extended until 2019 by the 2016 NASA Senior Review, and WISE, whose
NEOWISE-R phase is planned to extend until the end of 2017. In the near future, NASA's
JWST\footnote{\url{www.stsci.edu/jwst}}, expected to launch in 2018, will offer unprecedented imaging and spectroscopic
capabilities in the 0.6--28.3~$\mu$m wavelength range thanks to its
6.5m diameter primary mirror. Observations with the JWST will
probe with high angular resolution a number of targets that are not
accessible from the ground, likely having a major impact in our
understanding of AGN. The proposed joint JAXA/ESA mission 
SPICA\footnote{\url{www.ir.isas.jaxa.jp/SPICA/SPICA_HP/index-en.html}}, that will be launched in 2028 if approved by the two agencies, will have a 2.5m aperture and will provide
low-to-high resolution spectroscopy in the wavelength range between 35
and 240~$\mu$m, and imaging capabilities. 
Finally, the upcoming {\it{Euclid}} and {\it WFIRST} missions will have a
significant impact in NIR AGN identification by mapping large areas of
the sky to very faint depths. {\it{Euclid}}\footnote{\url{www.euclid-ec.org}}, expected to
launch in 2020, will have a 1.2m primary mirror and will survey
$15,000~\rm deg^{2}$ of extragalactic sky down to limiting AB
magnitudes of 24 in $Y$, $J$ and $H$, as well as 24.5 in a very wide
optical broad-band. It will also observe a smaller region
of $40~\rm deg^{2}$ to limiting AB magnitudes of 26 in the NIR bands
and 26.5 in the optical band. The {\it{WFIRST}} mission\footnote{\url{wfirst.gsfc.nasa.gov}}, expected to
launch by 2024, will, on the other hand, have a 2.4m primary mirror
and a similar set of bands, and is planned to image
2,227 $\rm deg^{2}$ down to a limiting AB magnitude of 26.7 in $J$
\citep{spergel15}. Both telescopes will also obtain NIR slitless grism
spectroscopy in their survey areas. Through their unique combination
of area and depth, both surveys will probe AGN activity during the
formation of the first galaxies in the Universe.

\section{Optically-selected AGN}\label{sec:optical}

This section discusses the selection and properties of
optically-selected AGN as contrasted with investigations at other
wavelengths. The focus here is on the more luminous subsets that would
typically be classified as quasars or Seyfert 1 galaxies. We will not
cover objects like the host-galaxy dominated AGN
\citep{Kauffmann_2003a}, the LINERs \citep{Heckman_1980}, or XBONGs
\citep{Hornschemeier_2005}.  While optical surveys are able to
identify quantitatively {\em more} AGN than other wavelengths (through
a combination of area and depth), this size comes with a bias towards
brighter, unobscured sources with $L/L_{\rm Edd} > 0.01$ (see 
Sect. \ref{sec:opt_Edd_masses}). Even though optical surveys are not
ideal for probing obscured AGN, we discuss how they can guide our
search for them.  The bias towards unobscured sources in the optical
is partially mitigated, however, by an increase in information content
for the sources that {\it are} identified---in the form of physics
probed by the combination of optical continuum, absorption, and
emission. An example is the ability to estimate the mass of the SMBHs
based on the optical/UV emission lines. We discuss the physical
mechanisms behind optical emission in Sect. \ref{sec:opt_physics} and
AGN identification, selection effects, information content, and the range of masses (and
accretion rates) probed by the optical band in
Sect. \ref{sec:opt_ident} -- \ref{sec:opt_Edd_masses}.  While the redshift/luminosity evolution of
optically-selected luminous quasars would seem to be well constrained
from today to redshift $\sim6$, we review evidence suggesting that the
\citet{Hopkins_2007} bolometric LF needs to be updated. How
next-generation surveys such as LSST can bridge the evolution of
luminous quasars to lower-luminosity AGN (that are typically better
probed at other wavelengths) will be addressed in
Sect. \ref{sec:opt_evol}. 
In this section we cover the $\sim 1,000 - 8,000~\AA$ range (rest-frame).

\subsection{Physical mechanisms behind optical emission}\label{sec:opt_physics}

AGN are believed to be powered by accretion onto a SMBH, that gives
rise to high X-ray-to-optical luminosities, a characteristic
rest-frame UV/optical power-law continuum (very distinct from the
continuum of non-active galaxies) as well as the so called ``big blue
bump'', and a break of this continuum blueward of $\approx1000$~\AA~
(Fig. \ref{fig:SED}).  Many models, usually assuming a geometrically
thin, optically thick accretion disk, have been developed in order to
explain this emission \citep[e.g.][and references therein]{sun89,
  laor89, hubeny01}. AGN that have a line of sight to the central
engine that is not obscured show broad emission lines with typical gas
velocities of a few 1000 km s$^{-1}$ covering a large range in
strength and profiles, whose properties correlate with the luminosity
of the AGN \citep[e.g.][]{baldwin77}. The source of the broad emission
lines is the BLR, believed to be located between the SMBH and the inner
wall of the dusty torus (see Sect. \ref{sec:IR}), with photoionized
gas that has been heated by the radiation coming from the accretion
disk around the SMBH. Finally, AGN display narrow emission lines, with
gas velocities between 300 and 1000 km s$^{-1}$, originating in the
Narrow Line Region (NLR) with sizes $\approx$ hundreds of parsec, located above
(and below) the plane of the dust.


\subsection{Photometric and spectroscopic identification}\label{sec:opt_ident}

The features described in the previous section are the basis of optical AGN identification, be it
photometric or spectroscopic. Broad band photometry is sensitive to
the presence of broad emission lines in the various filters as a
function of redshift, as they alter the otherwise very typical colours
of the AGN that separate them from the stellar locus \citep[see,
e.g.][Fig. 4]{Richards_2001}. Narrow band surveys such as COMBO17
\citep{wolf03}, ALHAMBRA \citep{moles08}, and now J-PAS
\citep{benitez14} make use of the spectral features to not only
identify AGN but to also estimate their (photometric) redshifts with a
precision that reaches below 1\% \citep[e.g.][]{Salvato_2009,Hsu_2014}, 
a great improvement with respect to
early attempts \citep{hatzimi00}. Finally, we are generally reliant on
optical {\em spectroscopy} to provide confirmation of a source as an
AGN and to determine its redshift, while the presence of narrow
emission lines in the spectra of galaxies and their ratios are
indicative of the presence of an AGN \citep[e.g.][]{feltre16}.

The problem with the optical band, as compared to, say, the hard X-rays, is that
bright optical sources are not necessarily AGN. 
The same is true for the radio (Sect. \ref{sec:FS_SS}) and IR to some extent (Sect. \ref{ssec:MIR_phot_sel})---the brightest
sources on the sky have a high probability of being an AGN. This
point is illustrated in the comparison of number counts between the X-ray
\citep[][Fig.~5]{Lehmer_2012} and optical
\citep[][Fig.~24]{Shanks_2015}: in the optical, stars far outnumber AGN at
typical survey depths.

\begin{figure}
\centering
\vspace{-0.5em}
\includegraphics[width=8.5cm]{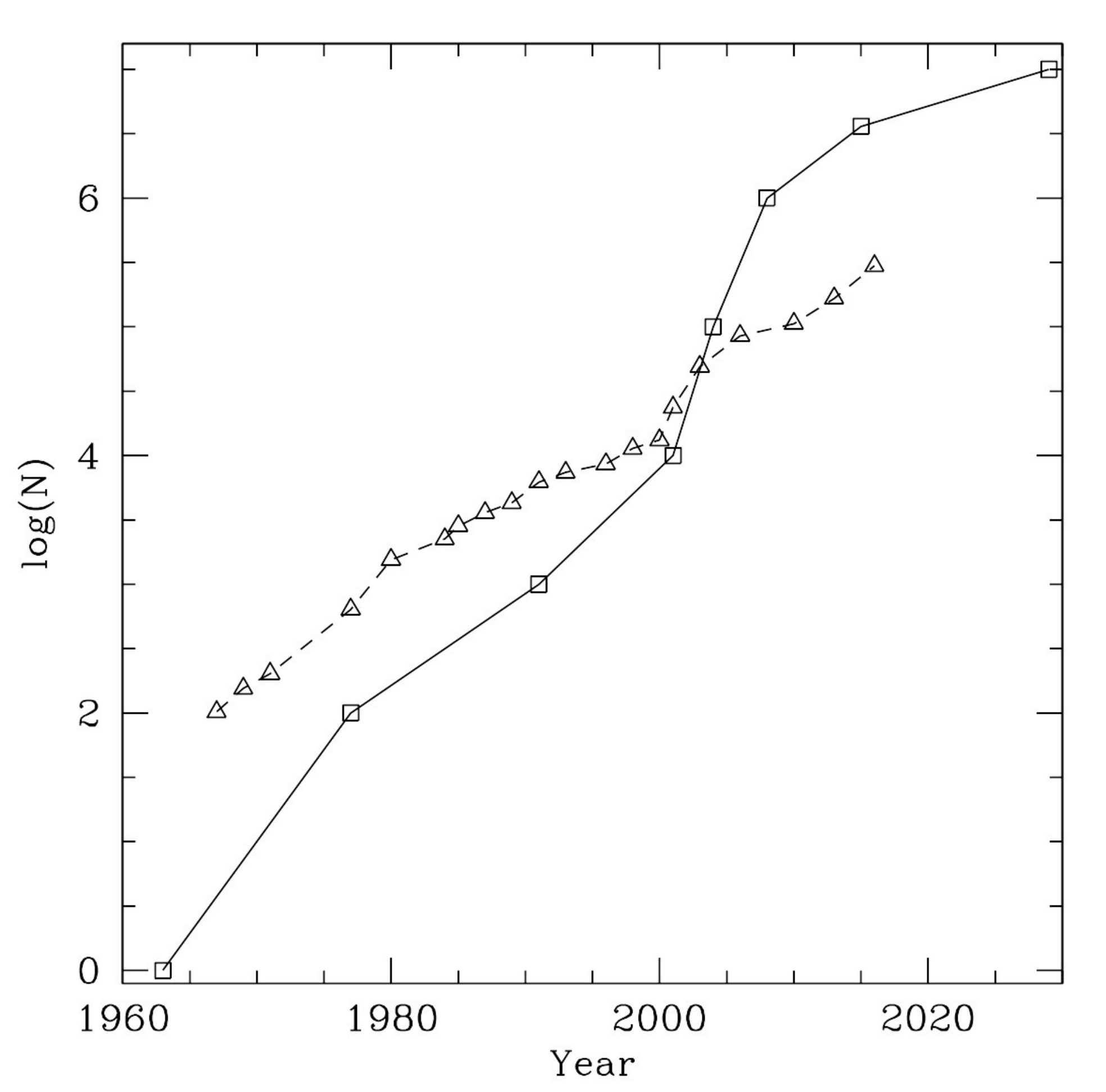}
\caption{Number of quasars as a function of time. The dashed line and triangles give the number of quasars
in the largest heterogeneous samples to date. The solid line and squares give the largest homogeneous quasar
samples (including photometric quasar candidates) to date---extrapolating 
to 10,000,000 expected AGN for LSST \citep{LSST}. Adapted from \cite{Richards_2009}; the 2015 photometric point is
from \cite{DiPompeo_2015}.}
\label{fig:qsoswtime}       
\end{figure}

Optical selection can make up for this (photometric) uncertainty with
sheer numbers. Deep, high-resolution X-ray, IR and radio fields provide a
much higher AGN density---up to $\sim 24,000$\,deg$^{-2}$
\citep[Sect. \ref{sec:X-ray_phys}; e.g.][see also 
Sect. \ref{sec:discussion}]{Luo_2017}, but only over minuscule areas.
The density of the largest-area photometric and spectroscopic quasar
samples is only $\sim 150$\,deg$^{-2}$, but over a large fraction of
the sky, which results in larger samples of AGN overall
\citep[e.g.][]{Richards_2009,DAbrusco_2009,Bovy_2011,Flesch_2015,Brescia_2015,Paris_2017}.
Figure~\ref{fig:qsoswtime} shows the growth in quasar numbers with
time for both heterogeneous and homogeneous quasar samples. We discuss
the impact of future facilities in Sect. \ref{sec:opt_evol}.

\subsection{Selection effects}\label{sec:opt_bias}

Paradoxically, many (optically) unobscured AGN are missed by optical surveys.
These are objects whose colours put them in (or close to) the stellar
locus.  Since luminous quasars are point sources, but are outnumbered
by stars in our galaxy by $\sim 100:1$ at the SDSS depth, it is very difficult to create
a complete sample of quasars at certain redshifts---without
considerable stellar contamination. The redshifts affected span a
large range around $z\sim2.6$ with very low completeness, and a
smaller range around $z\sim3.5$
\citep{Richards_2002,Richards_2006b,Worseck_2011}.  Even the
SDSS-III/BO\-SS quasar sample is only 60\% complete at these redshifts despite being
designed to identify $2.2<z<3.5$ quasars \citep{Ross_2013}.  It is
worth noting that most MIR selections have a similar redshift ``hole'' over
$4 \lesssim z \lesssim 5$ (Sect. \ref{ssec:MIR_broad_MIR})
but this can be filled in by using optical {\em and} MIR data
simultaneously \citep{Richards_2015}.

Lower-luminosity AGN are also a challenge for imaging-only optical
surveys (like the Dark Energy Survey [DES\footnote{\url{www.darkenergysurvey.org}}] and LSST\footnote{\url{www.lsst.org}}) for the reasons noted above: without
spectrosco\-py, it is difficult to distinguish a normal galaxy from an
active galaxy. Variability selection may not help all that much for
such sources considering that, while the amplitude of variability
increases at lower luminosity \citep[][see also Sect. \ref{sec:var_lum}]{VandenBerk_2004}, 
the fraction of optical emission from the central engine decreases. Yet this is the
population that we most need to probe, especially for comparison to
X-ray and MIR samples.

The biggest hole in the selection of AGN via optical
photometry is certainly obscured (or type 2) AGN. It does not 
matter whether the optical obscuration is by a smooth torus or a clumpy one (see Sect. \ref{ssec:IR_SED} for a full list of 
references on smooth and clumpy distributions):
the traditional AGN model has a region where most of the optical
emission is obscured. The irony is that 
a large fraction of known type 2 AGN are still selected in the optical
\citep{Zakamska_2003,Reyes_2008,Alexandroff_2013,Yuan_2016}.  This
result is due to a combination of effects: the host galaxy is not
(always) obscured and both strong emission lines and scattering can
result in non-negligible optical flux producing unusual (or even
AGN-like) colours, which can cause them to be identified as potential type 1 sources despite 
them being type 2. 

Thus the question of the relative fraction of obscured and unobscured
AGN is still much debated/investigated (see also Sect. \ref{ssec:IR_SED}), particularly as a function of
luminosity \citep[e.g.][]{Gilli_2007,ueda2014}. Crucially, differences in the
definition of ``obscured'' between the optical and X-ray
\citep[e.g.][]{Hickox_2007} make it more difficult than one might
imagine to paint a full picture (see also Sect. \ref{sec:X-ray_types}). 

\subsection{Information content}\label{sec:opt_info}

In Sect. \ref{sec:IR} we saw that IR-detected quasars are relatively
unbiased against type 2 AGN, whereas most of the optical light
comes from the central accretion disk, which is blocked in these
sources.  However, while the optical may be missing a crucial
component of the AGN zoo in terms of obscured AGN, it more than makes
up for that loss in terms of information content of those AGN that
{\em are} detected.  Moreover, the information content in the
continuum, emission lines, and absorption lines from optical
spectroscopy is particularly rich.  For example, BAL quasars
\citep{Weymann_1991} have provided us with direct evidence of winds in
AGN systems with outflow velocities extending to tens of thousands km s$^{-1}$ 
\citep[e.g.][]{Hamann_2011}.

While the BAL sub-class represents only $\sim 20\%$ of the population
of {\em luminous} quasars \citep{Hewett_2003}, the advent of principal
component analysis, both using photometry \citep{Boroson_1992} and
spectroscopy \citep{Francis_1992,Yip_2004} and large data sets from
SDSS \citep{Schneider_2010,Paris_2017}, has enabled the community to
treat quasars as diverse systems.  For example, \citet{Richards_2011}
argue that we can learn about winds using {\em emission} lines in
addition to absorption lines, which potentially turns every quasar
into powerful probe of AGN outflows.

\begin{figure}
\centering
\includegraphics[width=9.0cm]{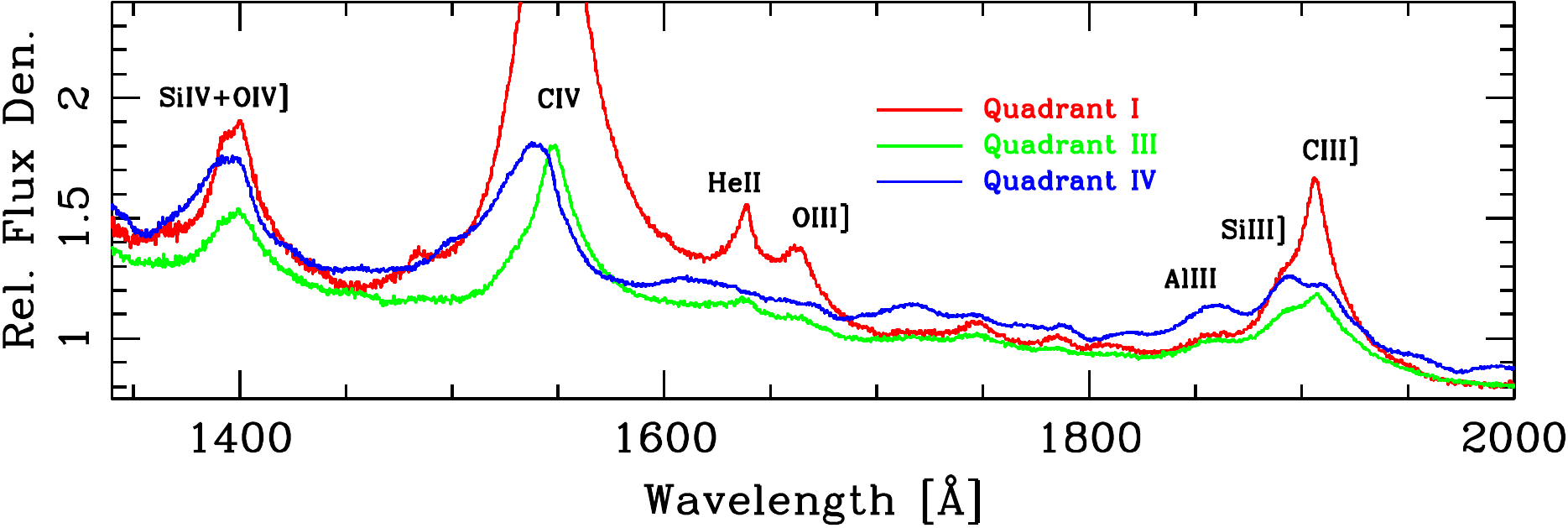}
 \caption{Diversity of UV emission line properties in SDSS quasars. Quadrants refer to the C \textsc{iv}  EW-blueshift plane with red objects having low C \textsc{iv}  blueshifts and large EW and blue objects having large C \textsc{iv}  blueshift and small EW.  These correlated features may be indicative of differences in the hardness of the spectral energy distributions \citep[e.g.][]{Leighly_2004}.}
\label{fig:emlines}       
\end{figure}

Indeed, one way to illustrate the diversity of quasars is presented in
Fig.~\ref{fig:emlines}, which shows how physical trends are manifested
in the different emission lines: quasars with harder spectra have
stronger emission lines and less ``blueshift'' of the C \textsc{iv}
emission line.  The probability of a quasar being radio detected or
having strong BAL troughs is a strong function of the appearance of
these emission lines \citep{Richards_2011}. The trends shown in
Fig.~\ref{fig:emlines} appear to be related to those that define the
``Eigenvector 1''\footnote{This is a set of correlations between properties observed 
in quasar spectra, which comes out of principal component analysis.}
 parameter space
\citep{Boroson_1992,Brotherton_1999,Sulentic_2000,Sulentic_2007},
which together highlight the great diversity of AGN even when
considering only those that are both optically selected and very
luminous.  Often overlooked is the fact that this diversity has
important implications for accurate determination of quasar redshifts
\citep{Hewett_2010}.

\subsection{$L/L_{\rm Edd}$ and $M_{\rm BH}$}\label{sec:opt_Edd_masses}
  
Arguably the best example of where optical provides additional
information content and makes up for selection effects is in our
ability to utilise BH mass {\em scaling relations} to estimate the
masses of the BHs powering quasars \citep[e.g.][]{Vestergaard_2006}.  Again,
this process requires optical spectroscopy\footnote{NIR spectroscopy
  can be used as well, but the sample size of objects with appropriate
  data is relatively small by comparison \citep[but
  see][]{Ricci_2017}.}. By: (1) knowing the width of the emission
lines; (2) assuming that that width is dominated by gravitational
effects; (3) having an estimate of the characteristic radius of the
emitting gas assuming $R\propto L^{\sim0.5}$ \citep{Bentz_2009}; and (4) calibrating this
information against the few dozen objects for which there exists
``reverberation mapping'' \citep{Peterson_1993} data, we can then estimate the masses of the
BH in every type 1 quasar.

Generally speaking the types of AGN discussed herein have masses of
$10^6$--$10^{10}~M_{\odot}$ and accretion rates (in terms of
$L/L_{\rm Edd}$) of 0.01--1
\citep{Greene_2007,Vestergaard_2008,Shen_2012,Trakhtenbrot_2012}.  See
\citet{Schulze_2015} for a recent, comprehensive analysis that
combines data from the VIMOS-VLT Deep Survey (VVDS), zCOSMOS, and SDSS.

Scaling relations for estimating the masses of BHs in relatively local
AGN (up to $z\sim0.7$) make use of the H$\beta$ emission line and are
thought to be relatively robust (at the level of $\sim 0.3$ dex).
Using the Mg \textsc{ii} emission line these scaling relations have
been extended to $z\sim1.9$.  However, beyond that redshift other
broad emission lines need to be used.  As such, attempts have been
made to calibrate C \textsc{iv} to produce BH mass estimates, though
it is becoming clear that the winds discussed above can significantly
bias the estimates for high-redshift quasars which rely on the C
\textsc{iv} emission line
\citep{Baskin_2005,Richards_2011,Denney_2012,Shen_2013}.  As the
sample size of high-redshift quasars with both optical and IR
spectroscopic coverage grows, corrections to this scheme might help
bringing the BH masses into alignment
\citep[e.g.][]{Runnoe_2013,Coatman_2016}. However, the uncertainties
associated with the determination of SMBH based on C \textsc{iv} are
not solely due to low spectral resolution and/or signal-to-noise ratio
(S/N) but rather point towards differences in the physics of the
BLR. In other words, either the C \textsc{iv}-emitting gas is
non-virialised or objects with low and high FWHM(H${\beta}$) have
different ionisation structure, since the FWHM(C \textsc{iv}) has only a loose
correlation, if any, with FWHM(H${\beta}$) \citep{Denney_2013,mejia16,Coatman_2017}.

\subsection{The evolution with redshift and the impact of future facilities}\label{sec:opt_evol}

The SDSS quasar LF \citep{Richards_2006b} was unique
not in the redshift range or luminosity that it probed, but rather
because it probed such a large range with just one
uniform data set with a large number of quasars. One of the
shortcomings of the wide, but shallow SDSS work was that it generally
only probed the bright end of the LF, whereas narrow,
but deep X-ray surveys were better able to probe the faint end.
\citet{Hopkins_2007} combined the best of both worlds from various
multi-wavelength surveys to create a bolometric LF.

Recent works, however \citep[e.g.][]{Assef_2011,Ross_2013,McGreer_2013}, 
have shown that the evolution of the break
luminosity (where the slope of the LF changes rapidly) may be very different from that predicted by
\citet{Hopkins_2007} and that the faintest SDSS quasars might start probing the faint end 
(rather than the bright end, as was assumed) of the LF at high $z$. 
What is needed here are both deeper surveys in the optical and wider
surveys in the X-ray (Sect. \ref{sec:X-ray_future}), or similarly deep/wide NIR surveys
(Sect. \ref{ssec:IRfuture}). But even
without new data there is a decade of observations that could be
incorporated into an updated bolometric LF.

Improving our knowledge of the quasar LF has
consequences beyond the study of quasars: it also has important
consequences for reionisation in the early Universe. For example, the
results of \citet{McGreer_2013} and \citet{Glikman_2011} suggest
photoionisation rates at $z > 4$  that do not fully agree. 

It is important to understand that the bias in the optical band is towards
the high luminosity end of the AGN distribution, i.e. bona fide
quasars. In part that is because that is the population that optical
surveys themselves are biased to. However, current and upcoming
experiments like Pan-STARRS\footnote{\url{pswww.ifa.hawaii.edu/pswww/}}, DES, and LSST (will) 
probe to fainter limits that are more compatible with the lower-luminosity AGN 
(where the host emission is more comparable to the central engine) that surveys
at other wavelengths most commonly identify. These new optical surveys will help
complete a multi-wavelength bridge that will allow AGN astronomers to
more fully sample luminosity-redshift space across the full
electromagnetic spectrum, as illustrated by
\citet[][Fig.~14]{LaMassa_2016}.

\section{X-ray-selected AGN}\label{sec:X-ray}

We discuss in this section X-ray-selected AGN. The physical mechanism behind 
X-ray emission is examined in Sect. \ref{sec:X-ray_phys}, while Sect. \ref{sec:X-ray_sel}
deals with AGN selection and its challenges. Sect. \ref{sec:X-ray_types} reviews the types of AGN selected in the
X-ray band, the range of $M_{\rm BH}$ and $L/L_{\rm Edd}$ they sample, and their evolution,
while Sect. \ref{sec:X-ray_future} gives an overview of the future of this field. 
In this review we define the X-ray band as covering the energy range of 0.2--200~keV. 

\subsection{Physical mechanism behind X-ray emission}\label{sec:X-ray_phys}

X-ray observations provide a near complete selection of AGN with low
contamination from non-AGN systems. The
primary reasons for this are: (1) X-ray emission from AGN appears to be
(near) universal; (2) X-rays are able to penetrate through large
column densities of gas and dust (particularly at high X-ray
energies); (3) X-ray emission from host-galaxy processes are
typically weak when compared to the AGN \citep[see also Sect. 1.1 
of][]{brandt2015}. On the basis of these advantages, the deepest
blank-field cosmic X-ray surveys have identified the largest reliable
AGN source density to date ($\approx$~24,000~deg$^{-2}$;
\citealt{Lehmer_2012,Luo_2017} and Sect. \ref{sec:discussion}).

The intrinsic X-ray emission from AGN is due to processes related to
the accretion disk (see \citealt{mushotzky1993,done2010,gilfanov2014}
for reviews; note that in jetted AGN the jet can make 
a major contribution in the X-ray band as well)\footnote{In this section we focus on high accretion-rate
  AGN (with an optically thick and geometrically thin accretion disk;
  i.e.\ $L/L_{\rm Edd} > 0.01$), which account for the
  majority of the BH growth in the Universe
  (e.g.\ \citealt{ueda2014,aird2015}). Low accretion-rate AGN (with
  an optically thin, geometrically thick, hot accretion flow;
  i.e.\ $L/L_{\rm Edd} < 0.01$) can also be selected
  at X-ray energies, although the accretion process is driven by
  advection of a hot plasma; see Sect. \ref{sec:LEG_HEG}, \cite{done2007}, and \cite{yuan2014}
  for more details.}. The primary process is thought to be inverse Compton
scattering of the accretion-disk photons to X-ray energies via the
accretion-disk ``corona'' (see Fig. \ref{fig:SED}; this is generally depicted as an atmosphere 
above the inner accretion disk, though its exact geometry is unknown). However, thermal X-ray emission due to the inner regions of
the accretion disk can also be produced at the lowest X-ray
energies \citep[e.g.][]{Sobolewska_2004}. The X-ray emission is then modified due to the interaction
with matter in the nuclear region (e.g.\ reflection, scattering, and
photo-electric absorption of photons from the accretion disk and/or
the obscuring AGN torus: see Sect. \ref{ssec:IR_SED}). The relative strength of these
components can vary quite significantly from source to source, mostly
due to differences in the geometry and inclination angle of the
torus to the line of sight, leading to a broad range of X-ray
spectral shapes. The intrinsic X-ray emission from the ``corona'' is
tightly connected to the accretion-disk emission (as parameterised by
$\alpha_{\rm OX}$; e.g.\ \citealt{steffen2006,lusso2016}) for almost
all systems (for exceptions see, e.g.\ \citealt{wu2011,luo2014}),
demonstrating a causal relationship and showing that X-ray emission
from AGN is (near) universal. 

A large number of X-ray observatories have been launched since the
first pioneering rocket flights of the 1960s (see
\citealt{giacconi2009} for a review). The majority of the results
presented here have been obtained from the most sensitive
X-ray observatories in operation, all of which employ
grazing-incidence optics to focus X-ray photons and achieve high
sensitivity at high spatial resolution: {\it Chandra}, {\it
  XMM-Newton}, and {\it NuSTAR}. {\it Chandra} (launched July 1999;
\citealt{weisskopf2002}) provides up-to sub-arcsecond imaging at
$\approx$~0.3--8~keV with sufficient collecting area for high
S/N X-ray spectra of bright X-ray sources. The low
background, high spatial resolution, and good collecting area allows
{\it Chandra} to detect sources three orders of magnitude fainter than
previous-generation X-ray observatories. {\it XMM-Newton} (laun\-ched
December 1999; \citealt{jansen2001}) provides lower spatial resolution
imaging than {\it Chandra} ($\approx 5^{\prime\prime}$ FWHM) 
over $\approx$~0.2--12~keV but with substantially larger
collecting area, allowing for higher S/N X-ray
spectra of bright X-ray sources than {\it Chandra}. {\it NuSTAR}
(launched June 2012; \citealt{harrison2013}) provides lower spatial
resolution imaging than {\it Chandra} and {\it XMM-Newton}
($\approx 18^{\prime\prime}$ FWHM and $\approx 58^{\prime\prime}$
half-power diameter) but is sensitive to much higher energy photons
($\approx$~3--79~keV). In comparison to previous generation
observatories with sensitivity at $>10$~keV {\it NuSTAR} is able to
detect sources two orders of magnitude fainter.

\begin{figure}
\centering
\includegraphics[width=8.4cm]{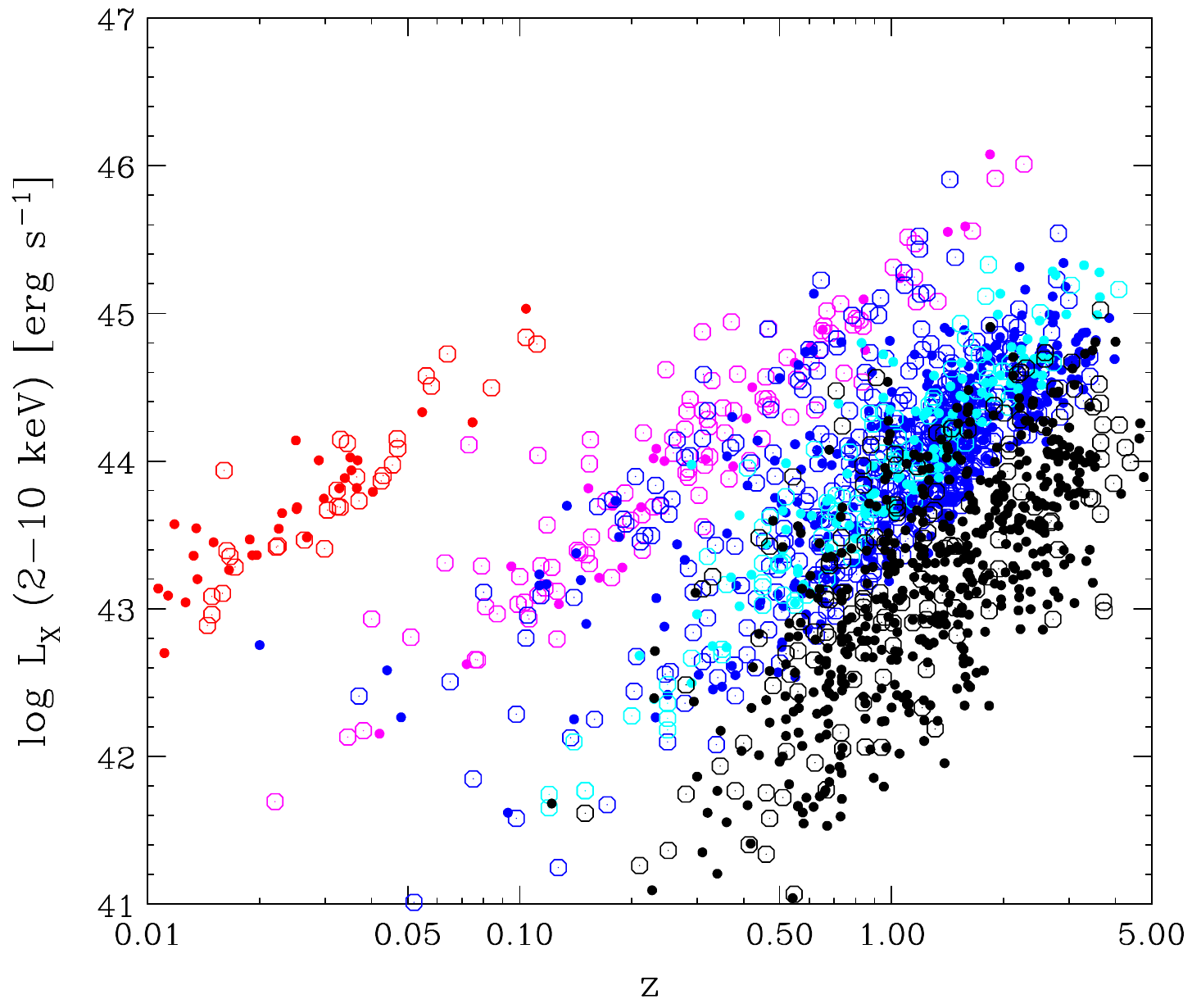}
\caption{X-ray luminosity versus redshift for sources detected in a
  selection of blank-field cosmic surveys undertaken with {\it Swift}-BAT
  (red), {\it ASCA} (magenta), {\it XMM-Newton} (blue), and {\it
    Chandra} (cyan and black). The open and filled symbols indicate
  X-ray unabsorbed ($N_{\rm H}<10^{22}$~cm$^{-2}$) and X-ray absorbed
  ($N_{\rm H}>10^{22}$~cm$^{-2}$) systems, respectively. \copyright~AAS. Figure reproduced from 
  \cite{ueda2014}, Fig. 3, with permission.}
\label{fig5_1}       
\end{figure}

\subsection{Selection of AGN in the X-ray band: identification challenges}\label{sec:X-ray_sel}

The (near) universality of X-ray emission from AGN, the high
penetrating power of X-rays, and the low contamination from
host-galaxy emission, mean that AGN selection effects at X-ray
energies are generally modest, particularly at high energies
(rest-frame energies $>10$~keV). Fig.~\ref{fig5_1} 
demonstrates the broad $L_{\rm X}$--$z$ plane coverage for AGN
selected from blank-field cosmic X-ray surveys: AGN with $L_{\rm
  X}>10^{44}$~erg~s$^{-1}$,  $L_{\rm X}=10^{42}$--$10^{44}$~erg~s$^{-1}$, 
  and $L_{\rm X}<10^{42}$~erg~s$^{-1}$ are typically classified as high luminosity
(broadly corresponding to ``quasars''), moderate luminosity (broadly corresponding 
to ``Seyfert galaxies''), and LLAGN respectively.

\begin{figure}
\centering
\includegraphics[width=8.4cm]{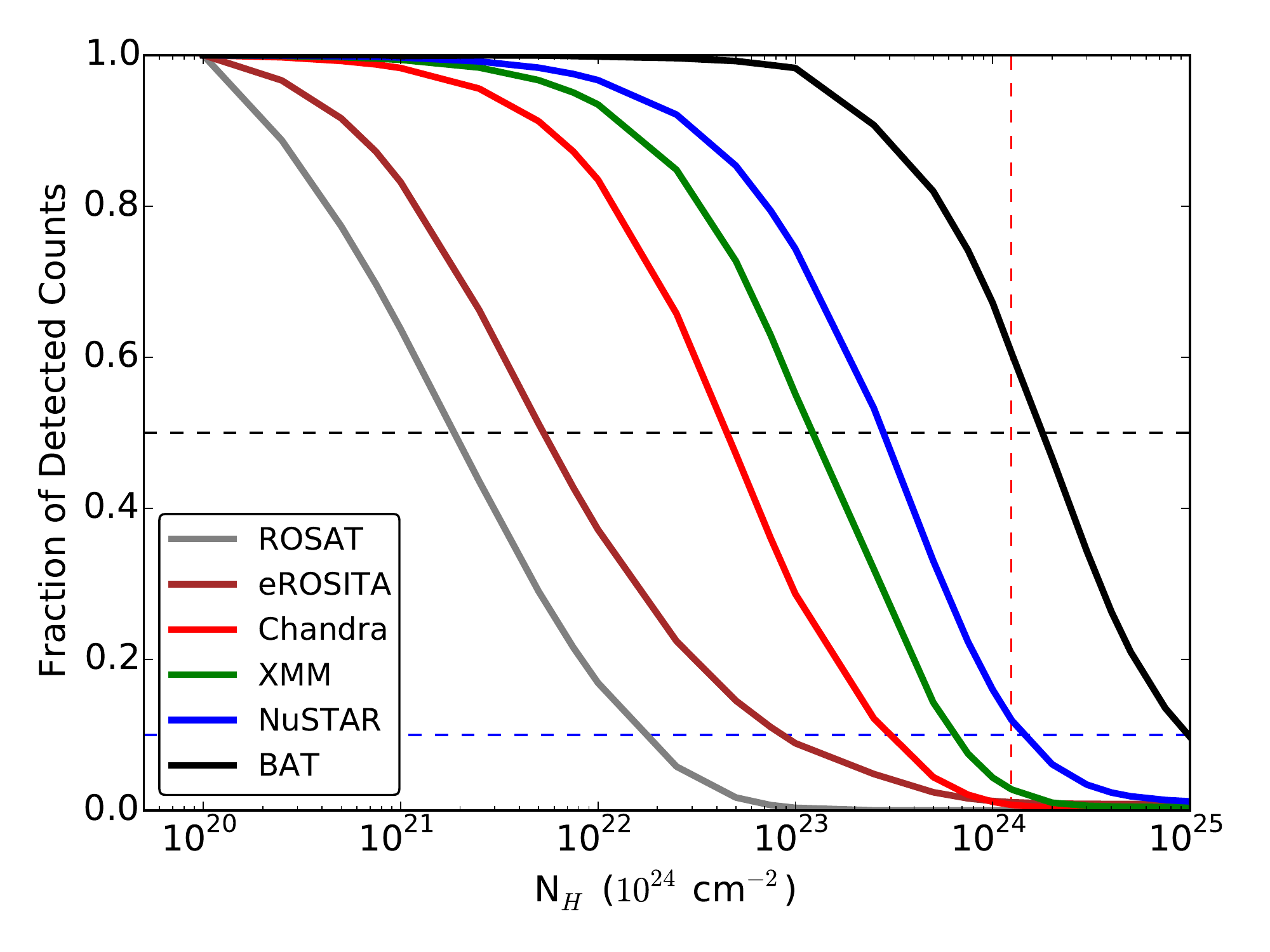}
\caption{The impact of varying amounts of absorption on the fraction
  of X-ray detected counts (with respect to the case for no
  absorption) for an AGN at $z=0$ for a variety of X-ray
  observatories: {\it ROSAT} (grey; $\approx$~0.2--2.4~keV), 
    eROSITA (burgundy; $\approx$~0.2--5~keV), {\it Chandra} (red;
  $\approx$~0.3--8~keV), {\it XMM-Newton} (green;
  $\approx$~0.2--12~keV), {\it NuSTAR} (blue: $\approx$~3--79~keV),
  and {\it Swift}-BAT (black: $\approx$~14--195~keV). A cutoff power law 
  with $\Gamma = 1.9$ and E$_c > 200$ keV has been assumed. \copyright~AAS. 
  Figure reproduced from 
  \cite{koss2016}, Fig. 2, with permission.}
  \label{fig5_2}       
\end{figure}

For the majority of AGN, the most significant selection effect is that
of absorption, which is a function of rest-frame X-ray energy: lower
energy X-rays are more easily absorbed than higher energy X-rays
(\citealt{wilms2000}). The impact of varying amounts of absorption on
the detection of X-ray photons for a $z=0$ AGN from a selection of
X-ray observatories is shown in Fig.~\ref{fig5_2}. The effect of
absorption on the fraction of X-ray detected counts is clear for all
of the X-ray observatories but is only significant at $>10$~keV for
heavily obscured AGN ($N_{\rm H}>3\times10^{23}$~cm$^{-2}$). Since an
increase in redshift leads to an increase in the rest-frame energies
probed within a given observed-frame band, the effect of absorption is
less significant at high redshift than low redshift. However,
CT AGN (i.e.\ AGN where the absorbing column density exceeds 
the inverse of the Thomson scattering cross section: $N_{\rm
  H}>1/{\sigma_{\rm T}}>1.5\times10^{24}$~cm$^{-2}$) are challenging
to detect even at high rest-frame energies due to Compton recoil and
subsequent absorption of the X-ray photons (see \citealt{comastri2004}
for a review). Other wavebands that are less sensitive to the effects
of heavy obscuration (e.g.\ the radio and IR wavebands; see Sects. \ref{sec:radio}
and \ref{sec:IR}) will be less biased towards the detection of CT 
AGN. However, the unambiguous identification of the signatures of
CT absorption and the measurement of absorbing column
densities requires X-ray observations. This is most effectively
achieved from broad-band X-ray spectral fitting (i.e.\ the
combination of [ideally simultaneous] data from, for example, {\it NuSTAR} and {\it Chandra}
or {\it XMM-Newton}), which can break the degeneracy between the
contributions from various X-ray emitting components
(e.g.\ photo-electric absorption; reflection; scattering;
\citealt{balokovic2014,DelMoro_2014,gandhi2014,Akylas_2016}).

For AGN with low observed X-ray luminosities (e.g.\ intrinsically LLAGN, 
including LEGs, or heavily obscured AGN), the host
galaxy can start to contaminate the emission from the AGN, leading to
challenges in the identification and characterisation of the AGN. The
dominant host-galaxy phenomenon at X-ray energies is commonly referred
to as X-ray binaries (see \citealt{fabbiano2006} and
\citealt{remillard2006} for reviews). The emission from X-ray binaries
is due to mass accretion onto a degenerate star (neutron star or BH) 
from a companion star in a binary system: X-ray binaries are sub-classified 
into low-mass X-ray binaries (LMXBs) and high-mass X-ray
binaries (HMXBs), depending on the mass of the companion star. The
emission from LMXBs is closely tied to the stellar mass of the galaxy
and can reach $L_{\rm X}\approx10^{41}$~erg~s$^{-1}$ for a massive
galaxy of $\approx10^{11}$~$M_{\odot}$
(e.g.\ \citealt{lehmer2010,boroson2011}) while the emission from
HMXBs is closely tied to the SFR and can reach $L_{\rm
  X}\approx3\times10^{42}$~erg~s$^{-1}$ for an extreme SFG with a SFR of
$\approx1000$~$M_{\odot}$~yr$^{-1}$
(e.g.\ \citealt{lehmer2010,mineo2012a}).
To accurately identify or characterise the AGN
requires taking account of both of these X-ray binary
components. However, since the X-ray emission from host-galaxy
processes rarely exceeds $L_{\rm X}>10^{42}$~erg~s$^{-1}$, and has not
been known to exceed $L_{\rm X}>10^{43}$~erg~s$^{-1}$
(e.g.\ \citealt{alexander2005,wang2013}), an X-ray source with
$L_{\rm X}>10^{42}$~erg~s$^{-1}$ is likely to be an AGN. Furthermore,
the integrated emission from a population of X-ray binaries is mostly
produced at low energies ($<10$~keV) and therefore AGN can be more
reliably identified and characterised at higher energies \citep[see Sect. 2.3 of]
[for a list of additional criteria to identify X-ray
emission from an AGN]{brandt2015}. 

Another potential component of non-AGN contamination at X-ray energies
is emission from hot gas, either from the host galaxy or a galaxy
cluster. The hot-gas component can be up-to $L_{\rm
  X}\approx10^{41}$--$10^{42}$~erg~s$^{-1}$ from galaxies
(e.g.\ \citealt{boroson2011,mineo2012b}) and can be as high as
$L_{\rm X}\approx10^{44}$--$10^{45}$~erg~s$^{-1}$ in the cores of
massive galaxy clusters
(e.g.\ \citealt{stanek2006,ebeling2010}). However, since the X-ray
emission from the hot gas is thermal, it is mostly produced at low
energies ($<$~2--5~keV) and therefore the AGN can be more reliably
identified and characterised at higher energies, even in the most
massive galaxy cluster cores.

On the basis of the aforementioned factors, it is clear that X-ray
observations provide an efficient and reliable selection of the
overall AGN population. Furthermore, the modest X-ray selection biases
are well understood and can be reliably modelled to allow for robust
measurements on the evolution of AGN and the growth of BHs
(e.g.\ \citealt{Gilli_2007,Georgantopoulos_2013,ueda2014,aird2015,buchner2015,miyaji2015,merloni2016}). However,
significant uncertainties remain on the contributions to the cosmic
BH growth from LLAGN and CT
AGN. Various complementary approaches can be utilised to make further
progress. For example, combining constraints in the X-ray band with
AGN and host-galaxy measurements in other wavebands (e.g.\ IR
and radio wavelengths) to assess the expected strength of the AGN and
host-galaxy components at X-ray energies
(e.g.\ \citealt{alexander2008,gandhi2009,georgantopoulos2011,DelMoro_2013,delmoro2016,asmus2015})
or utilising high spatial resolution X-ray imaging and spectroscopy to
directly identify the AGN and host-galaxy components in nearby
low-luminosity systems
(e.g.\ \citealt{annuar2015,annuar2017,ricci2016}).

\subsection{Selection of AGN in the X-ray band: AGN types, $M_{\rm BH}$, $L/L_{\rm Edd}$, 
and cosmic evolution}\label{sec:X-ray_types}

\begin{figure}
\centering
\includegraphics[width=8.4cm]{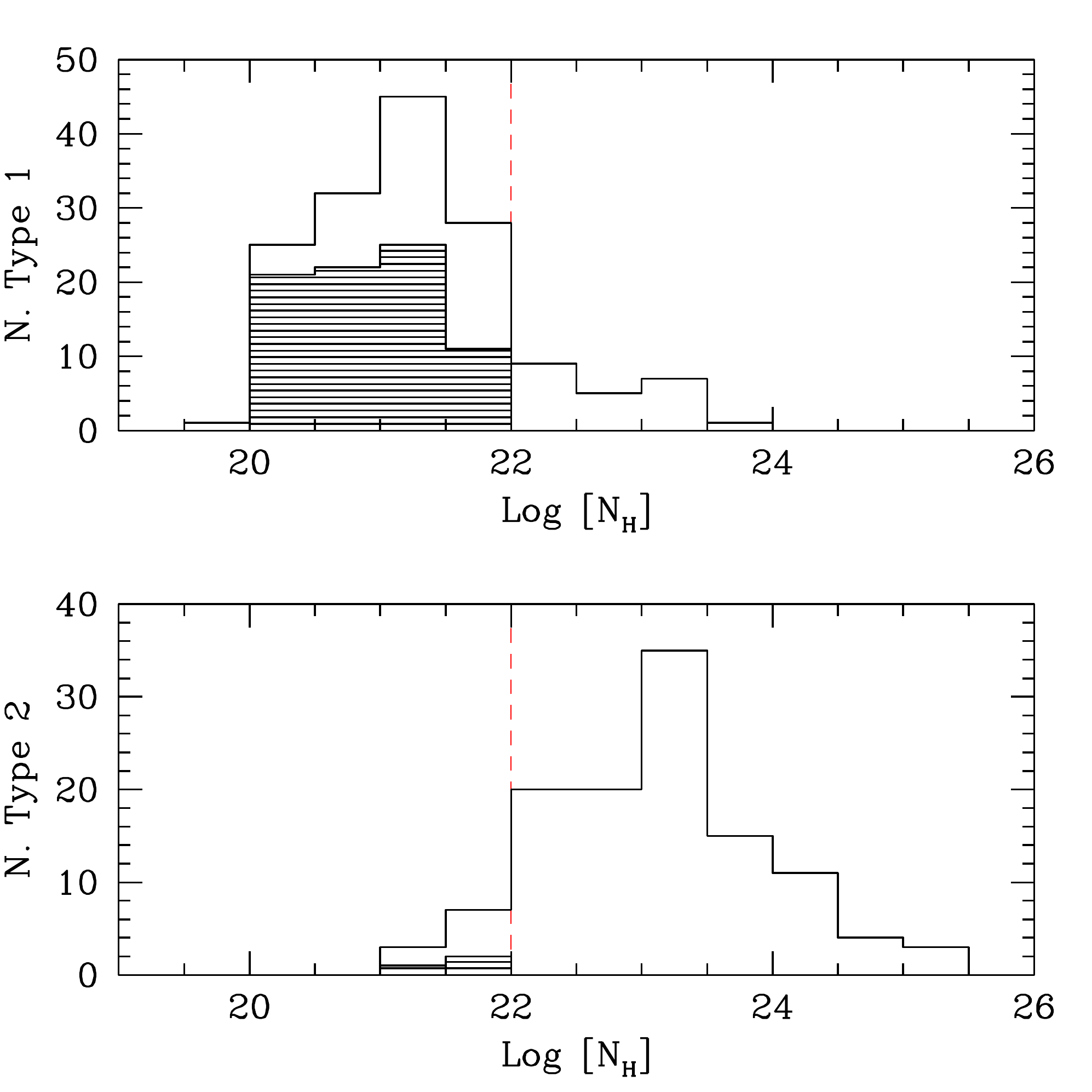}
\caption{Distribution of absorbing column densities for X-ray selected
  AGN from the {\it INTEGRAL} observatory (20--100~keV) with type 1
  (optically unobscured; top) and type 2 (optically obscured; bottom)
  optical spectral signatures. The dashed bins indicate column density
  upper limits and the dashed lines indicate the typical adopted
  column density threshold between X-ray absorbed and unabsorbed
  AGN. Figure reproduced from \cite{malizia2012}, Fig. 5, with permission. 
  \copyright~The Authors.} 
\label{fig5_4}       
\end{figure}

Essentially all types of AGN are selected at X-ray energies: absorbed
and unabsorbed AGN of low, moderate, and high luminosity (see
Fig.~\ref{fig5_1}). Overall there is also good agreement between the
optical and X-ray signatures of absorption
(e.g.\ \citealt{malizia2012,merloni2014,Burtscher_2016}): the
majority ($>$~80--90\%) of optical type 1 AGN are X-ray unabsorbed
($N_{\rm H}<10^{22}$~cm$^{-2}$) while the majority ($>$~80--90\%) of
optical type 2 AGN are X-ray absorbed ($N_{\rm H}>10^{22}$~cm$^{-2}$;
see Fig.~\ref{fig5_4}). Careful consideration of the thresholds in the
classification of X-ray and optical absorbed and unabsorbed AGN can
provide even closer agreement (\citealt{Burtscher_2016}). The overall
consistency between the X-ray and optical absorption indicators
provide some of the strongest observational support for the basic
unified AGN model (e.g.\ \citealt{Antonucci_1993,Urry_1995}).

However, the clear disagreements between absorption indicators for a
small subset of the AGN population (e.g.\ X-ray unabsorbed type 2 AGN
and X-ray absorbed type 1 AGN;
e.g.\ \citealt{panessa2002,bianchi2012}) also provide interesting
insight on the overall universality of the basic unified model (see
\citealt{Netzer_2015} for a review). Time-series observations have
shown that the X-ray and optical spectral properties of AGN can vary
on relatively short timescales as a result of changes in the
absorbing column density along the line of sight
(e.g.\ \citealt{risaliti2002,risaliti2005,matt2003,macleod2016,ruan2016}; see also Sect. \ref{sec:var_models}). Therefore,
some of the occasional disagreements between absorption indicators are
due to non-simultaneous X-ray and optical observations; however, AGN
variability does not explain the differences for all cases and some
AGN appear to genuinely depart from the basic unified model (see also Sect. \ref{ssec:IR_SED}). 

\begin{figure}
\centering
\includegraphics[width=8.4cm]{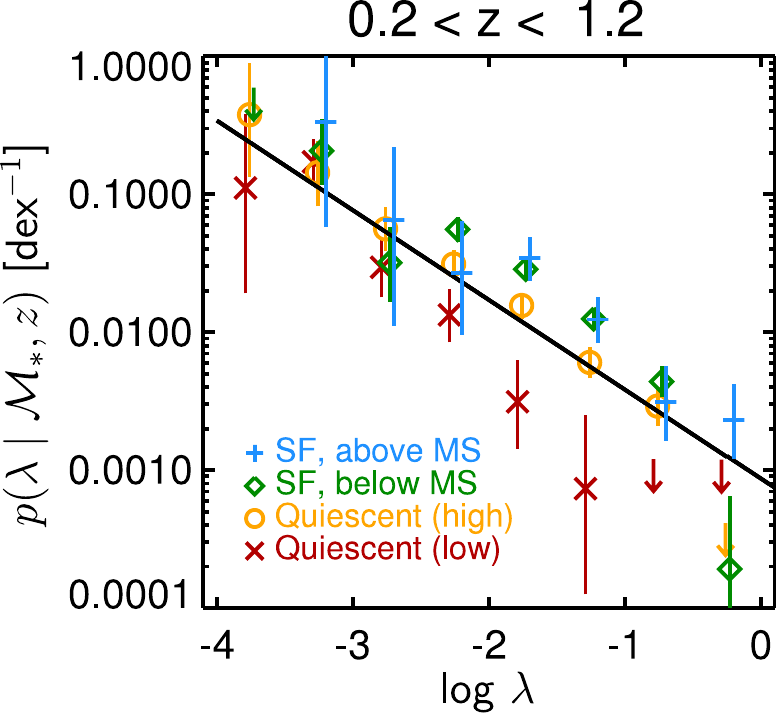}
\caption{Specific accretion rate (a proxy for $L/L_{\rm Edd}$)
  distributions for X-ray selected AGN at $z= 0.2 -1.2$ residing in a
  range of host galaxy types dividing the sample into four populations according to their epoch-normalized 
  specific SFRs (quiescent galaxies with low SFR, quiescent galaxies with higher SFR, SFGs below the star-forming main sequence (MS), 
  and SFGs above the star-forming MS). \copyright~AAS. Figure reproduced from \cite{azadi2015}, Fig. 12, with permission.}
\label{fig5_5}
\end{figure}

Due to the (near) universality of X-ray emission from AGN, its 
production should not be inherently biased
towards specific ranges in $M_{\rm BH}$ and $L/L_{\rm Edd}$. 
Indeed, X-ray emission has been detected from AGN with a broad
range of $M_{\rm BH}$ out to high redshifts
($\approx$~$10^{5}$--$10^{9}$~$M_{\odot}$;
e.g.\ \citealt{hickox2009,page2014,baldassare2015,chen2017}). Despite
the lack of an intrinsic bias towards detecting low-mass BHs
at X-ray energies, the majority of the AGN detected in blank-field
cosmic X-ray surveys are nevertheless found to reside in massive galaxies (stellar
masses $>10^{10}$~$M_{\odot}$, implying $M_{\rm BH} 
>10^{7}$~$M_{\odot}$; e.g.\ \citealt{brusa2009,hickox2009,xue2010};
see \citealt{brandt2015} for a review). However, the deficiency of
X-ray AGN in low-mass galaxies (i.e., a low AGN fraction) appears to be due to selection effects:
for a fixed X-ray flux limit (i.e.\ a luminosity limit at a given
redshift), the probability of detecting a low $L/L_{\rm Edd}$ AGN with
a massive BH is significantly higher than a high $L/L_{\rm Edd}$
AGN with a low-mass BH
(e.g.\ \citealt{aird2012,bongiorno2012}). Taking account of these
observational biases, the distribution of $L/L_{\rm Edd}$ for X-ray
AGN appears to be broadly consistent with a power law and is largely
mass independent (e.g.\ \citealt{aird2012,bongiorno2012}). However,
there is evidence for differences in the normalisation (and
potentially the shape) of the $L/L_{\rm Edd}$ distribution between
quiescent and star-forming host galaxies
(e.g.\ \citealt{aird2012,azadi2015}; see Fig.~\ref{fig5_5}).

\begin{figure}
\centering
\includegraphics[width=8.4cm]{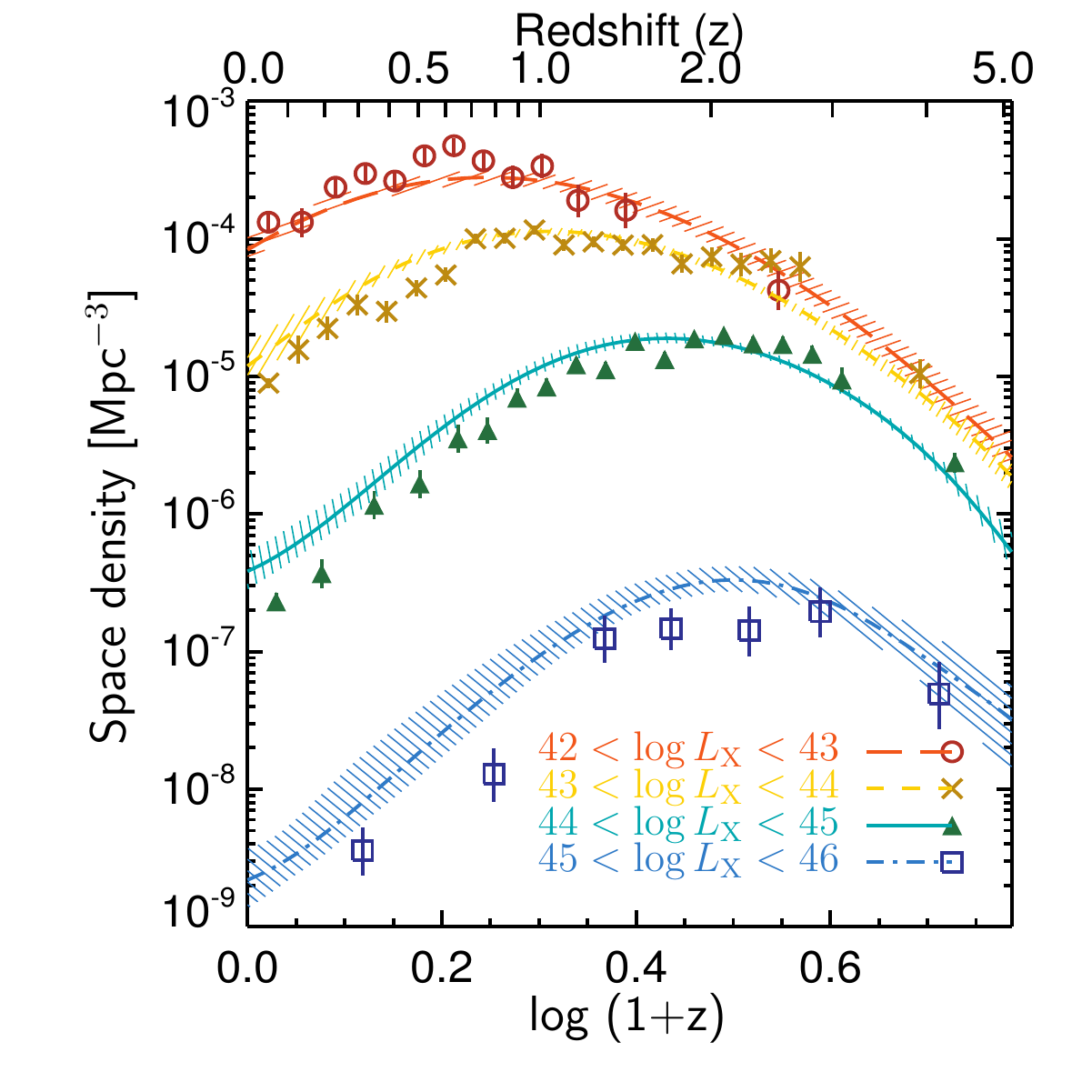}
\caption{Space density versus redshift for AGN selected across a wide
  range in X-ray luminosity. The symbols indicate the data for AGN
  from \cite{miyaji2015} in different X-ray luminosity bins while the
  curves and shaded regions indicate the models for the evolution of
  X-ray selected AGN from \cite{aird2015}. Figure reproduced from  
  \cite{aird2015}, Fig. 18, with permission. \copyright~The Authors.}
\label{fig5_6}
\end{figure}

Blank-field cosmic X-ray surveys have provided some of the most detailed and
sensitive constraints on the evolution of the AGN population. It is
now clear from a large suite of studies that the evolution in the
space density of X-ray selected AGN is dependent on luminosity:
moderate-luminosity AGN peak at $z\approx$~0.5--1 while
high-luminosity AGN peak at $z\approx$~2--3
(e.g.\ \citealt{ueda2014,aird2015,buchner2015,miyaji2015}); see
Fig.~\ref{fig5_6}. This luminosity dependent evolution is commonly
referred to as ``AGN downsizing'' (see also Sect. \ref{sec:radio_evolv}) and, to first order, is likely to be
driven by the availability of a cold-gas supply in the vicinity of the
accreting BH; however, the processes that influence the
availability of the gas are likely to be manifold (e.g.\ SF; 
AGN and stellar feedback; large-scale environment; see
\citealt{alexander2012} for a general review). In terms of the cosmic
BH growth (essentially the product of the space density and
AGN luminosity), the majority has occurred at $z\approx$~1--2, with
broadly similar amounts of growth at $z<1$ and $z>2$: high-luminosity
AGN dominate the BH growth density at $z>1.5$ ($L_{\rm
  X}\approx10^{44}$--$10^{45}$~erg~s$^{-1}$ AGN contribute the
majority) while moderate-luminosity AGN dominate at $z<0.5$ ($L_{\rm
  X}\approx10^{43}$--$10^{44}$~erg~s$^{-1}$ AGN contribute the
majority).

The fraction of X-ray absorbed AGN is found to be a function of
luminosity, with a decreasing fraction towards higher X-ray
luminosities, and appears to also increase with redshift
(e.g.\ \citealt{LaFranca_2005,ueda2014,aird2015,buchner2015}; but
see also Sect. \ref{ssec:IR_SED} for IR-selected AGN). The
redshift evolution in absorption may be a consequence of the
luminosity dependent X-ray absorbed AGN fraction shifting to higher
luminosities at higher redshifts
(e.g.\ \citealt{aird2015,buchner2015}). These results suggest that
the covering factor of the obscuring material (i.e.\ the AGN
torus) might change as a function of luminosity and redshift; however,
it is currently unclear what drives this behaviour.

\subsection{The future of AGN studies in the X-ray band}\label{sec:X-ray_future}

We now look towards the scientific gains that can be anticipated from several
future X-ray facilities; see Sect. 6.4 of \cite{brandt2015} for a more
detailed discussion of these observatories and other proposed X-ray
facilities.

eROSITA\footnote{\url{www.mpe.mpg.de/eROSITA}} (\citealt{merloni2012}) 
is a joint Russian-Ger\-man
mission planned for launch in 2018. The principle objective of 
  eROSITA is to undertake a sensitive all-sky survey, achieving
sensitivity limits $\approx$~20 and $\approx$~200 times deeper than
{\it ROSAT} (0.5--2~keV; \citealt{voges1999}) and {\it HEAO 1 A-2}
(2--10~keV; \citealt{piccinotti1982}), respectively. The great advance
that eROSITA will provide over the more sensitive {\it Chandra}
and {\it XMM-Newton} observatories is huge AGN statistics ($\approx$~3
million AGN out to $z\approx$~6), effectively providing an X-ray
equivalent of the SDSS (\citealt{york2000}) or WISE, to
explore the cosmic growth of BHs and large-scale structure at
X-ray energies.

The Advanced Telescope for High ENergy Astrophysics ({\it Athena}\footnote{\url{www.the-athena-x-ray-observatory.eu}}; 
\citealt{nandra2013}) is an ESA-led mission planned for
launch in 2028. {\it Athena} will be a revolutionary general-purpose
X-ray observatory with a wide field of view (FoV; 40$^{\prime} \times 40^{\prime}$), large collecting area
(2 m$^2$ at 1 keV),
good spatial resolution ($5^{\prime\prime}$), and excellent spectral resolution (2.5 eV). {\it
  Athena} has many key scientific aims but, from the point of view of
this review, two of the main advances will come from: (1)
excellent-quality X-ray spectroscopy (both in terms of unprecedented
sensitivity and spectral resolution), to elucidate the physics of AGN
activity (e.g.\ \citealt{cappi2013,dovciak2013}); (2) unsurpassed
sensitivity for wide--deep blank-field cosmic surveys, to construct a
near-complete census of AGN activity out to $z\approx$~1--3 and
identify moderately luminous AGN out to $z\approx$~6--10 (down to
$L_{\rm X}\approx10^{43}$~erg~s$^{-1}$;
e.g.\ \citealt{aird2013,georgakakis2013}).

Finally, the X-ray Imaging Polarimetry Explorer (XIPE\footnote{\url{www.isdc.unige.ch/xipe}}; \citealt{Soffitta_2016}) is a new mission concept
selected by ESA in 2015 to undergo a 2 year-long assessment phase in
the context of the Cosmic Vision competition. {\it XIPE} is dedicated
to undertake temporally, spatially, and spectrally resolved X-ray
polarimetry. From the point of view of AGN the polarised emission is
expected to be due to the scattering and reflection of photons and to
originate in the vicinity of the accretion disk, the absorbing region
(the AGN torus), and larger-scale ionisation ``cones''. Therefore,
{\it XIPE} will provide unique physical insight on the geometry and
connections between these different regions (\citealt{Goosmann_2011}).

\section{$\gamma$-ray-selected AGN}\label{sec:gamma}

We discuss in this section $\gamma$-ray-selected AGN. The $\gamma$-ray
band is conventionally split into the High Energy (HE) band, between 100 MeV and $\sim$100 GeV, 
and the Very High Energy (VHE) band, covering the $\sim$50 GeV to $\sim$10 TeV range. 
The types of AGN selected,
their SEDs, and the physical mechanism(s) behind $\gamma$-ray emission 
are detailed in Sect. \ref{sec:gamma_AGN}, while $\gamma$-ray detectors are
described in Sect. \ref{sec:gamma_detectors}. Sects. \ref{sec:gamma_HE} and Sect. \ref{sec:gamma_VHE} deal
with the HE and VHE bands respectively, while 
Sect. \ref{sec:gamma_pop} addresses the blazar population and its contribution to the
$\gamma$-ray background. Sect. \ref{sec:gamma_bias} discusses selection 
biases and, finally, Sect. \ref{sec:gamma_future} addresses the future of this field. 
\subsection{$\gamma$-ray AGN, their SEDs, and physical mechanism(s) 
behind $\gamma$-ray emission}\label{sec:gamma_AGN}

The $\gamma$-ray band is the most energetic part of the electromagnetic spectrum and, as such, is beyond the reach of most astronomical sources. This applies also to most 
types of extragalactic objects, including the non-jetted AGN detected in large numbers in IR, optical, and X-ray surveys thanks to the radiation resulting from accretion onto the central SMBH 
(Sects. \ref{sec:IR}, \ref{sec:optical}, and \ref{sec:X-ray}; see Fig. \ref{fig:SED}). 

Blazars, despite being intrinsically very rare (orders of magnitude less abundant than non-jetted 
AGN of the same optical magnitude), dominate the extragalactic $\gamma$-ray sky, whi\-ch include also  
a few other AGN, mostly nearby RGs (blazars are also prevalent in the bright radio sky: Sect. \ref{sec:FS_SS}). 
Non-jetted AGN have not been detected in the $\gamma$-rays\footnote{With
  the exception of NGC 1068 and NGC 4945, two Seyfert 2 galaxies in which
  the $\gamma$-ray emission is thought to be related to their starburst
  component \citep{Ackermann_2012b}.} \citep{Ackermann_2012a}.
The power output of blazars covers the entire electromagnetic spectrum and is dominated by non-thermal, blue-shifted, and Doppler boosted radiation arising in a relativistic jet pointed in the direction of the observer \citep[e.g.][see also Sect. 
\ref{sec:radio_unif}]{Urry_1995}. It is because of this intense non-thermal emission, and the very special geometrical conditions, that blazars reach the most extreme parts of the electromagnetic spectrum with fluxes well above the sensitivity of the 
instruments operating today. Blazars, as discussed in Sect. \ref{sec:FS_SS}, are further divided into two subgroups defined by their optical spectra, i.e. BL Lacs and FSRQs.

\begin{figure}[h]
\centering
\includegraphics[width=8.5cm]{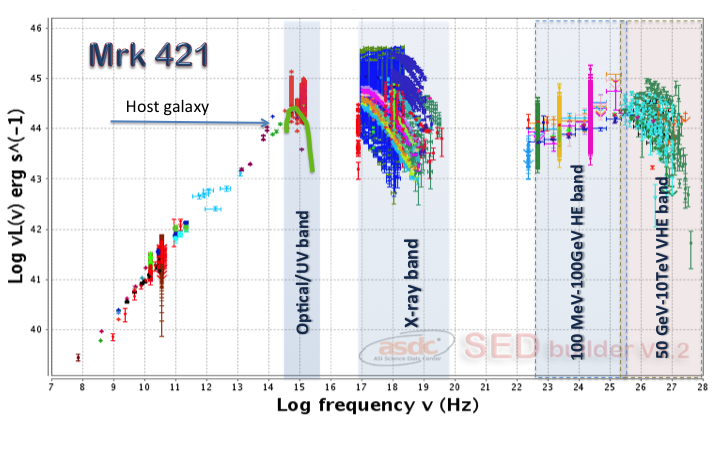}
\vspace{-2em}
\caption{The SED of  the BL Lac Mrk 421. Strong and variable emission is present at all energies, from the radio band to 
$\gamma$-rays. Variability is most pronounced in the X-rays and $\gamma$-rays where the two SED components peak. 
The green line indicated by the arrow represents the expected emission from a typical  blazar host galaxy. A hard spectrum is present in the HE band 
with variability of approximately a factor 50 in the 200 MeV (light green), 1 GeV (orange) and 10 GeV (purple) data. 
The $\gamma$-ray SED peaks in the VHE band, where large variability is also present. VHE data are from 
\cite{ARGO,BW2015,Aha2005,TACTIC2010,TACTIC2015,TIBET2003,MG2007}. The SEDs here and in Fig. \ref{fig:3c273} have been generated using the 
SED builder tool of the ASI Science Data Center (ASDC), available at \protect\url{tools.asdc.asi.it/SED}, while the light curves have been produced at the 
ICRANet site of Yerevan as part of a collaboration with ASDC, using {\it Fermi} public software and archival data.}
\label{fig:mkn421}       
\end{figure}

\begin{figure}[h]
\centering
\includegraphics[width=8.5cm]{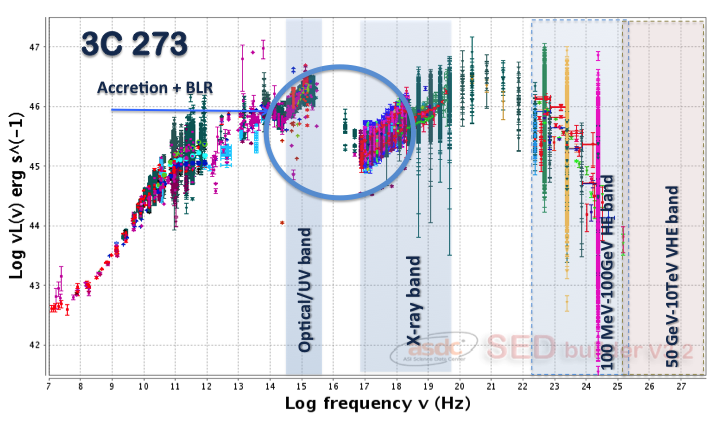}
\caption{The SED of the FSRQ 3C 273. The strong and highly variable non-thermal radiation from the jet encompasses the entire electromagnetic spectrum, but is not dominant in the 
optical-UV and the soft X-ray bands (indicated by the blue circle), where most of the emission is due to accretion onto the SMBH and to the BLR. 
The $\gamma$-ray spectrum is steep with extremely high variable intensity (up to a factor of 10,000: see the light green, orange and purple light curves at 0.2, 1.0 and 10.0 GeV, respectively) in the HE band, while very little or nothing is present in the VHE band.}
\label{fig:3c273}       
\end{figure}

The SED of blazars (see Figs. \ref{fig:mkn421} and \ref{fig:3c273} for two examples of well known objects) covers the entire electromagnetic spectrum, from radio waves to $\gamma$-ray energies, and is characterized by a typical ``double humped'' shape (in a $\nu L(\nu)$ vs. $\nu$ [or $\nu f(\nu)$ vs. $\nu$] space). 
The low energy component, peaking between the IR and the X-ray band, is generally attributed to synchrotron radiation produced by relativistic electrons moving in a magnetic field. 
Sources where this component peaks at low energies ($\nu_{\rm synch~peak} < 10^{14}$ Hz) are called LSPs, while objects 
with SED peaking at high energies ($\nu_{\rm synch~peak} >10^{15}$ Hz) are called HSPS \citep{Padovani_1995,Abdo10}. Objects with $\nu_{\rm synch~peak}$ located at intermediate energies are called ISPs.

The nature of the second SED component, that extends well into the $\gamma$-ray band, is still debated as two alternative (or complementary) interpretations are being considered. In leptonic models \citep[e.g.][]{leptonic} the emission is explained as inverse Compton scattering between the electrons in the jet and their own synchrotron emission (synchrotron self-Compton) or an external photon field (external inverse Compton). 
In hadronic scenarios \citep[e.g.][]{hadronic} $\gamma$-rays are instead assumed to originate from high-energy protons either loosing energy through synchrotron emission \citep{Aharonian} or through photo-meson interactions \citep{Mannheim}. 
In this case blazars would also be neutrino emitters (from the decay of charged pions) extending their SEDs outside the electromagnetic spectrum into newly explored multi-messenger scenarios, which might even include cosmic rays (CRs) 
\citep[e.g. ][]{Padovani_2016_nu,Resconi_2017}.

Figures \ref{fig:mkn421} and \ref{fig:3c273} show the SEDs of the HSP BL Lac Mrk 421 and of the LSP FSRQ 3C 273 built using large amounts of archival data from many space- and ground-based observatories covering 
almost all frequencies and a time span of several years.
The HE and VHE  $\gamma$-ray bands are highlighted. A clear difference in the $\gamma$-ray emission of the two objects is apparent: a flat spectrum in the HE band extending 
to VHE energies for Mrk 421, a steep spectrum in the HE band with almost no emission in the VHE band for 3C 273.  
Strong variability, another defining characteristics of blazars, is clearly visible in both objects at all energies, with the largest amplitude occurring in the X-rays for Mrk 421 and in the $\gamma$-ray band for 3C 273.

\subsection{$\gamma$-ray detectors}\label{sec:gamma_detectors}

Different $\gamma$-ray detectors operate in different bands. 
The HE band is where $\gamma$-rays are detected in electron pair-conversion telescopes. These instruments operate in space and are characterized by a very large FoV 
\citep[thousands of square degrees; e.g.][]{AGILE,LAT}. 
The VHE band is where the present detection capability is provided by Imaging Atmospheric Cherenkov Telescopes (IACTs) and Extensive Air Shower (EAS) observatories \citep[e.g.][]{VHEDetectors}. These instruments observe from the ground the particle showers that are produced by the impact of VHE $\gamma$-ray photons on the top layers of the atmosphere, either through the Cherenkov light they generate, or via the direct detection of the charged particles in the shower.
The small overlap between the two bands (between 50 and 100 GeV) allows for inter-calibration between space and ground-based observatories. This is particularly important as this is 
where spectral breaks occur in the most extreme astrophysical sources. 

\subsection{The HE band}\label{sec:gamma_HE}


Currently four $\gamma$-ray space telescopes or instruments sensitive to photons in the 100 MeV -- 100 GeV band are operational: the ``Astro-rivelatore Gamma a Immagini LEggero'' (AGILE), which was launched in 2007 and is the oldest in operation, {\it Fermi} (launched in 2008), the Alpha Magnetic Spectrometer (AMS-2, launched in 2011), and the DArk Matter Particle Explorer (DAMPE, launched in 2015).
Published results or publicly available data are, however, so far available only from AGILE and {\it Fermi}. The other facilities, although sensitive to cosmic $\gamma$-rays, 
have different prime objectives, such as the detailed measurement of charged CRs and dark matter, and have not contributed to the $\gamma$-ray astronomy literature so far.

The best sensitivity is provided by the Large Area Telescope on board {\it Fermi} \citep[{\it Fermi}-LAT,][]{LAT}, which 
is approximately two orders of magnitude more sensitive than the previous generation of $\gamma$-ray telescopes like the Energetic Gamma Ray Experiment Telescope (EGRET) 
on board of the Compton Gamma Ray Observatory  \citep{Thompson93}. 

The size of the point spread function 
and the effective area of {\it Fermi}-LAT strongly depend on energy, resulting in a sensitivity limit that 
significantly depends on the intrinsic source spectrum\footnote{See \url{www.slac.stanford.edu/exp/glast/groups/canda/lat_Performance.htm}}. This has a strong impact on the blazar types that are detected in different $\gamma$-ray bands, introducing selection biases, 
as discussed below (Sect. \ref{sec:gamma_bias}). 

\subsubsection{AGN in the HE $\gamma$-ray band}

Two main approaches have been followed to produce catalogues of $\gamma$-ray sources in the HE band. In the one adopted by the project teams the detection 
is based on $\gamma$-ray data only and is therefore ``blind'' with respect to any additional information about astrophysical sources. For this reason they use conservative statistical thresholds (e.g. AGILE: \citealt{AGILECat}; {\it Fermi}-LAT 0,1,2,3FGL: \citealt[][and references therein, where the numbers denote the n-th {\it Fermi}-LAT source catalogue]{3FGL}; {\it Fermi} 1,2,3FHL: \citealt[][and references therein, where the numbers denote the n-th catalogue of hard sources]{3fhl}). 
Other lists have instead been obtained by searching for $\gamma$-ray excesses at the positions of previously known blazars, or blazar candidates \citep[e.g.][]{ArsioliChang,arsioli2017}
in {\it Fermi}-LAT archival data. The two approaches are complementary as they make use of   
different information in the application of the statistical detection methods, use different thresholds and integration times, and are sensitive in different ways to source confusion, 
a problem that is becoming an issue as the sensitivity improves after many years of {\it Fermi}-LAT data.   

\begin{figure}[h]
\centering
\includegraphics[width=9.0cm]{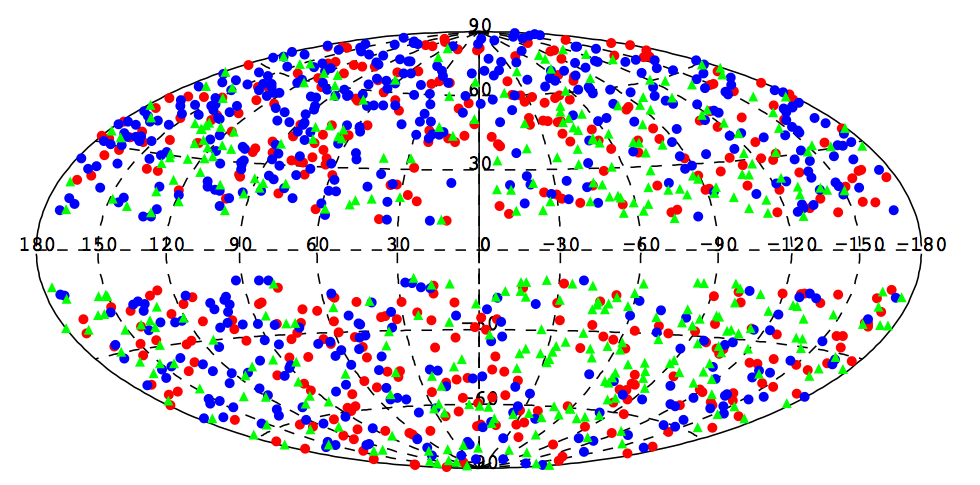} 
\caption{The third catalogue of AGN detected by the {\it Fermi}-LAT \citep[3LAC;][]{3LAC} plotted in Galactic coordinates. 
FSRQs (all of the LSP type) are plotted as red filled circles, BL Lacs (mostly of the HSP type) are 
shown as blue points; green filled circles represent blazars of uncertain type.}
\label{fig:3lac}       
\end{figure}

The largest catalogue of $\gamma$-ray sources published so far, the 3FGL, is based on the first 4 years of {\it Fermi}-LAT data and includes 3033 objects. 
Outside the Galactic plane almost all $\gamma$-ray detections that have been associated to known objects are jetted AGN, the large majority of them being blazars 
with only a handful of RGs.

Figure \ref{fig:3lac} shows a plot in Galactic coordinates of the subsample of 1563 high Galactic latitude ($|b|>10^{\rm \circ}$) sources in the 3FGL catalogue that have been firmly associated with AGN \citep{3LAC}. 
This includes 415 FSRQs (all of the LSP type), 657 BL Lacs (mostly of the HSP type), and 402 blazars of uncertain type (assumed to be blazars because of their radio to $\gamma$-ray SED but with no 
optical spectrum yet available to classify them as FSRQs or BL Lacs). 

An example of the second approach to the detection of $\gamma$-ray sources is the Brazil ICRANet Gamma-ray Blazar
catalogue \citep[1BIBG;][]{ArsioliChang} where the authors report 150 new $\gamma$-ray 
sources in the sample of multi-frequency selected Second WISE High Synchrotron Peaked (2WHSP) blazars.

A comprehensive list of AGN detected by {\it Fermi}, compiled at the ASDC from all published catalogues and other publications, is available on-line\footnote{\url{www.asdc.asi.it/fermiagn}}. 
At the time of writing it includes 1959 distinct sources, 536 of which are FSRQs, 687 are BL Lacs, 15 are RGs, and 75 are AGN of unknown type.

\subsection{The VHE band}\label{sec:gamma_VHE}

The VHE $\gamma$-ray band is where currently operating IACTs and EAS are sensitive, with {\it Fermi}-LAT also partly covering this band up to $\sim 2$ TeV. 
Indeed, the largest catalogue of VHE sources, the 3FHL \citep{3fhl}, is based on {\it Fermi}-LAT 
data. 

This energy region is particularly challenging for AGN observations as VHE $\gamma$-rays are subject to pair production interactions with the extragalactic background light \citep[EBL;][]{ebl} causing strong flux attenuations that are energy and redshift dependent, thus modifying the observed spectra and limiting the horizon to the relatively low redshift Universe.  
 
The present generation of IACTs (e.g. MAGIC, H.E.S.S., VERITAS) and the upcoming Cherenkov Telescope Array (CTA) are characterized by relatively small FoV (a few square degrees), and very good sensitivity. IACTs are excellent instruments for detecting fast variability, catching flaring states, and measure the status of the object during the (typically few hours) of observations.  Several years of observations led to the detection of many blazars, especially during large flares, often following pointings triggered by high states discovered in other wavebands.

Water Cherenkov/EAS detectors (e.g. HAWK, ARGO, TIBET, and the future Large High Altitude Air Shower Observatory [LHAASO]) are instead characterized by a large FoV (thousands of square degrees), moderate sensitivity, and operate nearly 100\% of the time (day and night).
These observatories are well suited to carry out long-integration surveys of large parts of the sky, with typical output being fluxes averaged over long integration periods or the discovery of strong flares in bright objects. 

\subsubsection{AGN in the VHE $\gamma$-ray band}

\begin{figure}[h]
\centering
\includegraphics[width=9.0cm]{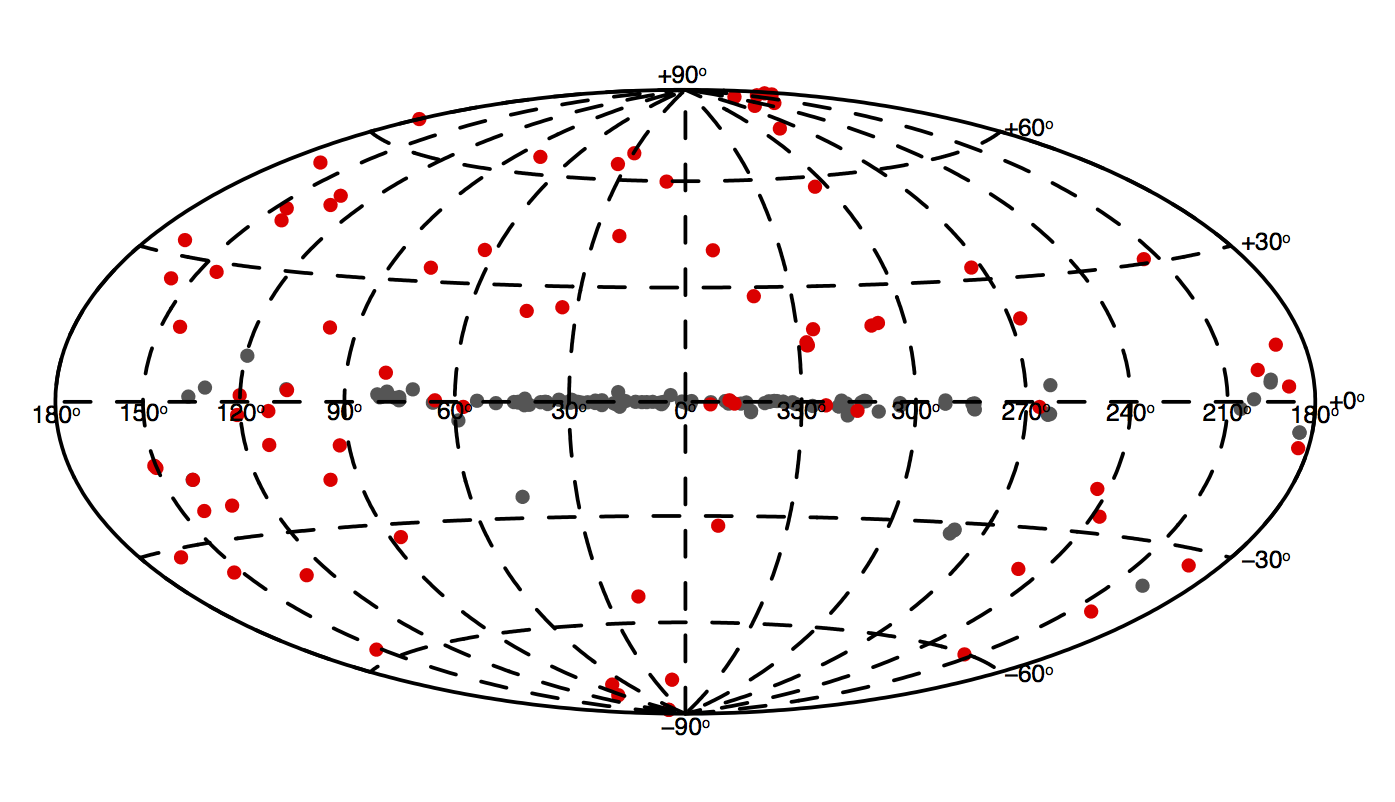} 
\caption{The sample of $\gamma$-ray sources detected in the VHE band as reported in TeVCat at the time of writing plotted in Galactic coordinates. Red and black points 
represent extragalactic and Galactic plus unidentified sources respectively.}
\label{fig:TevCat}       
\end{figure}

The main lists of objects in this energy band detected by IACTs are available on-line\footnote{\url{tevcat.uchicago.edu} and \url{www.asdc.asi.it/tgevcat}} as interactive tables 
that are updated periodically as new sources are detected. At present they include approximately 180  sources distributed in the Galaxy as shown in Fig. \ref{fig:TevCat}. These samples do not represent uniform surveys of the VHE sky as they only 
include sources detected during pointings of known sources, often during flaring states. Only a fraction of the high Galactic latitude sky has been observed so far.  

At high Galactic latitudes almost all sources in VHE catalogues are blazars, the majority of which (50 sources) are of the HSP or ISP type, while only 8 are of the LSP type (6 FSRQs and 2 BL Lacs); two of these, namely S3 0218+35 and PKS1441+25, are located at $z > 0.9$, a remarkably high value for this energy band. Four RGs are also detected (Centaurus A, NGC 1275, PKS 0625$-$35, and M 87).

The 2FHL catalogue \citep{2FHL} is based on 80 months of {\it Fermi}-LAT data and is a real full sky survey compiled with photons having energies between 50 GeV and 2 TeV. It includes 360 sources, many more that those detected by IATCs. In 
fact only about 25\% of 2FHL sources were previously detected by Cherenkov telescopes. Similarly to the {\it Fermi}-3FGL catalogue off the Galactic plane nearly all sources are blazars. However, in this case BL Lacs of the HSP type are the large 
majority; only 10 FSRQs have been found so far. 
This difference with respect to the HE band is easily explained as due to the much steeper $\gamma$-ray spectral slope of FRSQs (all of which are LSPs) and 
to EBL absorption that further steepens and reduces the intensity of the spectrum of the more distant FSRQs. 

To complement the 2FHL sample, and in an effort to construct a much larger list of targets for VHE observations, \cite{2whsp} assembled the 
2WHSP catalogue, a very large sample of confirmed and candidate blazars of the HSP type that are expected to emit in the VHE band. The sample was selected using radio, IR (WISE), optical, and X-ray survey data, 
imposing that the SED of the candidates is similar to that of known HSP blazars, which are the most abundant type of AGN found in the VHE band.
The 2WHSP sample includes 1691 sources and is available on-line\footnote{\url{www.asdc.asi.it/2whsp}}. 
The already mentioned study by \cite{ArsioliChang}, who found 150 $\gamma$-ray detections that were never reported before using over 7 years of {\it Fermi}-LAT data, 
confirms that 2WHSP blazars indeed constitute a very good reservoir of candidate VHE $\gamma$-ray sources for the next generation of detectors such as CTA and LHAASO.

\subsection{Blazar population properties and contribution to the $\gamma$-ray extragalactic background}\label{sec:gamma_pop}

The large samples that can be assembled from {\it Fermi}-LAT catalogues and other publications have been used to determine the population properties of blazars 
such as the $\gamma$-ray number counts, the LF and cosmological evolution for both BL Lacs and FSRQs \citep{Abdo2010b,Ajello2015}. 

\cite{Ajello2015} found that the blazar $\gamma$-ray LF can be represented by a broken power law and that its evolution with redshift is strong for all types of 
evolution models (luminosity, density and luminosity dependent density evolution) considered. 
By integrating the best fit LF, taking into account the estimated amount of cosmological evolution, the distribution of blazar spectral slopes 
and EBL attenuation, the authors estimated the contribution of AGN to the extragalactic $\gamma$-ray background \citep[EBG,][]{egb}. The results show  
that the integrated emission from blazars and RGs can explain both the intensity and the spectral shape of the extragalactic background in the 100 MeV -- 820 GeV energy band.  
In particular, above 100 GeV a very large fraction of the EGB is due to HSP blazars with SED similar to that shown in Fig. \ref{fig:mkn421}. 
The fact that AGN are responsible for most, if not all, of the EGB is particularly important also because it leaves little room to the contribution of diffuse components 
such as $\gamma$-rays from the annihilation of dark matter particles and from the interaction of ultra HE CRs with the cosmic microwave background.

\subsection{Blazars and selection biases in the $\gamma$-ray and other energy bands}\label{sec:gamma_bias}


The complex broad-band SEDs of blazars result from the superposition of many spectral components, such as the double humped non-thermal emission, light from the host galaxy, the BLR, and the accretion onto the SMBH. This mix, combined with different viewing angles and a wide range of maximum particle acceleration energies leads to SEDs with 
largely different shapes, causing very strong selection effects when looking at blazars in widely separated regions of the electromagnetic spectrum. 
For instance, blazars selected at radio or microwave frequencies are mostly of the FSRQ/LSP type, whereas X-ray selection leads to samples largely dominated by BL Lacs 
of the HSP type. In the $\gamma$-ray band selection effects are not less important, especially because the sensitivity of {\it Fermi}-LAT strongly depends on the spectral slope of the detected sources: steep $\gamma$-ray spectra FSRQs/LSPs (like 3C 273, see Fig. \ref{fig:3c273}) are therefore detected less efficiently than hard $\gamma$-ray spectra HSP BL Lacs (e.g. Fig. \ref{fig:mkn421}), leading to percentages of the two blazar types that do not represent the real cosmic abundance. The $\sim$40 to $\sim$60 \% mix between FSRQs and BL Lacs in the 3LAC sample and the over 95\% of HSP BL Lacs in the 2FHL catalogue are just the result of this effect in the HE and VHE $\gamma$-ray energy bands. 
To understand the intrinsic population properties of blazars is therefore essential to control all the selection biases. To this purpose Giommi, Padovani and collaborators published a series of papers \citep{bsv1,Giommi_2013,bsv3,bsv4}, where, through a detailed Monte Carlo approach, called the blazar simplified view (BSV), they showed that the widely different statistical population properties of blazars observed in the radio, X-ray and $\gamma$-ray bands can be reproduced in detail starting from simple minimal assumptions. 
In particular, the BSV correctly predicts the different composition of $\gamma$-ray catalogues in the HE and VHE bands, including percentages of FSRQs/LSPs and BL Lacs/HSPs, spectral slopes, redshift distributions, and contribution to the EGB. 

\subsection{The future of AGN studies in the $\gamma$-ray band}\label{sec:gamma_future}

Building on the technology of current generation IACTs, CTA\footnote{\url{www.cta-observatory.org}} will be ten times more sensitive and will have unprecedented accuracy in its detection of VHE $\gamma$ rays, 
with more than 100 telescopes located in the northern and southern hemispheres covering the 20 GeV -- 300 TeV energy range. Construction should start
in 2018, with the first telescopes on site in 2019. CTA will provide a systematic approach to blazar studies, as our knowledge of VHE $\gamma$-ray emission from blazars is very 
biased and patchy, since many of them have been detected only because they were in outburst (Sect. \ref{sec:gamma_VHE}). 

LHAASO\footnote{\url{english.ihep.cas.cn/ic/ip/LHAASO/}} \citep[][]{LHAASO_2016} is a new generation instrument, to be built at 4410 metres of altitude in the Sichuan province of China, with the aim of studying with unprecedented sensitivity the 
energy spectrum, the elemental composition and the anisotropy of CRs in the energy range between 10$^{12}$ and 10$^{17}$ eV, as well as to act simultaneously as a wide aperture 
($\sim$ 2 sr), continuously-operated $\gamma$-ray telescope in the energy range between 100 GeV and 1 PeV.

In the BSV paper dedicated to the VHE $\gamma$-ray band \cite{bsv3} estimated the number of blazars of the FSRQ and BL Lac type and their redshift distributions 
expected in future deep VHE surveys. The expectations are largely consistent with the 2FHL catalogue \citep{vheb}, which suggests that the BSV might also be a reliable predictor of the 
average VHE sky that will be probed by the upcoming large new VHE facilities like CTA and LHAASO.  

\begin{figure}
\centering
\includegraphics[width=9.0cm]{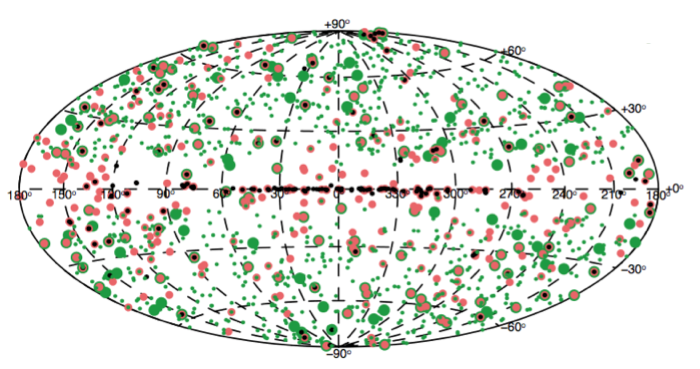}
\caption{The comparison of this plot with that of Fig.\ref{fig:TevCat} illustrates the rapid evolution of the VHE sky. 
VHE sources detected by IACTs are shown as black dots as in Fig. \ref{fig:TevCat}, those from the {\it Fermi} 2FHL catalogue (E $>$ 50 GeV) 
are shown as orange filled circles, while blazars in the 2WHSP sample that are predicted by Eq.  \ref{eq:gamma_1} to be VHE emitters with intensity larger 
than the sensitivity of current IACTs and that of CTA, appear as large and small green circles, respectively.
}
\label{fig:VHESky}       
\end{figure}

A study based on the subset of 2WHSP and 2FHL common sources, shows that the average (EBL unabsorbed) VHE flux F(E$>$ 50 GeV) of HSP blazars can be 
predicted with an uncertainty of the order of a factor two based on the shape and intensity of their SED synchrotron component.  
The following preliminary relationship 
\begin{equation}
F(E> 50 GeV) \approx 2 \times 10^{11}S_{\rm{peak~flux}}10^{(-0.154Log(\nu_{\rm{synch~peak}})-8.03)}
\label{eq:gamma_1}
\end{equation}
gives F(E$>$ 50 GeV), the VHE flux in units of photons cm$^{-2}$ s$^{-1}$, as a function of S$_{\rm{peak~flux}}$, which is the flux at the peak of the SED synchrotron hump in units of 
erg cm$^{-2}$ s$^{-1}$ and of $\nu_{\rm{synch~peak}}$ in units of Hz.

To emphasize the rapid evolution of the VHE AGN sky, Fig. \ref{fig:VHESky} shows the distribution of the currently known and predicted VHE AGN in galactic coordinates, including 
TevCAT sources and {\it Fermi} 2FHL objects. 2WHSP sources that are predicted (on the basis of Eq. \ref{eq:gamma_1}) to be at or above the sensitivity of current IACTs and also 
detectable by the  future CTA are also shown. 
    
\section{Variability-selected AGN}\label{sec:variability}

We discuss in this section variability-selected AGN. An over\-view of AGN variability in the 
local Universe is given in Sect. \ref{sec:var_overview}, while Sect. \ref{sec:var_surveys}
deals with AGN variability in extragalactic surveys. The future of this field is 
discussed in Sect. \ref{sec_var_future}. 

\subsection{An overview of AGN variability in the local Universe}\label{sec:var_overview}
 
Variability of the emitted multi-wavelength (from radio to $\gamma$-ray) radiation has been recognized as one of the main characteristics of AGN as a class (e.g. \citealt{Angione_1973}; \citealt{Marshall_1981}). Therefore, just like other diagnostics, variability can be used as a tool to select AGN in extragalactic surveys.

\begin{figure}
\centering
\includegraphics[width=8.5cm]{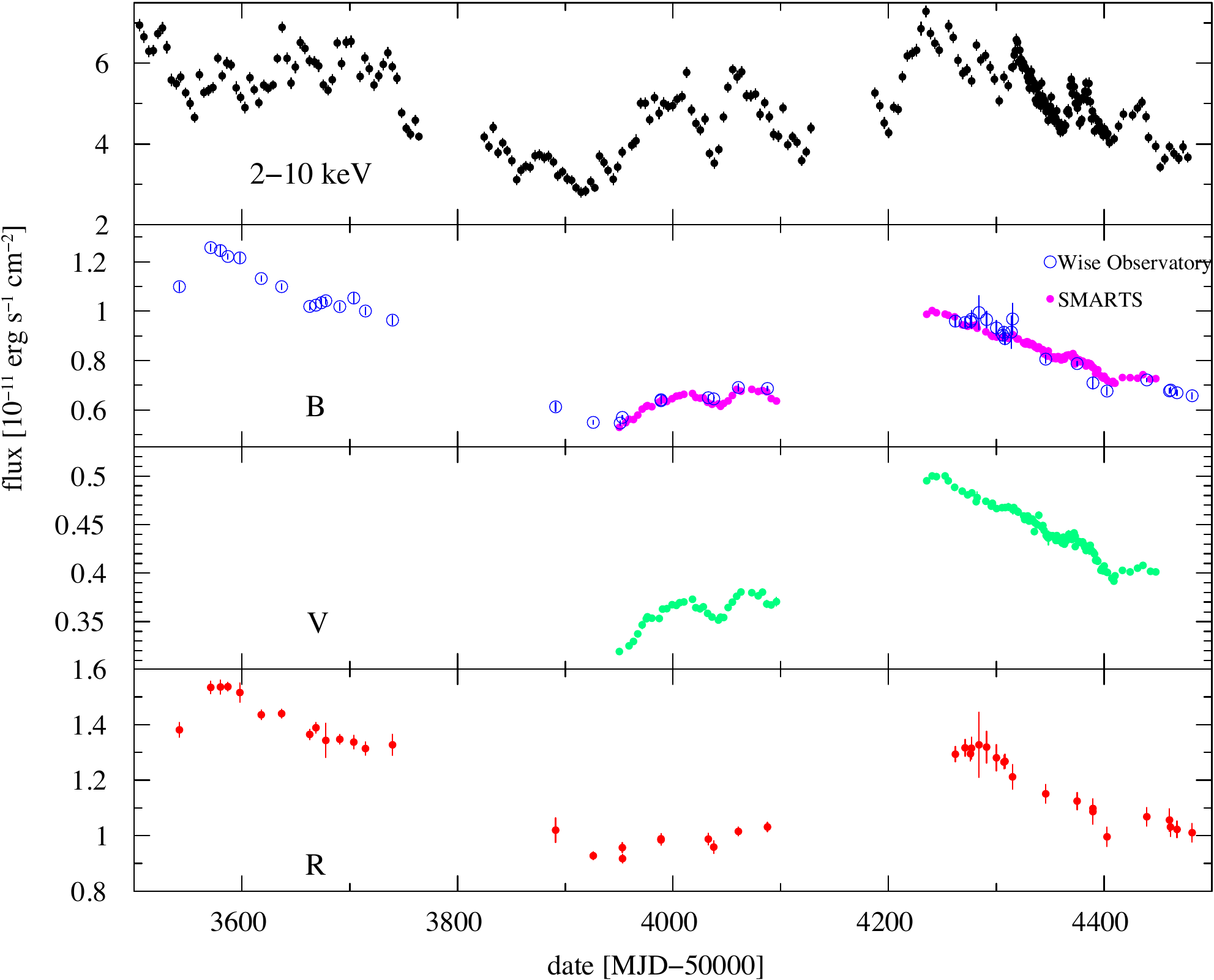}
\caption{Multiwavelength light curves of the non-jetted AGN MR 2251$-$178. From top to bottom: $2-10$ keV X-rays from {\it RXTE}; B band Wise Observatory data in blue open circles and B band SMARTS data in pink filled circles; V band SMARTS data, and R band Wise Observatory data. Figure reproduced from \citealt{Arevalo_2008}, Fig.1, with permission. \copyright~The Authors.}
\label{fig:arevalo}       
\end{figure}

AGN display erratic, aperiodic flux variability over a wide range of timescales (from years to minutes). The distribution of AGN variability power 
(i.e. the power spectral density [PSD] defined as the squared amplitude of the flux, e.g. \citealt{Uttley_2002}) over timescales 
(or, equivalently, temporal frequencies) strongly depends on the observing waveband. 
In other words, for example, much faster variability is observed in the X-ray band than in the optical band, 
where the same variability amplitude is reached only over longer timescales (Fig. \ref{fig:arevalo}). 
The minimum timescale of variability measured in a given waveband provides us with an estimate of the linear size of the source component emitting in that waveband (e.g. \citealt{Terrell_1967}). 
The X-ray band is where some of the most rapid (hours-minutes), largest-amplitude flux variations are measured. This variability is thought to originate in the innermost regions of the accretion flow (corona and inner disk). Moreover, it is responsible for driving (at least part of) the variability from the outer accretion disk, observed at longer wavelengths (UV and optical). Processes occurring in the jet can also contribute to the observed AGN variability (e.g. shocks or bulk injection of new particles; \citealt{MarscherGear_1985}; \citealt{Bottcher_2010}). Variability associated with the jet can dominate in the radio-to-$\gamma$-ray bands in jetted AGN (e.g. \citealt{Max-Moerbeck_2014}; see also Sect. \ref{sec:gamma_AGN}
and Figs. \ref{fig:mkn421} and \ref{fig:3c273}). These variations may ultimately originate from stochastic instabilities within the accretion flow (\citealt{Malzac_2014}). However, whether and how accretion-driven variability is transferred into the jet is not definitely known. In the following we will focus on discussing variability directly associated with the accretion process.

Note that one can indirectly probe AGN variability on much longer timescales by studying powerful past events in nearby galaxies through studies of extended emission line regions. 
These can then trace the history of AGN emission over the light travel time from the nucleus to the gas \citep[typically $10^4 - 10^5$ years: e.g.][]{Dadina_2010,Keel_2012,Gagne_2014,Davies_2015}. 

\subsubsection{X-ray variability}
\label{sec:var_xray}
The availability of long-term monitoring and continuous sampling X-ray observations has allowed us to make significant progress towards a good characterization of X-ray variability in low-redshift AGN. These studies played a key role in our general understanding of AGN variability.

\begin{figure*}
\centering
\includegraphics[width=6.5cm]{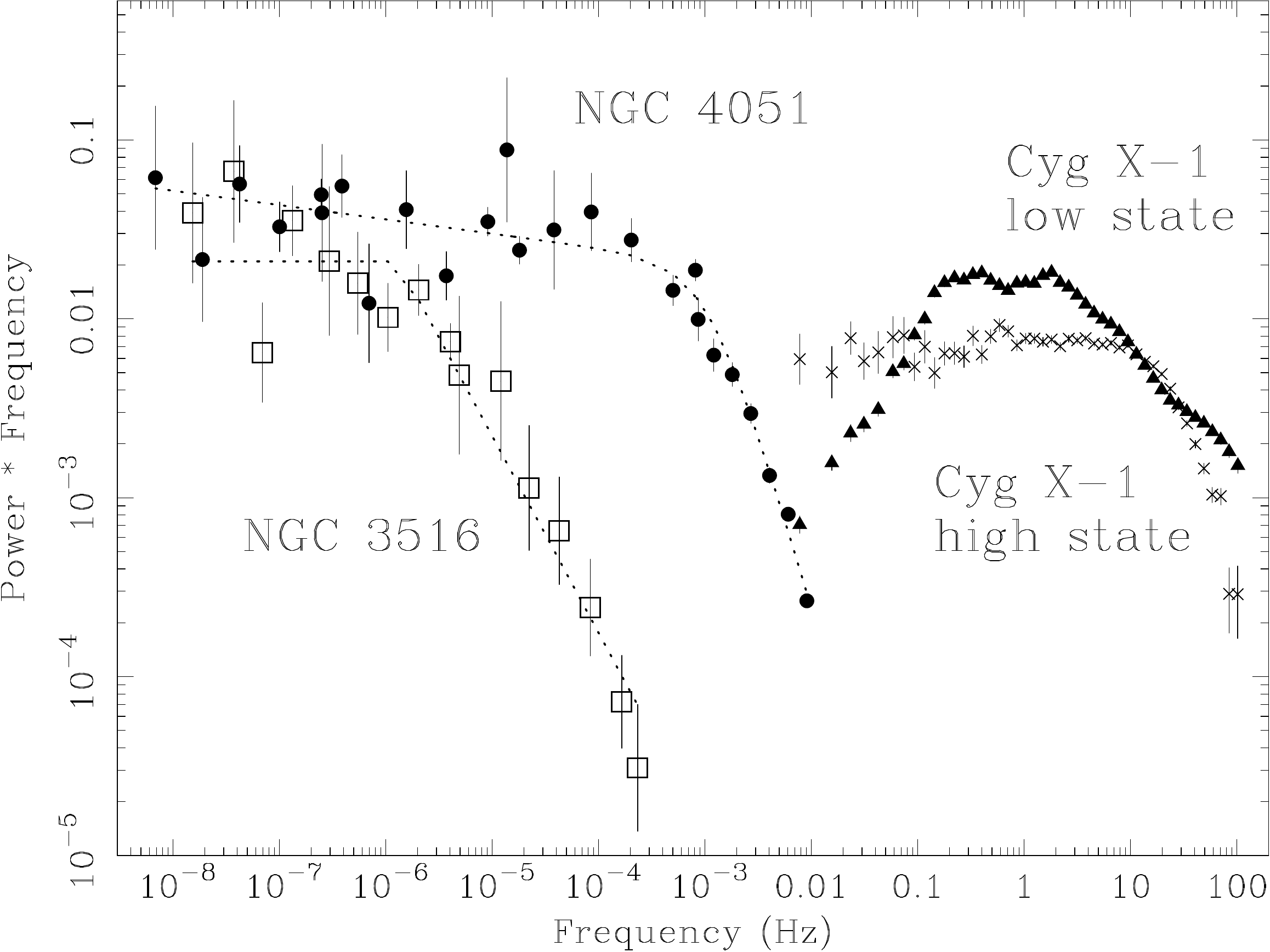}
\includegraphics[width=5.3cm]{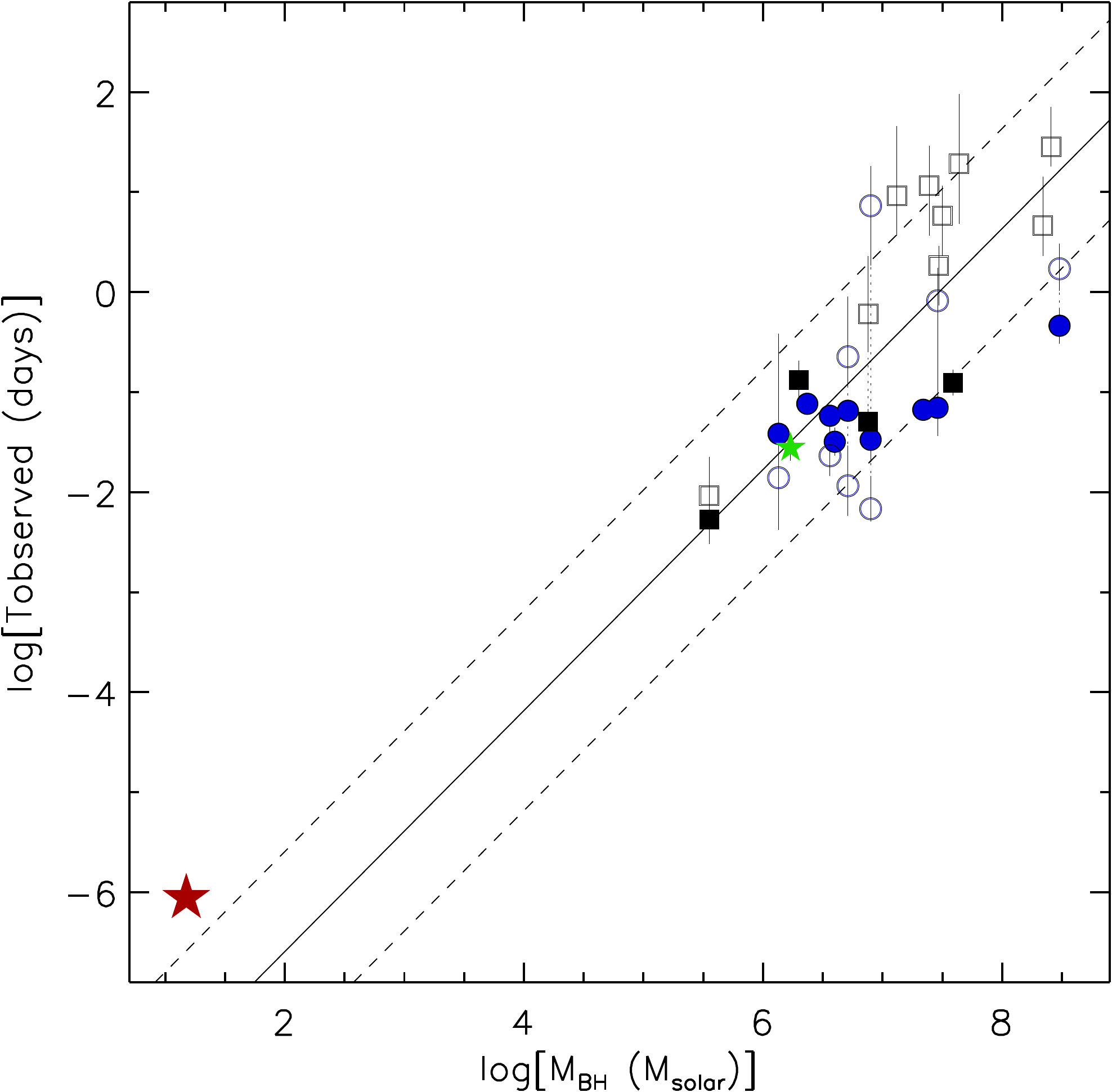}
\caption{\emph{Left panel:} X-ray PSDs from combined RXTE and  {\it XMM-Newton} observations of NGC 4051 (filled circles) and NGC 3516 (open squares), and comparison with the PSD of the BHXRB Cygnus X-1 in different accretion states.  Figure reproduced from \citealt{McHardy_2004}, Fig. 18, with permission. \emph{Right panel:} The scaling of the X-ray PSD break (expressed in terms of timescale $1/\nu_b$) as a function of $M_{\rm BH}$ in a sample of AGN observed by {\it XMM-Newton}. Open and filled circles are Narrow Line Seyferts, squares are Seyfert 1s, the green star is a Seyfert 2, the red star is Cygnus X-1. The continuous and dashed lines show the best-fitting linear model and the $\pm$1 dex region around this model. Figure reproduced from \citealt{GonzalezMartin_2012}, Fig. 5, with permission.}
\label{fig:varFig1}       
\end{figure*}

The main property of AGN variability is its ``red noise'' character, namely the occurrence of larger amplitude variations on longer timescales. 
This property became evident from early EXOSAT long-looks (e.g. \citealt{Lawrence_1987}). The more recent long monitorings performed by the {\it Rossi X-ray Timing Explorer }({\it RXTE}) enabled a detailed characterization of the X-ray 
PSD. Particularly when combined with continuous but shorter X-ray observations (e.g. by  {\it XMM-Newton}) they allowed sampling AGN variability over a broad range of timescales (from several years to seconds, Fig. \ref{fig:varFig1}, left panel). 
These studies revealed that the X-ray PSD of AGN has a universal shape, characterized by a steep (spectral index $\sim -2$) high-frequency (i.e. short-timescales) slope. Moreover, they showed that in many AGN the PSD flattens out to a slope of $\sim -1$ below a characteristic frequency, dubbed ``break frequency'', $\nu_{b}$ (e.g. \citealt{Edelson_1999}; \citealt{Uttley_2002}). 
Apart from this characteristic frequency, the PSD of AGN is mostly featureless over several decades in frequency\footnote{Notable exceptions are: Ark 564, which shows evidences of a second break at lower frequencies and a more structured PSD (\citealt{McHardy_2007}); REJ 1034+396 and MS 2254.9$-$3712, where the presence of quasi-periodic oscillations, QPO, has been reported (\citealt{Gierlinski_2008a}; \citealt{Alston_2015}).}.
The existence of a break in the PSD is indicative of $\nu_{b}$ corresponding to a physical timescale of the flow at some characteristic radius. There are a number of fundamental physical timescales that might be associated with the timescales of accretion flow variability (e.g. \citealt{Treves_1988}). However, these timescales depend also on some unknown parameters such as the viscosity and scale-height of the disk. Therefore, it is difficult to unequivocally associate $\nu_{b}$ with one of these timescales. Nonetheless, all the above mentioned timescales depend linearly on $M_{\rm BH}$. Early indications that also $\nu_{b}$ scales (inversely) with $M_{\rm BH}$ were reported in several studies (\citealt{Edelson_1999}; \citealt{McHardy_2004}) and finally confirmed by \citet[see also \citealt{Koerding_2007}]{McHardy_2006}, who showed that the correlation becomes tighter after correcting for the accretion rate/bolometric luminosity. Although the exact functional dependence on the accretion rate/bolometric luminosity is currently debated (e.g. \citealt{GonzalezMartin_2012}; \citealt{Ponti_2012}), 
it is widely accepted that stellar-mass BHs in X-ray binary systems (BH\-XRBs) lie on the extension of the $\nu_{b}-M_{\rm BH}$ relation, which holds for SMBHs 
(Fig. \ref{fig:varFig1}, right panel; a similar scaling occurs also in accreting neutron stars and white dwarfs: \citealt{Koerding_2007} and \citealt{Scaringi_2015}). 
Overall, the striking similarities of the X-ray timing properties of AGN and BH\-XRBs suggest the existence of a common physical process driving the observed variability.

\subsubsection{UV/optical variability} 
\label{sec:var_opt}
The UV/optical emission of AGN displays highly correlated variability, as expected if produced in the same physical region, the accretion disk. 
A detailed characterization of UV/\-optical variability is made difficult by the sparse and irregular sampling of ground-based observations, as well as the relatively long timescales associated with the regions of the disk emitting at these wavelengths (days-months). In recent years, dedicated ground-based campaigns (e.g. \citealt{Sanchez_2016}, \citealt{Caplar_2017}), the use of new powerful analysis approaches (\citealt{Kelly_2014}), and the increasing exploitation of data from space observatories (\emph{Kepler} and \emph{Swift}; e.g. \citealt{Smith_2015}) are opening the way to detailed studies of UV/optical variability of AGN. These studies have been revealing properties significantly different from those characterizing X-ray variability (e.g. \citealt{Mushotzky_2011}, \citealt{Simm_2016}), including steeper PSD high-frequency slopes (ranging between $\sim -3$ and $-4$) and the detection of the PSD break at systematically longer timescales than expected from X-rays ($\sim 100-300$ days) with no clear dependence on $M_{\rm BH}$.

An important breakthrough in the understanding of UV/\-optical variability came from the detection of inter-band, red lags (i.e. with longer wavelengths lagging behind shorter wavelengths) on day or sub-day timescales (e.g. \citealt{Sergeev_2005}). These lags are too short to be explained by some form of outwards diffusion in the flow. Therefore, heating from a central X-ray source is considered the most plausible mechanism. Indeed, the derived lag amplitude-wavelength dependence is consistent with that expected from reprocessing of X-rays in a standard disk (e.g. \citealt{Cackett_2007}; \citealt{Edelson_2015}). In this scenario the lags are dominated by the light crossing time from the X-ray source to the region of the disk emitting at a given wavelength, with longer (red) wavelengths coming from larger radii, thus producing longer lags. Recent analyses of simultaneous X-ray, UV/optical, and IR observations confirmed the disk reprocessing hypothesis, and extended it to explain also the variable emission from the surrounding dusty torus (e.g. \citealt{McHardy_2014}; \citealt{Vazquez_2015}).

Reprocessing of variable X-ray photons in the UV/optical emitting regions of the disk explains why, on timescales of hours-to-days, AGN are more variable in the X-rays than in the UV/optical. However, the availability of long X-ray observations and simultaneous optical monitoring has allowed the extension of these studies to longer timescales, revealing that on month-to-year timescales variations in the optical are more intense than in the X-rays (e.g. \citealt{Arevalo_2009}; \citealt{Breedt_2010}; \citealt{Uttley_2003}). This behaviour cannot be explained by X-ray reprocessing only, but it requires some additional source of intrinsic disk variability dominating on long timescales (Sect. \ref{sec:var_models}).

\subsubsection{What does variability tell us about the driving physical mechanism?} 
\label{sec:var_models}
The X-ray PSD of AGN has been extensively studied to gather information about the driving physical mechanism. 
However, the sole analysis and modelization of the PSD has not led to a clear identification of the underlying process. 
Indeed, most of the originally proposed models (such as shot-noise models: e.g. \citealt{Lehto_1989}) had difficulties in explaining other X-ray timing properties of both AGN and BHXRBs, such as the observed ``rms-flux'' relation,
whose existence was first shown by \cite{Uttley_2001} and later confirmed by several studies (e.g. \citealt{Vaughan_2003}; \citealt{Gaskell_2004}; \citealt{McHardy_2004}). 
This relation indicates that the absolute amplitude of X-ray variability (rms, e.g. \citealt{Vaughan_2003}) is linearly correlated with the flux level, meaning that the source is more variable when it is brighter. The existence of an rms-flux relation implies that the underlying physical process is multiplicative and that the long-term variations should modulate the short-term variations (\citealt{Gaskell_2004}; \citealt{Uttley_2005}).

Such findings argue against models (such as additive shot-noise models) invoking the presence of independent flares or active regions.
Rather, it appears that the class of ``propagating-fluctuation'' models, first introduced by \cite{Lyubarskii_1997}, are more suitable to explain a vast range of X-ray timing properties, including the rms-flux relation (\citealt{Arevalo_2006}; \citealt{Ingram_2013}). These models assume the emergence of local perturbations of disk parameters triggering variations of the accretion rate, with long timescale perturbations produced at larger radii. If the time\-scale of the perturbations is longer than the radial diffusion time, the perturbations can propagate inward, combining multiplicatively with perturbations produced at smaller radii, and reach the innermost zones where most of the energy is released (e.g. \citealt{Churazov_2001}; \citealt{Hogg_2016}).

These models can explain the presence of large amplitude X-ray variability on a wide range of timescales, even orders of magnitude longer than the viscous-timescale of the compact X-ray emitting regions of the flow. Moreover, the perturbations produced throughout the disk could be responsible for driving intrinsic UV/optical disk variability \citep{Uttley_2003}. This component would explain the excess long-timescales optical variability detected in some AGN (Sect. \ref{sec:var_opt}). 
Propagating-fluc\-tu\-a\-tion models are also invoked to explain the common detection of ``hard'' X-ray lags (time-delays of the hard X-ray band variations with respect to soft X-ray band variations) in both AGN and BHXRBs (e.g. \citealt{Miyamoto_1988}; \citealt{McHardy_2004}; \citealt{Uttley_2011}; \citealp{DeMarco_2013,DeMarco_2015}). These lags have large amplitudes, typically of the order of 1\% of the variability timescale. By interpreting the lags as due to light-travel time effects, the inferred size of the emitting regions would be too large ($\sim 10^3$ gravitational radii) to be plausible (e.g. \citealt{Nowak_1999a}). However, if due to the slower, viscous propagation in the flow the large lags can be easily recovered in standard disk-corona geometries (e.g. \citealt{Kotov_2001}; \citealt{Arevalo_2006}).

Variations of photoelectric absorption, e.g. associated with temporary obscuration of the central engine as a consequence of gas clouds located in the BLR or the torus crossing the observer's line of sight can also contribute to the observed AGN variability (\citealt{Risaliti_2011}; \citealt{Cappi_2016}). However, this is unlikely to be responsible for the bulk of it, as several arguments and observational evidences rather favour an intrinsic origin. These include: the ubiquity of AGN variability; the universal shape of the PSD; the scaling of characteristic timescales with $M_{\rm BH}$; the similarities of timing properties in AGN and BHXRBs; the fact that the inner/outer disk and the BLR directly respond to continuum variations as inferred from results of X-ray (\citealt{Fabian_2009}; \citealt{Zoghbi_2012}; \citealt{DeMarco_2013}; \citealt{Kara_2016}) and UV/optical reverberation mapping (e.g. \citealt{Peterson_2004}; \citealt{Edelson_2015}). Moreover, recent studies of long-term ($\sim$15 years) X-ray variability of distant AGN \citep[$z = 0.6-3.1$:][]{Yang_2016} showed that intrinsic X-ray variability prevails over absorption variability (see also \citealt{Hernandez_2015}; \citealt{Soldi_2014}), further supporting this conclusion.

\subsection{AGN variability in extragalactic surveys}\label{sec:var_surveys}
\subsubsection{Selecting AGN by variability}

Extragalactic surveys are typically characterized by uneven sampling, with large gaps between consecutive cycles. This makes PSD techniques unsuitable for studying variability in these datasets. The identification of variable sources requires assessing whether the variability observed between observations exceeds the variations due to statistical fluctuations. This is usually done by computing the quantity 
$X^2 = \sum_{i=1}^{N_{obs}} (x_{i}-\bar{x})^2/\sigma^2_{err,i}$, where $x_{i}$ denotes the photon flux (e.g. \citealt{Lanzuisi_2014}; \citealt{DeCicco_2015}; \citealt{Simm_2016}; \citealt{Kozlowski_2016}) during each observation, 
$\bar{x}$ is the average photon flux, and $\sigma^2_{err,i}$ is the measurement error.
The $X^2$ statistics tends to a $\chi^2$ distribution for large photon fluxes (in a low-photon flux regime assessing whether a source is significantly variable requires carrying out MonteCarlo simulations; e.g. \citealt{Paolillo_2004}; \citealt{Young_2012}).  
Thus, the ability to detect significant variability in single sources depends on the error of each measurement. At a given source flux and exposure, this in turn depends on the collecting area of the detector and the background contribution, which affects the uncertainty on the flux measurements (e.g. \citealt{Lanzuisi_2014}).

\begin{figure*}
\centering
\includegraphics[width=5.8cm]{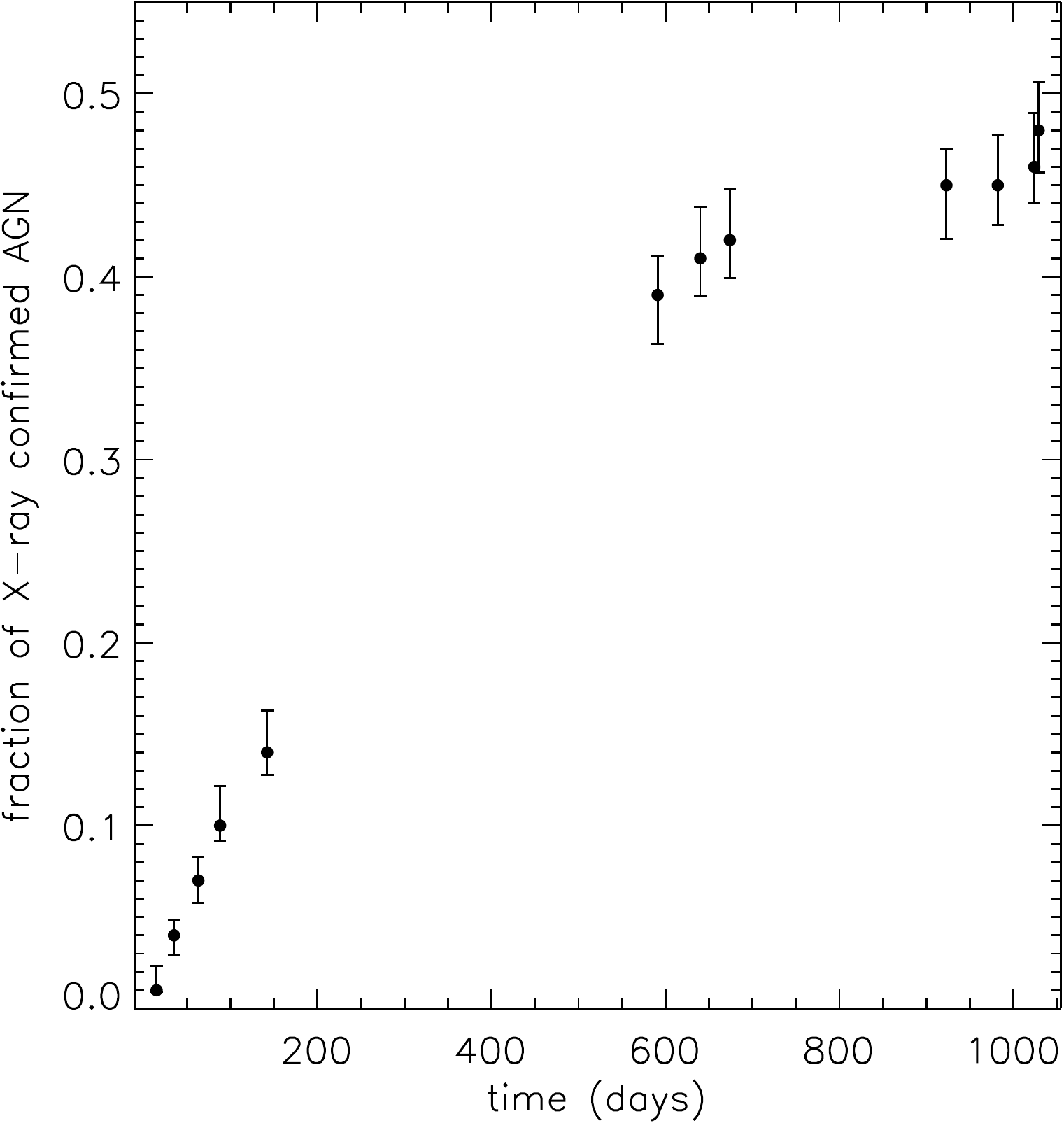}
\hspace{0.5cm}
\includegraphics[width=5.4cm]{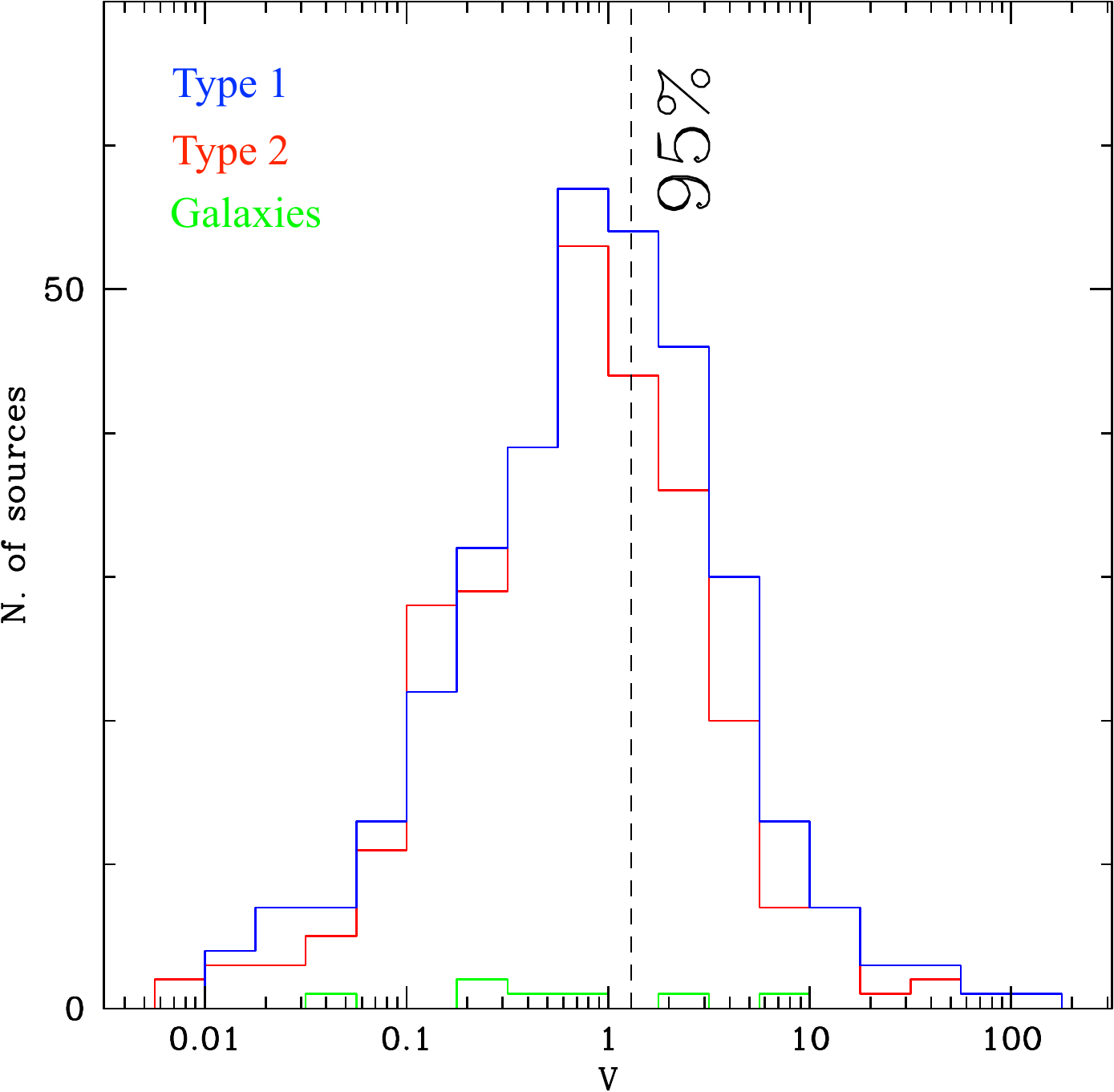}
\caption{\emph{Left panel:} Fraction of optically variable AGN in the VST-SUDARE/VOICE survey of the COSMOS field as a function of temporal baseline. Figure reproduced from De Cicco et al. 2017, in prep., with permission. \emph{Right panel:} Distribution of X-ray variability index V (which parametrizes the confidence level of variability) for sources selected from the XMM-COSMOS survey divided by class. 
The dashed vertical line marks the value V$=$1.3 used to discriminate between variable and non-variable sources. Sources with V$>$1.3 are intrinsically variable at the 95\% confidence level. \copyright~AAS. 
Figure reproduced from \citealt{Lanzuisi_2014}, Fig. 5, with permission.}
\label{fig:varFig2}       
\end{figure*}

Selection on the basis of variability can return a relatively high number of AGN candidates. The latest estimates from optical variability over a temporal baseline of 3 years (De Cicco et al. 2017, in prep.) are 275 deg$^{-2}$ AGN at a magnitude limit $r<23$. This corresponds to $\sim 48$\% completeness with respect to X-ray confirmed AGN at the depth reached in the COSMOS field
(Fig. \ref{fig:varFig2}, left panel). 
However, the resulting AGN density strongly depends on the characteristics of the survey - mainly the spanned temporal baseline and sampling cadence, the observing wavelength, the depth of the single exposures - as well as on the AGN type. Some of these aspects are discussed in Sect. \ref{sec:opt_sel_eff}. 

\subsubsection{Selection effects}
\label{sec:opt_sel_eff}

As the bulk of AGN variability appears to be intrinsically associated with the accretion process, any AGN can in principle be selected by variability
provided a significant fraction of the emission from the nucleus is directly observable. 
This requirement makes the X-ray band particularly suitable for the identification of AGN, independently of their classification, thanks to the penetrating power of X-rays and the 
relatively weak host galaxy emission (Sect. \ref{sec:X-ray_sel}). 
As a matter of fact, the number of X-ray variable AGN appears to be type-independent (see Fig. \ref{fig:varFig2}, right panel), with X-ray variable type 2 AGN typically showing less variability at longer wavelengths and on similar time scales (e.g. \citealt{Lanzuisi_2014}, \citealt{Hernandez_2015}). Indeed, in the optical band the completeness of AGN samples selected through variability decreases significantly for type 2 AGN (\citealt{DeCicco_2015} estimate 25\% and 6\% completeness respectively for type 1 and type 2 AGN over a temporal baseline of 5 months), while in the X-ray band it is similar for the two classes (e.g. $\sim$40\% and $\sim$32\% completeness respectively for type 1 and type 2 AGN, \citealt{Lanzuisi_2014}).
Nonetheless, it is important to note that high column densities of circumnuclear absorbing material can significantly affect the fraction of detected X-ray variable AGN, e.g. as a consequence of the increased contribution from a large scale, constant reflector (\citealt{Paolillo_2004}; \citealt{Hernandez_2015}).

X-ray variability is also suitable for identifying LLAGN, which are missed by other selection criteria (\citealt{Young_2012}).
Systematic X-ray variability studies showed that, gi\-ven good statistics, variability can be detected in a high fraction of AGN (e.g. $80-90\%$ \citealt{Paolillo_2004}; \citealt{Lanzuisi_2014}; \citealt{Soldi_2014}; \citealt{Yang_2016}) independently of the optical classification of the object or its radio loudness (note that in jetted AGN the X-ray variability can contain a very significant contribution from the jet: e.g. \citealt{Lichti_2008}; see also Figs. \ref{fig:mkn421}  and \ref{fig:3c273}).
 
The effectiveness of variability as a tool to select AGN is strongly related to the temporal baseline spanned by the observations and the depth of a single frame. Indeed, given the red-noise character of the PSD of AGN, the ability of detecting variability is higher at longer timescales (see Fig. \ref{fig:varFig2}, left panel), particularly when the low-frequency part of the PSD below $\nu_b$ is sampled. 
In the X-ray band, $\nu_b$ scales inversely with $M_{\rm BH}$ and directly with the accretion rate/bolometric luminosity (but see discussion in Sect. \ref{sec:var_xray}), thus the sampling of longer timescales also allows detection of variability over a wider range of 
$M_{\rm BH}$ and $L/L_{\rm Edd}$. In the UV/optical this scaling is not clearly observed (\citealt{Simm_2016}), but the break is measured at systematically longer timescales than in the X-rays (Sect. \ref{sec:var_opt}). Therefore, the longer the maximum sampled timescales, the higher the detection fraction and, at least as far as the X-rays are concerned, the larger the range of $M_{\rm BH}$ and $L/L_{\rm Edd}$ probed. Note that time dilation effects reduce the effective range of sampled time\-scales. Therefore, given a total exposure time, the more distant a quasar is the more likely it is to intercept the high-frequency part of the PSD, where the variability power drops rapidly, thus making detection of variability more difficult and limited by statistics.

\subsubsection{Variability dependence on luminosity and redshift}
\label{sec:var_lum}

In the presence of uneven, gappy data, variability can be quantified using the so-called normalized excess variance ($\sigma^2_{NXV}$; e.g. \citealt{Nandra_1997}; \citealt{Vaughan_2003}; \citealt{Ponti_2012}; see also \citealt{Kelly_2014}), which provides an estimate of the amplitude of intrinsic source variability over the sampled timescales. 
Biases associated with measurements of $\sigma^2_{NXV}$ have been extensively studied by \citet[see also \citealt{Almaini_2000}]{Allevato_2013}. These biases can be greatly reduced choosing an appropriate observing strategy (e.g. planning observations separated by constant temporal gaps), while ``ensemble'' $\sigma^2_{NXV}$ estimates (i.e. obtained by averaging over single $\sigma^2_{NXV}$ estimates from multiple light curves of the same source, or from single light curves of many sources with similar properties) should be preferred.

Deriving a reliable estimate of $\sigma^2_{NXV}$ is important in order to properly analyze global variability properties of different AGN classes, to study correlations with fundamental parameters, and to investigate the evolution of variability with redshift. 
These studies have shown that variability is anti-correlated with luminosity in several wavebands (e.g. in the X-rays: \citealt{Ponti_2012}; \citealt{Lanzuisi_2014} and references therein; in the optical/MIR: \citealt{Hook_1994}; \citealt{Kelly_2009}; \citealt{MacLeod_2010}; \citealt{Zuo_2012}; \citealt{Kozlowski_2016}; \citealt{Simm_2016}). The origin of this anti-correlation is currently unknown, but, at least in the X-rays and on relatively short timescales (shorter than $\sim 1/\nu_b$), this correlation seems to be the byproduct of a more fundamental anti-correlation of variability with $M_{\rm BH}$ ultimately driven by the scaling of the PSD break (e.g. \citealt{Czerny_2001}; \citealt{Papadakis_2004}; \citealt{ONeill_2005}; \citealt{Nikolajuk_2006}; \citealt{Zhou_2010}; \citealt{Ponti_2012}; \citealt{Lanzuisi_2014}).  
The $\sigma^2_{NXV}$-$M_{\rm BH}$ relation is very tight, with an estimated scatter of the order of the uncertainties on $M_{\rm BH}$ (\citealt{Zhou_2010}; \citealt{Ponti_2012}). These findings reinforce early suggestions that $\sigma^2_{NXV}$ could be used to estimate $M_{\rm BH}$ (\citealt{Czerny_2001}; \citealp{Nikolajuk_2004,Nikolajuk_2006}, \citealt{Gierlinski_2008b}).

Several studies also showed that both X-ray and optical variability tend to be stronger at higher redshifts, but this appears to be a consequence of selection effects (e.g. \citealt{CidFernandes_1996}; \citealt{Almaini_2000}; \citealt{Manners_2002}; \citealt{Paolillo_2004}; \citealt{Lanzuisi_2014}; \citealt{Morganson_2014}; \citealt{Simm_2016}; \citealt{Yang_2016}; \citealt{Paolillo_2017}). Therefore, to date there are no clear indications of an evolution of AGN variability with redshift.

\subsection{The future of AGN variability studies}\label{sec_var_future} 

Variability is a defining property of AGN, therefore a fundamental tool to understand the AGN engine and map its close environments. Due to its diagnostic power and complementarity to other probes, variability is gaining increasing importance in the census of the AGN population. 
The recognized relevance of AGN variability studies is mirrored by the growing number of future missions and facilities that include the investigation of the variable Universe amongst their science goals. 
Among these is eROSITA (\citealt{merloni2012}; see also Sect. \ref{sec:X-ray_future}), due to launch in spring 2018. During its four years of operation eROSITA will perform a deep survey of the entire sky at X-ray wavelengths. The planned monitoring strategy (eight all-sky surveys and a cadence of six months, with the ecliptic poles being monitored at higher pace) will provide repeated observations of the same portions of the sky. Ultimately, this will allow sampling AGN X-ray variability over time scales from 250 s (corresponding to the daily eROSITA exposure) up to 4 years (corresponding to the scheduled duration of the survey phase). Based on AGN X-ray variability characteristics and scaling properties (Sect. \ref{sec:var_xray}), eROSITA is expected to detect around 60,000 {\it variable} AGN in the full sky (and many more overall: Sect. \ref{sec:X-ray_future}). 

In the far future, {\it Athena} (\citealt{nandra2013}; see also Sect. \ref{sec:X-ray_future}) is expected to greatly advance our knowledge of AGN X-ray variability. Thanks to its large collecting area, the sensitivity of X-ray variability measurements will be significantly enhanced. This will have a great impact on our understanding of the accretion process variability, it will allow the mapping of the close environments of AGN and will assess the role of AGN winds and outflows and their influence on the observed X-ray variability properties.
 
The exploration of the transient and variable sky is one of the driving science theme of several future facilities that will operate at longer wavelengths. In the optical band these include the Zwicky Transient Facility (ZTF\footnote{\url{www.ptf.caltech.edu/ztf}}), a high-cade\-nce, 3-year, time-domain survey, starting in summer 2017. However, the currently most ambitious planned survey in the optical band is the LSST, starting operations in 2022 (\citealt{LSST}). LSST is designed to uniformly observe an area of $\sim 20,000$ deg$^2$, with about $50-200$ visits per source for each of the six filters, over an operation period of ten years. This cadence will probe optical variability on time scales ranging between 1 minute and one decade, allowing selection of tens of millions of AGN and the construction of highly-complete catalogues with minimum contamination. Finally, variability studies of radio-emitting AGN will be one of the main science goals of the SKA (Sect. \ref{sec:radio_future}), which is expected to provide a deluge of AGN radio variability information. 

\section{Summary}\label{sec:discussion}

\begin{figure}
\centering
\includegraphics[width=9.0cm]{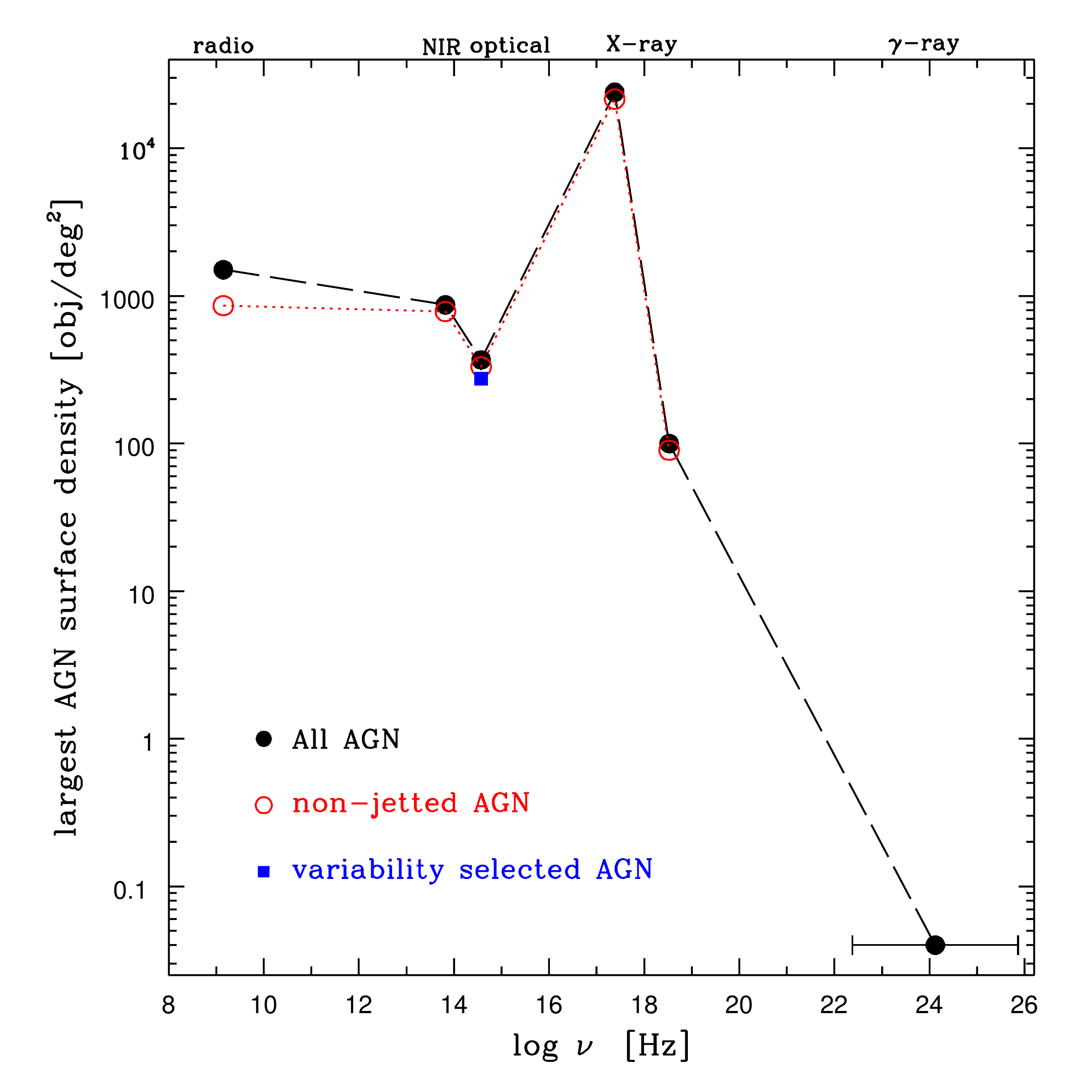}
\caption{The largest AGN surface density over the
  whole electromagnetic spectrum. Black filled points refer to all AGN,
  open red points are for non-jetted AGN. The latter are actually directly measured
  only in the radio band, while in the NIR to X-ray bands they have been
  derived by simply subtracting $10\%$ from the total values. Non-jetted AGN
  have not been detected in the $\gamma$-ray band. The blue square indicates 
  variability selected AGN (Sect. \ref{sec:var_surveys}). Updated from Fig. 11 of \cite{Padovani_2016},
  where one can find the references to the relevant samples and facilities, 
  to include variability selected AGN and the results of the CDFS 7 Ms sample \citep{Luo_2017}.}
\label{fig:agn_density}       
\end{figure}

This review highlights the extraordinary diversity of observational signatures produced by growing SMBHs, and illustrates the challenges of placing all these observations into a coherent picture. AGN are complex systems exhibiting a wide range of 
phenomena, and different classes of AGN can be selected across the full range of the electromagnetic spectrum (Fig.~\ref{fig:agn_density}). The efficiency of selection of AGN in various wavebands is the result of: (1) variations in the underlying 
physical properties of AGN; (2) observational capabilities; (3) selection effects. 

The first consideration relates to physical effects (introduced in 
Sect. \ref{sec:intro} and discussed throughout this review). The most straightforward of these is the presence or absence of a 
strong relativistic jet, leading to the distinction between jetted and non-jetted AGN as discussed in Sect. \ref{sec:radio}. 
As detailed in Sect. \ref{sec:RL_RQ} and shown in Fig.~\ref{fig:agn_density}, the latter class is fainter in the radio and {\em absent} in the $\gamma$-ray band (Sect. \ref{sec:gamma}), 
highlighting how AGN selection can be affected by variations in the physical properties of different AGN populations. Other physical effects include changes in the accretion flow with 
$L/L_{\rm Edd}$ (which can modify the shape of the intrinsic SED, in particular the relative outputs in the optical/UV and X-ray bands; e.g. \citealt{vasu09sed, yuan2014}), and 
obscuration by gas and dust that can highly suppress emission in the NIR through soft X-rays (Sects. \ref{sec:IR} -- \ref{sec:X-ray}). 

Second, observations in some wavebands are more sensitive than others relative to the typical AGN SED. For example, the current generation of radio telescopes (Sect. \ref{sec:radio}) 
can probe AGN with far smaller bolometric fluxes, compared to contemporary $\gamma$-ray observatories (Sect. \ref{sec:gamma}), yielding a far higher surface density of detected AGN. 

Third, AGN selection techniques (due to luminosity, co\-lour, morphology, variability, etc.) have different levels of efficiency in different wavebands. Of particular importance is the relative 
emission of the AGN compared to its host galaxy. At wavelengths at which stellar processes in the host galaxy can be particularly luminous (e.g. optical, IR, and to some extent the radio), 
it is challenging to select extremely faint AGN even in the deepest observations (although well-designed techniques using colours or emission line characteristics can help minimize contamination). 
By contrast, host contamination is particularly low in the X-ray band (Sect. \ref{sec:X-ray}), and cutting-edge X-ray observatories are capable of extremely deep observations. Simple X-ray luminosity 
selection can therefore probe AGN at very faint fluxes, leading the detected AGN surface density to be highest in this band (Fig. \ref{fig:agn_density}). 

We note that by multiplying the largest entry in Fig. \ref{fig:agn_density} by the area of the sky we 
estimate that there are {\em at least} $\approx 1$ billion AGN in the Universe that could be detected 
with current technology. This number needs to be compared with the number of currently known AGN, 
of the order of a million (Fig. \ref{fig:qsoswtime}), which shows the exciting potential for discovery in 
future AGN surveys (as detailed in Sects. \ref{sec:radio_future}, \ref{ssec:IRfuture}, \ref{sec:opt_evol}, \ref{sec:X-ray_future}, \ref{sec:gamma_future}, and 
\ref{sec_var_future}). We can also compare it with the number of currently observable galaxies in the 
Universe, $\approx 200$ billion \citep{Conselice_2016}, indicating that bright AGN are found in 
approximately 1\% of galaxies.

\begin{table*}
\centering
\caption{A multi-wavelength overview of AGN highlighting the different selection biases (weaknesses) and key capabilities (strengths). The definitions of some of the terms 
used in the bias and capability columns are as follows: {\em Efficiency}: 
ability to identify a large number of AGN with relative small total exposure times (this is thus a combination of the nature of AGN emission 
and the capabilities of current telescopes in a given band). {\em Reliability}: the fraction of sources that are identified as AGN using  
typical criteria that are truly AGN. {\em Completeness}: the ability to detect as much as possible of the full underlying population of AGN.}     

\label{tab:summary}      
\begin{tabular}{|  l  |  l  | l |  c | c |}  \hline
{\bf Band} &  {\bf Type} & {\bf Physics} & {\bf Selection biases/weaknesses} & {\bf Key capabilities/strengths} \\ \hline
Radio, $f_{\rm r} \gtrsim 1$ mJy & jetted & jet & non-jetted sources  & high efficiency, \\ 
 &  &  &   & no obscuration bias \\ \hline
Radio, $f_{\rm r}  \lesssim 1$ mJy & jetted and non-jetted & jet and SF & host contamination & completeness, \\ 
 & &  &  & no obscuration bias\\ \hline
IR  & type 1 and 2 & hot dust & completeness, reliability,  & weak obscuration bias,  \\ 
  &  & and SF & host contamination, no dust & high efficiency \\ \hline
Optical & type 1 & disk & completeness, low-luminosity, & high efficiency,  \\ 
 & &  &obscured sources, host contamination & detailed physics from lines \\ \hline
X-ray & type 1 and (most) 2 & corona & very low-luminosity,  & completeness, low  \\ 
          &                                 &  & heavy obscuration & host contamination \\ \hline
$\gamma$-ray & jetted & jet & non-jetted, & high reliability\\ 
 & & & unbeamed sources & \\ \hline
Variability & all (in principle) & corona, & host contamination, obscuration,  & low-luminosity   \\ 
 &  &  disk, jet  &cadence and depth of observations & \\ \hline
\end{tabular}

\end{table*}

Table \ref{tab:summary} summarizes the selection biases and key capabilities (in other words, the weaknesses and the strengths) in different bands, as described above and in detail throughout this review. Despite the significant differences between AGN classes detected in some bands, the observations highlighted in this review illustrate that AGN share some common components over a wide dynamic range (up to 12 orders of magnitude) in physical scales. In what follows, we will describe this basic ``architecture'' of an AGN in the context of the observational signatures described in the review. We will then discuss our present understanding of AGN unification models, highlighting the need for a more complex picture than the classic ``strict'' unification, and discussing the importance of understanding relevant timescales in interpreting observations. Finally, we will present a broad framework for an emerging picture of AGN and their host galaxies in light of the exciting recent advances such as those presented in this review and will conclude by highlighting
some future prospects and open questions. 

\subsection{Fitting together the components of an AGN}
\label{sec:components}

Based on the current observational knowledge of AGN as discussed throughout this review, we can now assemble the different components, covering a wide range of physical sca\-les, that together make up the complete picture of the AGN phenomenon.

{\em Black hole ($10^{-7}$--$10^{-3}$ pc)--} The fundamental element common to all AGN is of course the central BH itself. Astrophysical BHs are simple objects whose only distinguishing characteristics are mass and spin, but as we have seen throughout this review, these two parameters are critically important in determining the observable features of an AGN. It is a nearly universal feature of AGN selection that more massive BHs are easier to detect, because the mass of the BH sets the Eddington limit, and so massive BHs are more luminous at a given $L/L_{\rm Edd}$ (Sect. \ref{ssec:MIR_phot_ledd} and \ref{sec:X-ray_types}). (The exception here is selection via X-ray variability, for which characteristic frequencies vary inversely with $M_{\rm BH}$, so that detecting more massive BHs requires longer timescale observations, as described in Sect. \ref{sec:var_xray}.) The effects of BH spin are far more difficult to understand observationally, but rapidly spinning BHs may have accretion disks that extend closer to the event horizon, producing a bluer SED (see below) with a higher radiative efficiency, 
 and may influence the ability of accreting BHs to launch relativistic jets and radiation line-driven winds.\citep[e.g.][]{tche11jet, nara14jet}. 
 
{\em Accretion flow ($10^{-7}$--$1$ pc)--} The accretion flow onto the BH is the ultimate energy source in AGN and the most important component in determining their observational characteristics. Indeed, turning a normal galaxy into an AGN merely requires a SMBH accreting 
at a relatively high accretion rate. 
There is compelling theoretical and observational evidence that the structure of the flow changes with $L/L_{\rm Edd}$ \citep[e.g.][]{Narayan_1994, vasu09sed,yuan2014}, with higher $L/L_{\rm Edd}$ systems producing stronger emission from an optically-thick accretion disk in the optical and UV \citep[e.g.][]{Slone_2012,Netzer_2014,Capellupo_2015,CastelloMor_2016,Bertemes_2016}
 and stronger radiation line-driven winds \citep[e.g.][]{Laor_2014}. 

The structure of the flow also appears to influence emission from the magnetic corona (with lower $L/L_{\rm Edd}$ systems having stronger X-rays relative to the disk emission: Sect. \ref{sec:X-ray_phys}) and is also connected to the launching of relativistic jets, with luminous radio emission observed to be more common in lower $L/L_{\rm Edd}$ systems (see references in Sect. \ref{sec:radio}). Changes in the accretion flow are also the source of the bulk of observed variability in AGN. As discussed in Sect. \ref{sec:variability}, the variability is observed over a wide range of timescales from 
the optical to the X-rays, likely corresponding to characteristic (e.g. viscous) time scales at various radii in the accretion disk (although there is also some interplay due to reprocessing of emission from different regions). The nature of the variability changes with $L/L_{\rm Edd}$ further illustrating the dependence on the structure of the flow.
  
{\em Torus ($1$--$10$ pc)--} The putative torus of gas and dust surrounding the central engine in the AGN is critical to understanding two key observational phenomena: emission from the AGN in the IR, and obscuration of the accretion disk and corona emission (see also \citealt{Netzer_2015} for a review). As discussed in Sect. \ref{ssec:IR_SED}, the majority of the MIR light from luminous AGN is emission from hot dust. 
The geometric distribution (smooth or clumpy, axisymmetric or polar) and kinematics (static, inflowing, or outflowing) of the dusty material are however still uncertain, as is their connection to the accretion flow on larger and smaller scales. At minimum, it is now well-established that the structure of the torus is not universal but vary significantly for AGN types and even within an individual class (Sect. \ref{sssec:IR_hdps}). This variation in the dusty torus also has implications for AGN obscuration, as the covering factors of obscuring material can differ significantly for different AGN, and variations in torus structure can affect the observed column densities, extent, and time variability of the obscuration. We can ultimately conclude that the dusty torus is a common component of AGN that is critical for understanding observations, but its physical nature is uncertain, variable, and complex. 

{\em Jet ($10^{-7}$--$10^6$ pc)--} Another important but poorly understood component of the AGN phenomenon is the presence and strength of a relativistic jet. Jets dominate the emission from AGN in the $\gamma$-rays and often in 
 the radio band (as discussed in Sects.~\ref{sec:radio} and \ref{sec:gamma}), and can make important contributions in the X-rays and (in the case of blazars) across the whole SED. The structure and orientation of the jet can influence a range of rich observational phenomena, particularly in the radio and $\gamma$-rays; indeed the majority of the AGN classes presented in Tab. 1 relate in some way to the presence of a jet. However, despite jet emission being among the best- (and earliest-) studied aspects of AGN, the physical nature of AGN jets and the prevalence of relativistic outflows, particularly low-luminosity jets on small scales, remain uncertain. 

{\em Host galaxy and dark matter halo ($10$--$10^6$ pc)--} In understanding the observational signatures of AGN it is essential to account for the host galaxies and large-scale structures in which the BHs reside (see \citealt{hick16iau} for a review). One important consideration is the effect of host galaxy dilution (or equivalently, ``contamination'') on AGN selection, which are spelled out in detail in this review, particularly in the radio, IR, optical, and X-ray bands (Sects.~\ref{sec:radio} -- \ref{sec:X-ray}). In all cases, this contamination leads to uncertainties in the population of faint AGN. It also produces a bias in AGN selection toward higher $L/L_{\rm Edd}$, an effect which is strongest in the IR and optical bands, and serves as motivation for observations at high angular resolution that can separate the AGN emission from the stellar processes in the surrounding galaxy \citep[e.g.][]{gandhi2009, simm11edd}. The properties of gas in host galaxies can also have a physical influence on the fueling and emission from AGN. Some AGN obscuration originates on galaxy scales much larger than the putative dusty torus \citep[e.g.][]{chen15qsosf,buch17obs}, and connections between physical AGN properties (for example the distinction between HEG and LEG line classifications, indicating different $L/L_{\rm Edd}$: Sect. \ref{sec:LEG_HEG}) and the stellar populations of host galaxies indicate that the gas supply for BH accretion can be connected to the larger-scale reservoir from which stars form in the host galaxy \citep[e.g.][]{hickox2009, Smolcic_2009a, Smolcic_2009b, azadi2015}. Finally, the interaction between AGN and gas in surrounding dark matter halos can have important effects on observed phenomena, particularly with hot spots, lobes, and wide-angle tails in large-scale relativistic jets \citep[e.g.][]{hard15jet}.

\subsection{Unification and the importance of timescales}

Piecing together our current understanding of the various components of an AGN and their observational signatures, we can begin to re-assess the extent to which the full AGN population can be explained by a ``unified'' model that reduces the complex AGN ``zoo'' described in Sect. \ref{sec:intro} (Tab. \ref{tab:agn_zoo}) to a relatively small set of parameters. 

\begin{figure}
\centering
\includegraphics[width=10.cm]{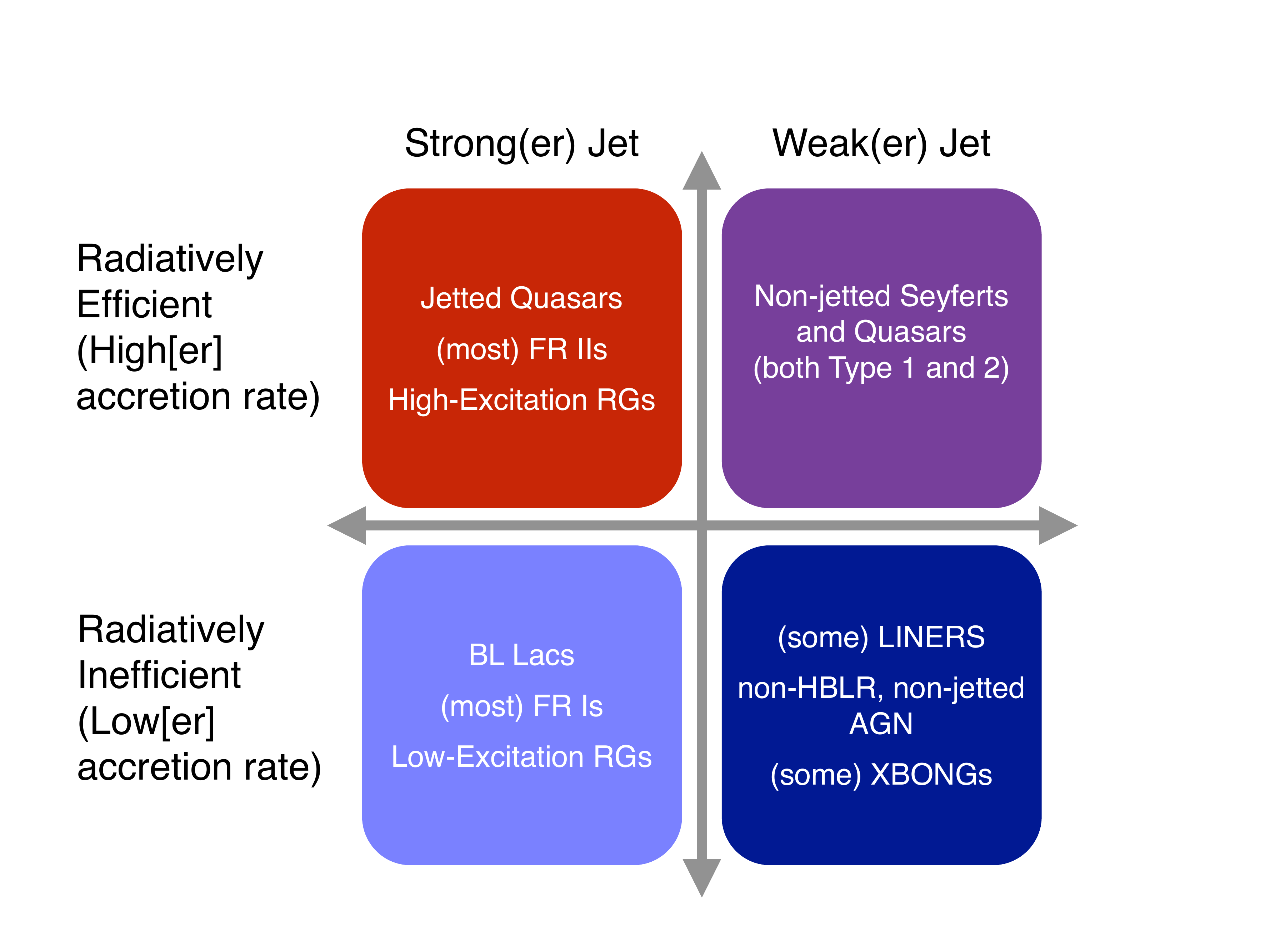}
\caption{Observational signatures in a ``weak'' unification model with two parameters (radiative efficiency [related to $L/L_{\rm Edd}$], and relativistic jet strength), showing the classes associated with each broad range in this parameter space. This schematic illustrates the need for parameters beyond orientation to explain the wide range of AGN properties. A theme that emerges throughout this review is that these classifications do not represent distinct groups of AGN, but rather general regions in a continuous distribution of AGN properties. Adapted from schematic by P.~Hopkins.}
\label{fig:schematic1}       
\end{figure}

In the most straightforward ``strict'' unified model, the only parameter is orientation relative to an axisymmetric dusty torus and relativistic jet \citep{Antonucci_1993,Urry_1995}. In broad terms this model has been remarkably successful in connecting type 1 and 2 Seyferts, and FSRQs and HEGs. In addition to the presence or absence broad lines, orientation can also help explain the details of the line profiles; for example, \citet{Shen_2014} suggest FWHM(H${\beta}$) is more of an indicator of orientation than mass. However, the evidence presented in this review and elsewhere (see \citealt{Netzer_2015} for a comprehensive overview) suggests that a picture based primarily on orientation and obscuring material is incomplete.  While the structure of AGN clearly deviates from spherical symmetry and so there is some orientation dependence on observed properties, other parameters also have key effects. Of particular importance is $L/L_{\rm Edd}$, which can change the structure and emission properties of the accretion flow and thus the SED shape and excitation properties of surrounding gas (Sect. \ref{sec:components}), and can also impact the detectability of AGN over emission from the host galaxy. 

Even at a fixed $L/L_{\rm Edd}$, the structure of the AGN is not universal. For example, the dusty torus can have a range of covering factors (producing a statistical difference in the intrinsic properties of objects that are observed to be type 2 as opposed to type 1: Sect. \ref{sssec:IR_hdps}), and the strength of a relativistic jet can vary, yielding very different multi-wavelength signatures particularly in the radio and $\gamma$-rays (as discussed in Sects.~\ref{sec:radio} and \ref{sec:gamma}). Further, obscuration of the AGN or dilution of stellar processes can be produced in the host galaxy, independent of the small-scale orientation of the central engine \citep[e.g.][]{chen15qsosf,buch17obs}. A more complete ``weak'' unified model is one in which all these components ($L/L_{\rm Edd}$, torus covering factor, jet strength, and host galaxy properties as well as orientation) come together to produce the observed signatures of an AGN. A schematic showing the effects of just two of these parameters, BH accretion rate (BHAR) and jet strength, is shown in Fig.~\ref{fig:schematic1}. An important distinction is that these parameters, as well as the others described above, comprise a {\em continuous} distribution. Therefore the separate AGN classes sample different parts of this distribution rather than representing truly distinct populations, extending the picture for the dominance of radio jets presented by \citet{Padovani_2016}.

\begin{figure}
\centering
\includegraphics[width=8.4cm]{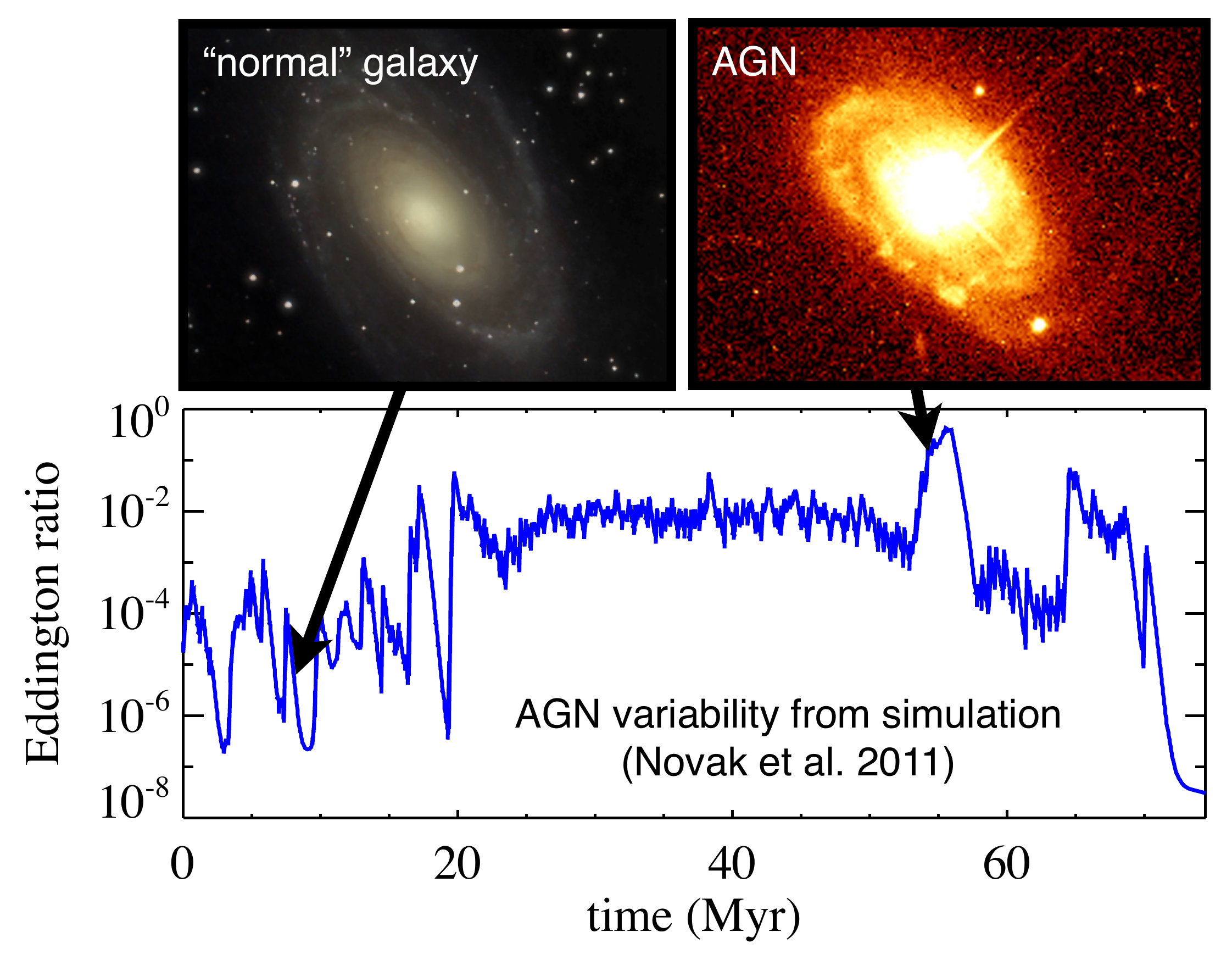}
\caption{An illustration of the stochastic fluctuations in AGN accretion rates (taken from the hydro simulations of \citealt{nova11bhsim}) on timescales shorter than those for typical galaxy evolution processes. This schematic highlights the weak connection between the instantaneous properties of the AGN and the evolution of the larger host galaxy. \copyright~AAS. Figure reproduced from  \cite{hick14agnsf}, Fig. 1, with permission.}
\label{fig:stochastic}       
\end{figure}

It is also important to note that the parameters that determine AGN properties will naturally be expected to vary with time. This has motivated another sort of ``unification'', in which AGN represent the same underlying objects viewed at different points in their evolution. One example of this is the classic merger evolution scenario, in which a galaxy merger triggers gas flows that fuel an AGN, starting off obscured until feedback clear the surroundings of the black hole producing an unobscured object \citep[e.g.][]{sanders88, hopkins08}.  While there is compelling evidence for time evolution in some cases (for example large-scale radio jets are likely to have evolved from more compact relativistic outflows) it is challenging to construct a successful unification picture based on time evolution alone. This difficulty arises from the relative time and spatial scales: while galaxy evolution processes can take $10^8$ years or more and on scales of kpc, variations in nuclear structures, gas flows, and BHARs can occur on timescales many orders of magnitude shorter (e.g. \citealt{hick14agnsf}; Fig.~\ref{fig:stochastic}) as evidenced by hydrodynamic simulations \citep[e.g.][]{nova11bhsim} and AGN light echoes \citep{Keel_2012, Schawinski_2015} that point toward large-amplitude fluctuations in AGN luminosity on timescales of $\sim 10^5$--$10^6$ years. 
Thus an AGN cannot directly ``know'' about a galaxy merger or other processes that happen on large galactic scales. While there may be some connection between galaxy-scale processes and the instantaneous observed signatures of an AGN, these will be primarily statistical in nature \citep[e.g.][]{hickox2009,goul14agn,azadi2015}.

\subsection{The emerging technicolor picture}

\begin{figure}[h]
\centering
\includegraphics[width=8.4cm]{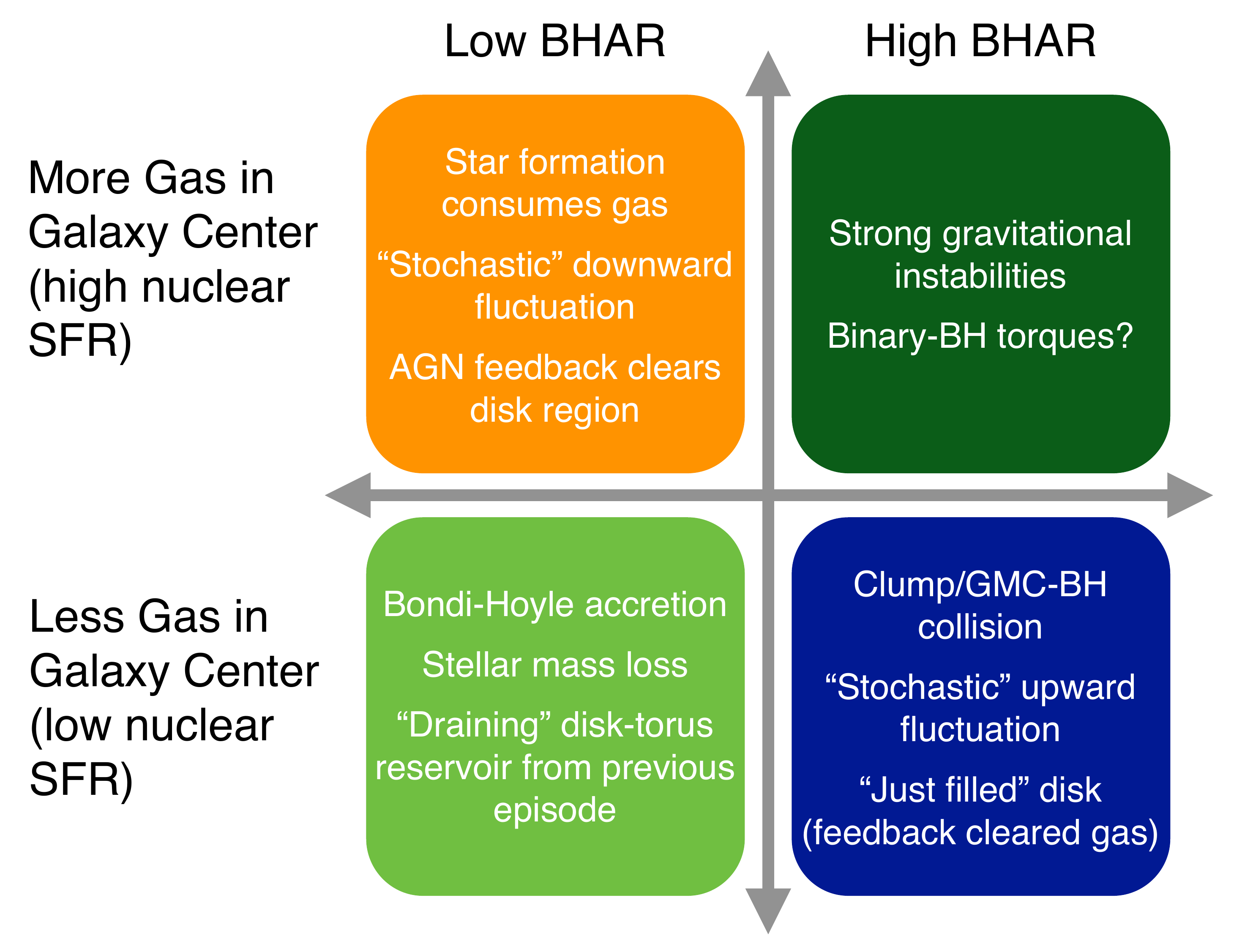}
\caption{The range of physical processes involved with BH fuelling, in different regimes of BHAR and gas supply in the host galaxy. Many observed properties of AGN can be attributed to the different physics operating in these various regimes. Adapted from schematic by P.~Hopkins.}
\label{fig:schematic2}       
\end{figure}

With this reasoning in mind, there is emerging a general picture to describe the ``technicolor'' properties of AGN and their host galaxies that is based on the statistical relationships between BH and galaxy evolution. In this picture, BHARs are broadly correlated with the supply of gas (and thus SFR) in the centres of galaxies, but can vary dramatically over several orders of magnitude in response to processes (instabilities, feedback, etc.) that occur on small scales. A schematic showing the range of physical processes involved in different regimes of BHAR and a galaxy's central gas supply is shown in Fig.~\ref{fig:schematic2}.  

\begin{figure*}
\centering
\includegraphics[width=1.0\textwidth]{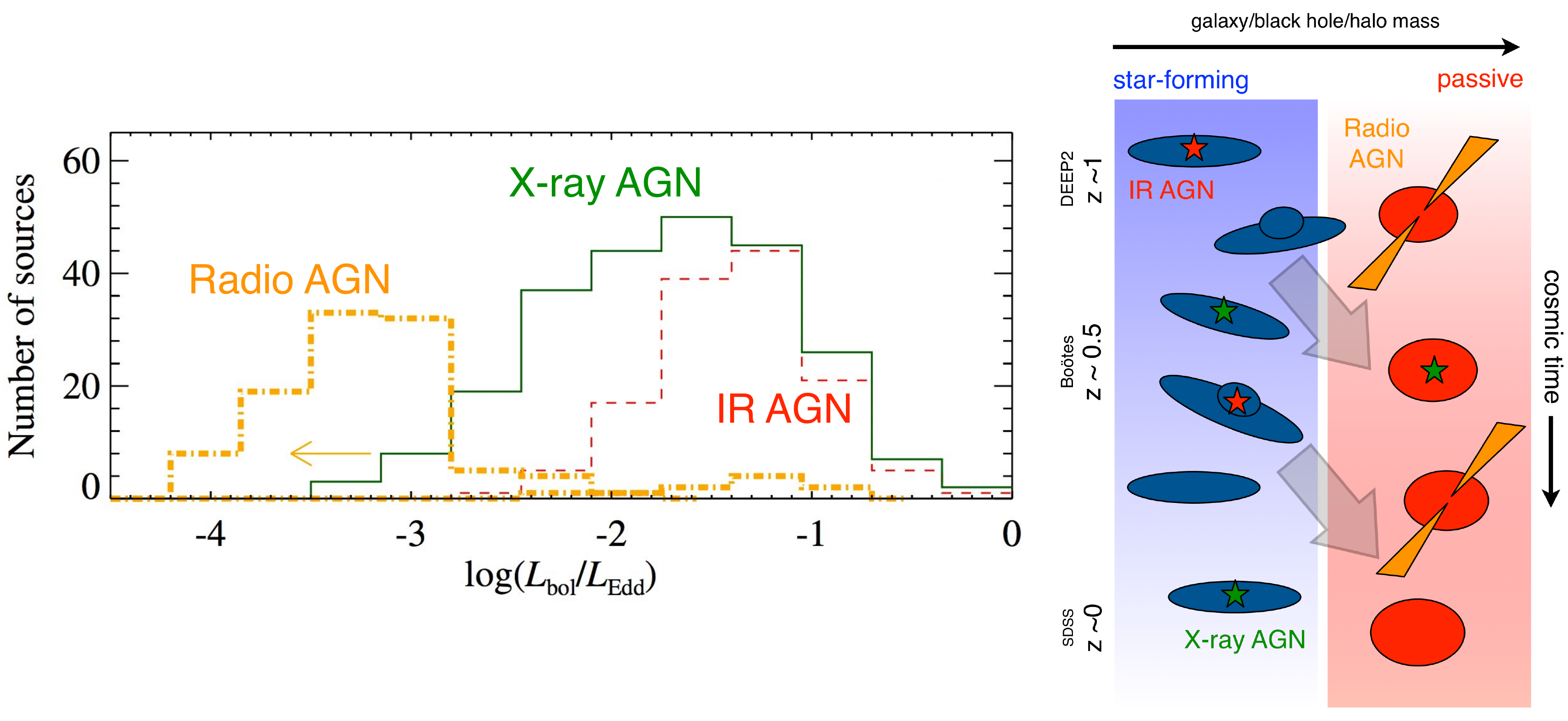}
\caption{{\em Left panel:} $L/L_{\rm Edd}$ distributions for radio, IR, and X-ray selected AGN. Adapted from \cite{hickox2009}, where one can find details on the area and depth of each survey. {\em Right panel:} A schematic picture for evolution 
of galaxies and their observed AGN classes, based on observations of AGN populations to $z\sim 1$. All galaxies appear to begin as star-forming blue-cloud systems and end as passive red-sequence sources, once their dark matter halos have 
grown sufficiently. However, we show that those galaxies hosting IR, X-ray, and/or radio AGN appear to follow a similar evolutionary path: radiatively efficient rapid BH growth (IR/X-ray AGN) appears to be linked with those galaxies 
with large supplies of cool gas, while mechanically dominated (radio) accretion is associated with passive galaxies, which may also be responsible for preventing late SF. The red and green stars represent schematically the subset of galaxies that are observed as IR- and X-ray bright AGN, respectively, in the different phases of evolution. \copyright~AAS. Figure reproduced from  \cite{goul14agn}, Fig. 7, with permission.}
\label{fig:schematic3}       
\end{figure*}

The key point is that {\em any} type of galaxy can in principle host AGN with a wide range of observational signatures, but broad correlations between gas supply and BHAR introduce some important general trends (e.g. high $L/L_{\rm Edd}$, IR-selected AGN being found in SFGs, while low $L/L_{\rm Edd}$ LERGs are found primarily in passive galaxies; Fig.~\ref{fig:schematic3}).

\subsection{The future and open questions}\label{sec:summary_future}

There are exciting prospects for future progress in studies of AGN over the coming decades. Many of the potential breakthroughs come from new observational resources, in particular observations at high resolution and sensitivity. On the theoretical side, increasingly sophisticated and complex simulations of AGN will inform the nature of the central engine and of AGN feedback. Here we briefly discuss each of these avenues for progress, and conclude with a survey of interesting open questions in the field.

{\em New observations at high resolution and sensitivity.--} One critical observational capability in studying the AGN central engine is the ability to directly resolve the relevant small scales with correspondingly high sensitivity. Recently, NIR adaptive optics and particularly interferometric observations have provided new insights about the structure of the AGN torus (Sect. \ref{ssec:IR_SED}). ALMA is already resolving the innermost parts of nearby AGN, 
such as NGC 1068 \citep{Gallimore_2016,Garcia-Burillo_2016, Imanishi_2016} and NGC 1097 (Hatziminaoglou et al. in prep., Izumi et al. in prep.) down to a few parsec scales, providing new insights into 
the dynamics near the SMBH by resolving its sphere of influence, the existence of the torus and the feeding of the SMBH, and is promising more ground-breaking results once the high frequency bands 
are offered in the longest baselines ($\sim 16$km).
ALMA, in collaboration with the Global mm-Very Large Baseline Interferometry (VLBI) Array\footnote{\url{http://www3.mpifr-bonn.mpg.de/div/vlbi/globalmm/}} and the Event Horizon Telescope\footnote{\url{www.eventhorizontelescope.org}} consortia, is also trying to directly observe for the first time the immediate environment of the SMBH in, amongst others, M 87, OJ 287, and Cen A, with angular resolution comparable to the event horizon, as part of the Cycle 4 mm-VLBI campaign,
building on previous work \citep[e.g.][]{john15ehtpol, fish16ehtagn}.
With the launch of {\em JWST} and the advent of $>$30m class ground-based telescopes, we will see a dramatic improvement in the ability to study the morphologies and spectral properties of the central engine on small scales. In the radio and submm, advances in interferometry (including along Earth-scale baselines) have the potential to bring further insights into the nature of AGN jet acceleration, mechanical feedback, and gas content and kinematics. Planned or large space missions or concepts in the X-rays (e.g. {\em Athena}, {\em Lynx}\footnote{\url{wwwastro.msfc.nasa.gov/lynx/}}) and optical/UV (e.g. HabEx\footnote{\url{www.jpl.nasa.gov/habex/}}, LUVOIR\footnote{\url{asd.gsfc.nasa.gov/luvoir/}}), along with a range of other potential missions (some of which are discussed throughout this review), will enable the study of AGN phenomena with high sensitivity and spectral and/or spatial resolution. 

{\em Large sky survey data sets.--} All areas of astronomy are being revolutionized by increasingly larger and deeper sky surveys at all wavelengths, and this flood of data has the potential for major breakthroughs in our understanding of AGN. As shown in Fig.~\ref{fig:agn_density}, X-ray surveys are the most efficient for detecting a high surface density of AGN with minimal host galaxy contamination. The next major X-ray survey will be the eRosita All Sky Survey (eRASS) by the eROSITA instrument 
(Sect. \ref{sec:X-ray_future}), which will detect $\approx$~3 million AGN over most of the sky, with ground-based optical spectroscopic follow-up from multiple planned surveys. Further deep, wide surveys with {\em Athena}, {\em Lynx}, and other future X-ray missions would probe large numbers of AGN over a wider parameter space in redshift and luminosity. 

In the optical and NIR, the {\it Euclid} and {\it WFIRST} satellite missions will perform wide-area surveys characterizing millions of AGN (Sect. \ref{ssec:IRfuture} and \ref{sec:opt_evol}), while dedicated ground-based surveys (for example from the LSST, the Subaru Prime Focus Spectrograph\footnote{\url{pfs.ipmu.jp/}}, and the Multi Object Optical and Near-infrared Spectrograph for the VLT [MOONS]\footnote{\url{www.eso.org/sci/facilities/develop/instruments/MOONS.html}}) will build on the legacy of ongoing surveys (for example DES and SDSS-IV). At longer wavelengths, the growing number of sensitive, wide-area radio telescopes (leading ultimately toward the SKA) will provide increasing numbers (in the tens of millions: Sect. \ref{sec:radio_future}) of radio-selected AGN across most of the sky. Together, these surveys will produce massive data sets with huge statistical samples of images, spectra, and time variability of AGN over a wide range in redshift and luminosity, and will enable statistical analyses of AGN with precision comparable to what is currently possible for ``normal'' galaxies. Harnessing these large data sets represents a major challenge and opportunity for the future of AGN science.

{\em Hydrodynamic simulations.--} Major theoretical challenges in understanding the AGN arise from the complexity of the physical processes involved and the enormous dynamic range in the relevant physical scales. With continual advances in computing power and improvements in astrophysical codes, theorists will continue to approach the ultimate goal of resolving the physics of AGN on all the scales described in Sect.~\ref{sec:components}, the structure and kinematics of the torus and central engine \citep[e.g.][]{chan16torus, name16torus, wada16torus}, the physics of the inner accretion flow and the launching of winds and relativistic jets \citep[e.g.][]{tche15jet, wate16wind}, and the impact of this feedback on host galaxies and halos \citep[e.g.][]{dubo16horizon, mcal17eagle, wein17agn}. Coupled with an improved understanding of the underlying physics, emission mechanisms, and prescriptions for radiative transfer \citep[e.g.][]{jud17disc, hopk17fire}, these simulations will be critical for interpreting the high-resolution, high-sensitivity observations described above.

{\em Forward modelling of selection effects.--} As discussed in depth throughout this review, an important difficulty in piecing together the AGN population is understanding the wide array of selection effects due to physical variation among AGN, host galaxy contamination, and observational limitation. Many previous studies have attempted to invert these selection effects to recover the intrinsic AGN population studied in a particular waveband. However, with a proliferation of multi-wavelength data on large samples of AGN, it is increasingly challenging to successfully account for selection effects across multiple bands simultaneously. There is therefore great potential for progress from {\em forward}-modelling these effects by producing a simulated population of AGN with various physical characteristics (e.g. luminosities, redshifts, $L/L_{\rm Edd}$, dust obscuration, and host galaxy properties) and then producing simulated SEDs and other observables for direct comparisons to data (for recent examples of this method see \citealt{jone16sdss}). 

This approach is made particularly powerful in combination with large hydrodynamic or semi-analytic simulations of galaxy formation \citep[for just a few examples, see][]{hirs14sim, scha15eagle, dubo16horizon, dima17bluetides}, which provide realistic populations of galaxies that enable a robust accounting for host contamination and connections between AGN and galaxy properties. This forward-modelling approach, informed by a better understanding of AGN physics and future high-resolution and high-sensitivity AGN observations, will move us closer to a complete ``unification'' of AGN that can account for the properties of the large AGN populations to be discovered in the next generation of surveys.

These exciting observational and theoretical resources in the coming decades will be critical in answering the many interesting open questions that remain about the AGN phenomenon. Following the structure of this review, we discuss some outstanding problems related to AGN observed using each technique, plus general questions regarding the full AGN population:

\begin{itemize}

\item {\em Radio:} What is the physical driver of AGN outflows, and in particular relativistic jets? How does the nature of these outflows change as a function of BH mass and accretion rate, and what processes (for example spin) explain the existence of jetted AGN that otherwise look (almost) identical to non-jetted AGN?

\item {\em IR:} What is the composition, geometry, and morphology of the AGN torus, and how is its structure related to the accretion flow and/or AGN feedback?

\item {\em Optical:} What is the physics of AGN emission lines and how can they be used to probe the physics of outflows? What are the emission line characteristics of high-redshift obscured quasars, and why are these currently rare in optical surveys? 

 \item {\em X-rays:} What is the origin, geometrical configuration, and energy source of the corona? How do the AGN detected in X-ray surveys (which are the most efficient method for uncovering AGN) trace the underlying full population of AGN?
 
 \item {\em $\gamma$-rays:} What is the connection between $\gamma$-ray emission from AGN and high-energy particles (neutrinos and CRs), and how can these observations constrain theoretical models for particle acceleration in AGN? 
 
 \item {\em Variability:} What is the physical driver of AGN variability as observed in different wavebands? Can X-ray variability be used as an independent indicator of BH mass?
 
 \item {\em General:} What is the complete census of the AGN population as observed at all wavebands (and those still undetected)? How does the cosmic history of BH accretion, as traced by the complete AGN population, compare to the history of SF? What are the physical connections between the evolution of BHs and that of their host galaxies and halos, and are these driven by common sources of gas, feedback processes, or other effects? 

\end{itemize}

In summary, the future of AGN studies is very bright and we will soon be flooded with exciting new data. We need to be ready to extract as much information as possible from these observations by asking the right questions and 
using the appropriate tools, so that we can most effectively piece together the physical nature of AGN and uncover the full cosmic evolution of growing black holes.

\begin{acknowledgements}
We thank the participants of the AGN Workshop ``Active Galactic Nuclei:
what's in a name?'' (see full list at \url{www.eso.org/sci/meetings/2016/AGN2016/participants.html}) 
for their presentations, Lisa Kewley, John Silverman, and Sylvain Veilleux for their role in the
SOC, Phil Hopkins for his summary talk, Chris Harrison for producing Fig. \ref{fig:SED}, and Sarah Gallagher, Darshan Kakkad, 
Andrea Merloni, Kevin Schawinski, and an anonymous referee for reading the paper and providing helpful comments. 
DMA thanks the Science and Technology Facilities Council (STFC) for support through grant
ST/L00075X/1, and James Aird and Bret Lehmer for useful critical
feedback. RJA was supported by FONDECYT grant number 1151408. 
BDM thanks D. De Cicco for helpful discussions and acknowledges support from the European Union's Horizon 
2020 research and innovation programme under the Marie Sk\l{}odowska-Curie grant agreement No. 665778 via the 
Polish National Science Center grant Polonez UMO-2016/21/P/ST9/04025. 
GTR is grateful for the support of NASA-ADAP grant NNX12AI49G, NSF grant 1411773, and the Alexander von Humboldt Foundation. 
VS acknowledges support from the European Union's Seventh Framework Programme under grant agreement 337595 (ERC Starting Grant, 'CoSMass').
\end{acknowledgements}

\addcontentsline{toc}{section}{References}

\small


\end{document}